\documentclass[11pt,eps]{report}
\usepackage{graphicx}
\usepackage{amsmath}
\usepackage{epsfig}
\usepackage{amsfonts}
\usepackage{amssymb}
\newtheorem{theorem}{Theorem}

\newtheorem{corollary}[theorem]{Corollary}

\newtheorem{definition}[theorem]{Definition}

\newcommand{\be}{\begin{equation}}
\newcommand{\ee}{\end{equation}}

\newcommand{\beq}{\begin{equation}}
\newcommand{\eeq}{\end{equation}}
\newcommand{\beqa}{\begin{eqnarray}}
\newcommand{\eeqa}{\end{eqnarray}}
\newcommand{\ba}{\begin{array}}
\newcommand{\ea}{\end{array}}

\begin{document}
%

\evensidemargin=-1truecm \textwidth=16truecm \textheight=22truecm

\begin{titlepage}

\vspace*{1.5cm}
\begin{center}
 { \LARGE WAVE PROPAGATION AND IR/UV MIXING IN NONCOMMUTATIVE SPACETIMES\footnote{PhD thesis, defended at the Physics Department of the University
  of Rome ``La Sapienza" on 20 January 2003, supervised by G. Amelino-Camelia and K. Yoshida.}}
\end{center}

\vskip0.9 cm

\begin{center}
\textbf{Gianluca Mandanici} \\
\vskip0.6 cm

\textit{Dipartimento di Fisica, Universit\`{a} di Roma ``La
Sapienza'', }

\textit{P.le A. Moro 2, I-00185 Roma, Italy}
\end{center}

\vspace{2.5cm}

\begin{center}
\textbf{ABSTRACT}
\end{center}

{\leftskip=0.6in \rightskip=0.6in\noindent}

In this thesis I study various aspects of theories in the two most
studied examples of noncommutative spacetimes: canonical spacetime
($[x_{\mu},x_{\nu}]=\theta_{\mu\nu}$) and $\kappa$-Minkowski
spacetime ($[x_{i},t]=\kappa^{-1} x_{i}$). In the first part of
the thesis I consider the description of the propagation of
``classical" waves in these spacetimes. In the case of
$\kappa$-Minkowski this description is rather nontrivial,
 and its phenomenological implications are rather striking.
 In the second part of the thesis I examine the structure
 of quantum field theory in noncommutative spacetime, with emphasis
 on the simple case of the canonical spacetime. I find that the so-called IR/UV mixing can affect significantly
 the phase structure of a quantum field theory and also forces us upon
 a certain revision of the strategies used in particle-physics
 phenomenology to constrain the parameters of a model.

\end{titlepage}


\evensidemargin=-1truecm \textwidth=16truecm \textheight=22truecm

\textwidth=5truecm

\begin{titlepage}
\vspace*{3.5cm}

{\LARGE{\textbf{Acknowledgements}}}

I thank my supervisors Giovanni Amelino-Camelia and Kensuke
Yoshida for their constant support and the advise during the
preparation of this thesis. I am especially grateful to Giovanni
Amelino-Camelia for having accepted the challenge of supervising
my thesis; his contribution has revealed crucial. A thanks goes
also to all the people with whom I had the fortune to discuss
about physics during my Ph.D. studies. Among them in particular I
wish to thank  Alessandra Agostini, Francesco D'Andrea, Luisa
Doplicher and Dario Francia. Finally I would like to thank the
referee of this thesis, Carlo Rovelli, for his very valuable
suggestions
\end{titlepage}

\begin{titlepage}
\vspace*{6.5cm} \hspace*{11.1cm} \emph{\large{ To Daniela}}
\end{titlepage}

\tableofcontents
\hoffset=-.0truecm \voffset=-1.2truecm
\baselineskip6.6mm\headheight=12pt \headsep=36pt \footskip=48pt
\textwidth=15truecm \textheight=22truecm
\setlength{\oddsidemargin}{0pt} \setlength{\textwidth}{16.2cm}
\setlength {\topmargin}{-0.35in} \setlength{\textheight}{22.6cm}
\textwidth=16truecm
\textheight=24truecm%



\chapter*{Introduction}

\addcontentsline{toc}{chapter}{Introduction}

\qquad It is widely believed that spacetime at lengths scales of
the order of the Planck length $(L_{p}=\sqrt{\hslash
G/c^{3}}\simeq10^{-33}cm)$\textbf{ }is no longer describable as a
smooth manifold. Nonclassical properties of spacetime are expected
to affect processes involving particles of ultrahigh energy. There
are in principle, at least, two ways to address the issue of
nonclassical properties of spacetime and their observable effects.
A first way, which is in a sense more fundamental, is the one of
trying to construct a whole quantum theory of spacetime according
to some picture of the unification of General Relativity with
Quantum Mechanics, and then look at its low-energy predictions.
This is, for instance, the strategy adopted by the most studied
and, presently, most promising approaches to Quantum Gravity, such
as string theory and (canonical) loop quantum gravity. A possible
reason of concern for these type of approaches is that one has to
make a correct guess of the laws of Nature at an energy scale
$(E_{p}=\frac{1}{L_{p}}\simeq10^{19}Gev)$ that is very far away
from the energy scales currently explored in the laboratory
$(E_{\exp}\simeq10^{2}Gev)$. This is a difficult task as the one
of trying to grasp the details of weak interaction
$(E_{W}\simeq10^{2}Gev)$ just by studying the properties of the
common macroscopic objects of everyday life$.$

A second, more humble way to approach research on nonclassical
properties of spacetime is the one of effective theories. One
tries to model some nonclassical spacetime effects without
necessarily assuming full knowledge of the short-distance
structure of spacetime. Among these proposals there has been
strong interest in the idea that it might be fundamentally
impossible to fully specify the position of a particle. This can
formalized through a
spacetime uncertainty principle of the form%
\begin{equation}
\Delta x_{\mu}\Delta x_{\nu}\geq\theta_{\mu\nu},
\end{equation}
that would introduce in spacetime an uncertainty relation which is
analogous to the Heisenberg phase-space uncertainty relation
$\Delta x\Delta p>\hslash/2$. At the formal level the Heisenberg
uncertainty principle turns out to be described by Hermitian
operators satisfying the noncommutativity relation
$[x,p]=i\hslash.$ Following the analogy one is led to consider
similar commutation relations in the spacetime sector%
\begin{equation}
\left[  x_{\mu},x_{\nu}\right]  =i\theta_{\mu\nu}(x), \label{gst}%
\end{equation}
which imply noncommutativity of spacetime.

In the last few years noncommutative spacetimes have attracted
interest from many authors not only for the reasons we have
outlined, which make them interesting on their own, but also since
they emerge as a possible description of spacetime in theory
constructed without assuming, a priori, noncommutativity of
spacetime. In particular the so-called canonical
spacetimes\footnote{Here $\theta_{\mu\nu}$ is a
coordinate-independent antisymmetric matrix.}
\begin{equation}
\left[  x_{\mu},x_{\nu}\right]  =i\theta_{\mu\nu},\label{gst1}%
\end{equation}
emerge in the descriptions of string theory in presence of
external fields~\cite{hep-th/9711162,hep-th/9908142}, and it is
also a useful-alternative tool in the description of electronic
systems in external-magnetic
field~\cite{peierls,physics/0209108,dunnejack}. While, certain
Lie-algebra noncommutative spacetime ($\left[
x_{\mu},x_{\nu}\right] =iC_{\mu\nu}^{\alpha}x_{\alpha}$), emerge
in the framework of some approaches to Quantum Gravity that
predict a minimum wavelength. This is, for instance,
the case of the so-called $\kappa$-Minkowski spacetime%
\begin{align}
\left[  x_{i},x_{0}\right]   &  =i\dfrac{1}{\kappa}x_{i},\label{mnc}\\
\lbrack x_{i},x_{j}] &  =0,\nonumber
\end{align}
that is connected by duality relation to a deformation of the
Poincar\'{e} group as a quantum group known as
$\kappa$-Poincar\'{e}. This $\kappa $-Poincar\'{e} quantum group
has been extensively studied in
literature~\cite{lnrt,hep-th/9312153,hep-th/9405107,lukie,hep-th/9907110,hep-th/0207022}
especially since it involves not only an invariant velocity scale
$c,$ but also an invariant length scale
$\lambda=\dfrac{1}{\kappa}\simeq L_{p}$.

In this thesis we will analyze the popular noncommutative
spacetimes of Eqs.(\ref{gst1}) and (\ref{mnc}), focusing on some
key theoretical issues and their phenomenological implications. We
will start our study from the problem of wave construction and
propagation in these types of noncommutative spacetime. The
analysis of waves is a key element for planned experimental
studies which hope to detect nonclassical effects of spacetime. It
is expected in
fact~\cite{gr-qc/9910089,astro-ph/0004225,gr-qc/0204051,gr-qc/0204092,gr-qc/0205121}
that spacetime noncommutativity might manifest with detectable
modifications of the usual laws of particle production and
propagation. Moreover, over the last few years, there has been a
sharp increase in the interest toward experimental investigations
of Planck-scale effects (see, \textit{e.g.},
Refs.~\cite{astro-ph/9712103,gr-qc/9810044,gr-qc/9808029,astro-ph/9904164,
astro-ph/0008107,gr-qc/0107086,gr-qc/0006061,gr-qc/0110034}). In
particular, studies such as the ones planned for the GLAST space
telescope~\cite{astro-ph/0009271} would be sensitive to small,
Planck-scale suppressed, deviations from the special-relativistic
relation between group velocity and momentum. Within the framework
of Planck-scale spacetime noncommutativity such modifications of
the relation between group velocity and momentum are often
encountered. In this respect, while canonical spacetime is
expected to play a minor role (alterations of in-vacuum
propagation seem to be relevant mainly for polarization-connected
effects~\cite{hep-th/0109191}), such effects are expected to have
deeper implications in $\kappa$-Minkowski spacetime, where they
are believed to manifest in the deformation of in-vacuum
dispersion relations and (related) modifications of the energy
thresholds for certain particle-production
processes~\cite{astro-ph/0008107, gr-qc/0107086}. The effect of
spacetime discreteness should be significant for particles of high
energy. Deviations from standard propagation are expected to be
suppressed as powers of the ratio between Planck length and
particle wavelength, but in spite of this huge suppression the
mentioned planned experiment should be able to see them. In
preparation for these planned experimental studies, on the theory
side there has been an intense debate on the proper description of
the notion of velocity in a noncommutative spacetime. Here (in
Chapter 2) we analyze velocity in $\kappa$-Minkowski spacetime.
This notion of velocity had been already discussed in several
studies (see Refs.~\cite{hep-th/9312153,hep-th/9907110} and
references therein), under the working assumption that the
relation $v=dE(p)/dp$, which holds in Galilei spacetime and
Minkowski spacetime, would also hold in $\kappa$-Minkowski. This
leads to interesting predictions as a result of the fact that,
upon identification of the noncommutativity scale $1/\kappa$ with
the Planck length $L_{p}$, the dispersion relation $E(p)$ that
holds in $\kappa$-Minkowski is characterized by
Planck-length-suppressed deviations from its conventional
commutative counterpart. Recently the validity of $v=dE(p)/dp$ in
$\kappa$-Minkowski has been questioned in the studies reported in
Refs.~\cite{hep-th/0107054,gr-qc/0111056}; moreover in the study
reported in Ref.~\cite{hep-th/0207022} the relation $v=dE(p)/dp$
was considered on the same footing as some alternative relations.
Especially in light of the plans for experimental studies, this
technical issue appears to be rather significant. We approach the
study of $\kappa$-Minkowski adopting the line of analysis proposed
in Ref.~\cite{hep-th/9907110}. We argue that key ingredients for
the correct derivation of the relation between group velocity and
momentum are: a fully developed $\kappa$-Minkowski differential
calculus, and a proper description of energy-momentum in terms of
generators of translations. Our analysis provides support for the
adoption of the formula $v=dE(p)/dp$, already assumed in most of
the $\kappa$-Minkowski literature. We discuss the \textit{ad hoc}
assumptions which led to alternatives to $v=dE(p)/dp$ in
Refs.~\cite{hep-th/0107054,gr-qc/0111056}, and we find that the
analysis in Ref.~\cite{gr-qc/0111056} was based on erroneous
implementation of the $\kappa$-Minkowski differential calculus,
while the analysis in Ref.~\cite{hep-th/0107054} interpreted as
momenta some quantities which cannot be properly described in
terms of translation generators.

We also discuss the proposals of construction of field theories on
noncommutative spacetimes based on the Weyl-Moyal map. We focus
mainly on the canonical spacetime which up to now has been the
most extensively studied. A key characteristic of field theories
on canonical spacetimes, which originates from the commutation
relation, is nonlocality. At least in the case of space/space
noncommutativity ($\theta_{0i}=0$), to which we limit our analysis
for simplicity\footnote{The case of space/time noncommutativity
($\theta _{0i}\neq0$) is not necessarily void of
interest~\cite{hep-th/0005129,hep-th/0005040,hep-th/0201222,hep-th/0206011,hep-th/0201144}%
, but it is more delicate, especially in light of possible
concerns for unitarity. Since our analysis is not focusing on this
point we will simply assume that $\theta_{0i}=0$.}, this
nonlocality is still tractable although it induces an intriguing
mixing of the ultraviolet and infrared sectors of the theory. This
IR/UV mixing has wide implications both for the phenomenology and
for the theoretical understanding of these models. One of the
manifestations of the IR/UV mixing is the emergence of infrared
(zero-momentum) poles in the one-loop two-point functions. In
particular one finds a quadratic pole for integer-spin particles
in non-SUSY theories~\cite{hep-th/9912072}, while in SUSY theories
the poles, if at all present, are
logarithmic~\cite{hep-th/0002075,hep-th/0005272,hep-th/0011218}.
It is noteworthy that these infrared singularities are introduced
by loop corrections and originate from the ultraviolet part of the
loop integration: at tree level the two-point functions are
unmodified, but loop corrections involve the interaction vertices,
which are modified already at tree level.

There has been considerable work attempting to set limits on the
noncommutativity parameters $\theta$ by exploiting the
modifications of the dressed/full
propagators~\cite{hep-th/0109191,hep-th/0105082} and, even more,
the modifications of the interaction
vertices~\cite{hep-ph/0112003,hep-ph/0106356,hep-ph/0205040}. Most
of these analyses rely on our readily available low-energy data.
The comparison between theoretical predictions and experimental
data is usually done using a standard strategy (the methods of
analysis which have served us well in the study of conventional
theories in commutative spacetime). We are mainly interested in
understanding whether one should take into account some of the
implications of the IR/UV mixing also at the level of comparing
theoretical predictions and data. It appears plausible that the
way in which low-energy data are used to constrain the
noncommutativity parameters may be affected by the IR/UV mixing.
These limits on the entries of the $\theta$ matrix might not have
the usual interpretation: they could be seen only as ``conditional
limits'', conditioned by the assumption that no contributions
relevant for the analysis are induced by the ultraviolet. The
study we report here is relevant for this delicate issue. By
analyzing a simple noncommutative Wess-Zumino-type model, with
soft supersymmetry breaking, we explore the implications of
ultraviolet supersymmetry on low-energy phenomenology. Based on
this analysis, and on the intuition it provides about other
possible features of ultraviolet physics, we provide a
characterization of low-energy limits on the noncommutativity
parameters.

To explore the consequences of the IR/UV mixing using a
nonperturbative technique (effectively resumming infinite series
of 1PI Feynman diagrams) we present an application of the
Cornwall-Jackiw-Tomboulis (CJT)~\cite{CJT} formalism to the
noncommutative scalar theory. The CJT formalism has proven to be a
powerful nonperturbative approach to the problem of phase
transition in QFT in commutative spacetime (see
Refs.~\cite{CJT,hep-ph/9311324}) and in Thermal-Quantum-Field
Theories~\cite{hep-ph/9311324,hep-ph/9211211,hep-ph/9305222}.
These theories suffer from severe infrared problems which recall
those related to the IR/UV mixing in noncommutative theories. We
analyze in the CJT\ formalism the issues of phase transitions and
renormalizability of (canonical) noncommutative
scalar-$\varphi^{4}$ theory. We discuss the applicability of the
CJT formalism in a \ noncommutative framework. Then we focus on
the so-called Hartree approximation that is equivalent to a
selective resummation of the diagrams of the common perturbative
expansion, in particular summing all the so-called ``daisy'' and
``super daisy'' diagrams. In the Hartree approximation, and under
the hypothesis of translational invariance of the vacuum, we
calculate the effective potential that will be expressed in terms
of a mass parameter to be determined as solution of a gap
equation. In particular, we analyze the renormalizability of the
gap equation and of the potential. We find that whereas in the
symmetric phase (characterized by a zero vacuum-expectation value
of the scalar field) both the gap equation and the potential are
renormalizable, in the broken-symmetry phase (characterized by a
non-zero vacuum-expectation value of the scalar field) the gap
equation and the potential do not renormalize. These results
appear to reinforce the hypothesis that in noncommutative theories
because of the IR/UV mixing there might be a stable (or
quasi-stable) translation-non-invariant vacuum.

This thesis is structured as follows. In Chapter~1, that is a
review chapter, we introduce spacetime noncommutativity focusing
on canonical noncommutativity and $\kappa$-Minkowski
noncommutativity. We discuss the physical frameworks from which
they emerge and study their symmetries, and we also briefly review
the (rather technical) mathematical structures that are used in
the rest of the thesis.

In Chapter~2, that is based on Ref.~\cite{hep-th/0211022}, we
analyze wave propagation both in canonical and $\kappa$-Minkowski
spacetimes. We focus on the concept of group velocity in view of
its relevance for planned experimental studies and stress the
differences between the two noncommutative spaces.

In Chapter~3, that is a review chapter, we introduce the quantum
field theories in noncommutative spacetime. We outline the
standard strategy of quantization, based on the Weyl-Moyal map,
and discuss its application to the noncommutative spacetimes.
While this procedure of quantization can be successful implemented
in canonical-noncommutative spacetime, in $\kappa $-Minkowski
spacetime it is still not well understood. Then, for the rest of
the chapter (and of the thesis) we focus only on canonical
spacetime and especially on the problems related to the IR/UV
mixing.

In Chapter~4, that is based on Ref.~\cite{hep-th/0209254}, we
study the implications of the IR/UV mixing for the phenomenology.
We discuss how the IR/UV mixing renders the low-energy prediction
sensitive to the (unknown) structure of the UV sector. We
illustrate our point in a Wess-Zumino model with soft
supersymmetry breaking. We show how ultraviolet supersymmetry can
modify drastically the low-energy predictions of the theory, and
analyze the implications of these observations for studies
attempting to constrain the non-commutativity parameters using
low-energy data. We conclude on the implication of this analysis
for the selection of reliable theoretical models from
low-energy-experimental data.

In Chapter~5, that is based on Ref.~\cite{gmgacld}, we review the
CJT formalism and apply it to the noncommutative scalar theory. We
analyze the effective potential and discuss the problem of its
renormalization both in the symmetric phase and in the
uniformly-broken phase. We focus on the so-called ``bubble
approximation''. We study under the hypothesis of translational
invariance of the vacuum, the gap equation and the effective
potential discussing also the planar limit (strong
noncommutativity) and the commutative limit. Then we discuss the
renormalization of both the gap equation and the potential in the
symmetric and in the broken-symmetry phase. In Chapter~6 we
present our conclusions.




\chapter{Noncommutativity in physics}

\footnotetext{* In this Chapter we review the issue of
noncommutativity in physics.}*\qquad The idea of extending to the
spacetime sector noncommutativity of the phase space is rather
old. The first paper on this subject was published by Snyder in
1947~\cite{sny} although the first proposal of noncommutativity
among coordinates is attributed to Heisenberg in the late 1930s.
Heisenberg hoped that noncommutativity would improve the
short-distance singularities typical of the quantum field theories
while extending the uncertainty relations to the coordinate
sector. Apparently~\cite{physics/0209108} Heisenberg suggested
this idea to Peierles who used it in a phenomenological approach
to the study of electronic systems in an external
field~\cite{peierls}. Then Peierles told about Heisenberg's idea
to Pauli, who in turn told to Oppenheimer, who told it to his
student Snyder.

In this chapter we briefly review the main arguments that, beyond
Heisenberg's idea, support the introduction of noncommutativity in
the spacetime. Then we will focus on canonical spacetime and
$\kappa$-Minkowski spacetime.

The interest for the canonical spacetime is motivated by the fact
that, since it involves coordinate-independent commutators, it is
the simplest noncommutative spacetime and can also be seen as the
zeroth-order approximation of a very general class of
noncommutative spacetimes. We briefly discuss how canonical
noncommutativity plays a role in the description of electronic
systems in strong magnetic field and how it emerges in certain
string theories in strong external field.

The interest in $\kappa$-Minkowski spacetime comes from the fact
that it is dual to a quantum deformation of the Poincar\'{e} group
which has recently attracted a lot of interest especially for the
relevance in Double Special Relativity theories. We discuss the
basic features of such Double Special Relativity (DSR) theories.
We analyze how DSR theories find precise realization in the
mathematical scheme of $\kappa$-deformed (quantum) Poincar\'{e}
group and we discuss the duality relation that leads to $\kappa
$-Minkowski spacetime. Then we also discuss how $\kappa$-Minkowski
spacetime and canonical spacetime emerge as the only possible flat
noncommutative spacetimes that can be constructed from a general
Lie-algebra spacetime and from the hypothesis of invariance under
undeformed spatial rotations ($\kappa $-Minkowski spacetime) or
undeformed translations (canonical spacetime).

\section{General arguments for noncommutativity in spacetime}

\qquad As mentioned the first studies of spacetime
noncommutativity were motivated by the analogy with quantum
mechanics and by the idea that one could use noncommutativity of
spacetime coordinates to introduce an effective cutoff to cure the
divergences appearing in quantum field theories. We will analyze
the issue of the regularization of the divergences in the next
chapters when a formulation of the quantum field theory will have
been introduced. Here we consider the analogy with quantum
mechanics.

Starting from the classical phase space in which a point can be
localized without any limit of precision, one can define the
quantum phase space by substituting canonical positions and
momenta $x_{i}$ and $p_{i}$ with Hermitian operators which obey to
the Heisenberg commutation relations
$[x_{i},p_{j}]=i\hslash\delta_{ij}.$ From the uncertainty relations%
\begin{equation}
\Delta x_{i}\Delta p_{j}\geq\frac{1}{2}\hslash\delta_{ij}\label{hi}%
\end{equation}
follows that in a quantized phase-space there exists a maximum in
the precision with which a point can be localized. Upon
quantization the classical phase-space becomes smeared out and the
notion of point must be replaced by that of Planck cell whose
characteristic size is $O(\hslash)$. One expects to become
sensitive to the quantization of the phase space when the action
of the system under consideration, which has the same dimensions
of $\hslash,$ is of the size of the Planck constant\footnote{In
the path integral language one obtains sensitivity to the quantum
phase-space when $S/\hslash\sim1$ and also the paths different
from the classical one ($\delta_{x}S=0$) give contributions to the
dynamics.}.

By analogy with the phase-space quantization one can attempt to
represent spacetime by replacing classical coordinates $x_{i}$
with the Hermitian generators of a noncommutative algebra of
functions satisfying the commutation
relation%
\begin{equation}
\left[  x_{\mu},x_{\nu}\right]  =i\theta_{\mu\nu}. \label{ngg}%
\end{equation}
Since coordinates do not commutate they cannot be simultaneously
diagonalized and one can expect spacetime uncertainty relations of
the form
\begin{equation}
\Delta x_{i}\Delta x_{j}\geq\frac{1}{2}|\theta_{ij}| \label{crs}%
\end{equation}
that, as already emphasized in the Introduction, may be
appropriate for a description of quantum spacetime fluctuations.

In this case the notion of spacetime point loses its meaning.
Spacetime points are replaced by cells of area of size
$|\theta_{ij}|.$ The quantum fluctuations of the spacetime prevent
the exact localization of the events inside this area. Now we
could expect to acquire sensitivity to the quantum structure of
spacetime when the size of the system, or of the probe, is of
order of $\sqrt{|\theta_{ij}|}$\footnote{The nature of the
uncertainty principle implied by (\ref{ngg}) is still subject of
study. These commutation relations don't imply by themselves a
minimal length, in the same way in which (\ref{hi}) doesn't
prevent exact measures of position or momentum. Each edge of the
quantum cell can be arbitrarily small if the other is accordingly
large. Only their product is bounded from the below. For this
reason in the case of (\ref{ngg}) one should perhaps speak of a
minimal area.}.

Expectations similar to (\ref{crs}) for the description of
spacetime at extremely small distances come also from string
theory presently one of the best candidates for a quantum theory
of gravity. Strings possess an intrinsic length-scale $l_{s}$, and
using string states as a probe it is not possible to observe
distances smaller than $l_{s}$. It is not surprising that in
certain string theories~\cite{ven,grmen,amcive} (but also in many
other approaches to quantum gravity,
see~\cite{gr-qc/9403008,hep-th/9301067}) modified Heisenberg
uncertainty relation has been found of the form%
\begin{equation}
\Delta x\gtrsim\frac{\hslash}{2}\left(  \frac{1}{\Delta
p}+l_{s}^{2}\Delta
p\right)  \label{SHIR}%
\end{equation}

If one minimizes (\ref{SHIR}) with respect to $\Delta p$ one
obtains a lower bound on the measurability of lengths in
spacetime: $\Delta x\simeq l_{s}.$ Therefore also from the point
of view of string theory the space of the configurations is
smeared out and the notion of point is meaningless. More generally
uncertainty relation has been
postulated~\cite{dofrero,hep-th/0004074} of the form
\begin{equation}
\Delta x_{i}\Delta x_{j}\geq l_{s}^{2}%
\end{equation}
that directly follows from noncommutativity of spacetime
(\ref{ngg}).

\section{Examples of noncommutative canonical spacetimes}

\qquad Beyond these general arguments there are explicit examples
in physics in which noncommutativity plays an important role and
some theoretical frameworks in which noncommutativity is not
assumed a priori, but eventually follows from the analysis. In
this section we want to discuss some of these examples showing how
canonical noncommutativity of coordinates arises.

\subsection{Canonical noncommutativity in condensed matter systems}

\qquad A first physical system which exhibits noncommutativity
comes from condensed matter (see, e.g.,~\cite{hep-th/0110057}).
Let us consider a point-particle moving on a 2-d plane\footnote{In
condensed matter certain systems are effectively 2-d in space.} in
presence of an external magnetic field $\vec{B}$ perpendicular to
the plane. The equation of motion for the
particle is%
\begin{equation}
m\text{
}\dot{v}^{i}=\frac{e}{c}\varepsilon^{ij}v^{j}|\vec{B}|+f^{i}(\vec
{r})\label{eq1}%
\end{equation}
where $\vec{r}\equiv(x,y)$ gives the position of the particle,
$\vec{v}$ is the particle velocity and $\vec{f}$ are some external
conservative forces: $\vec{f}=-\vec{\bigtriangledown}V.$

In the limit of strong magnetic field $|\vec{B}|\gg m$ the
equation
(\ref{eq1}) becomes%
\begin{equation}
v^{i}=\frac{c}{e|\vec{B}|}\varepsilon^{ij}f^{j}(\vec{r}).
\end{equation}

This last equation predicts that the particle moves always in a
direction orthogonal to that of the external force (for example if
the force is constant along the $x$ axis, the particle moves with
a constant velocity along the $y$ axis. If the force vanishes, the
particle stops).

Equation (\ref{eq1}) is also simply obtained from the observation
that the Lagrangian of the system is
\begin{equation}
\mathcal{L}=\frac{1}{2}mv^{2}+\frac{e}{c}\vec{v}\cdot\vec{A}-V.
\end{equation}

In the gauge $\vec{A}=(0,Bx)$, setting $m=0$ we have
\begin{equation}
\mathcal{L}=\frac{e}{c}x\dot{y}-V(x,y)
\end{equation}

This expression is of the form $\mathcal{L=}p\dot{q}-h(p,q)$ where
($\frac {e}{c}x,y$) are a canonical pair. This implies that the
system can be
equivalently described by an Hamiltonian%
\begin{equation}
H_{0}=V
\end{equation}

and variables satisfying (Poisson) relations%
\begin{equation}
\left\{  x^{i},x^{j}\right\}
=\frac{c}{e|\vec{B}|}\varepsilon^{ij}.
\label{parp}%
\end{equation}

We see that, in the strong field limit, the original system is
equivalent to another system in which no external field is present
but with (Poisson) noncommuting coordinates. What happens in the
strong field limit is that the Lagrangian becomes of the first
order in time derivatives so that the original-commutative
coordinate space can be viewed as an effective-noncommutative
phase-space (\ref{parp}).

These arguments can be extended to the quantum level simply
substituting the
Poisson brackets with commutators. In this case (\ref{parp}) becomes%
\begin{equation}
\lbrack x^{i},x^{j}]=\frac{i\hslash c}{e|\vec{B}|}\varepsilon^{ij}%
\end{equation}
and the strong magnetic field limit corresponds to the projection
on the lowest Landau level. This is a first example of canonical
noncommutativity which involves constant (spacetime independent)
commutation relations.

\subsection{Canonical noncommutativity in string theories}

\qquad Besides being relevant for electronic systems in external
magnetic field, canonical noncommutativity has recently attracted
a lot of interest since it emerges also from certain limits of
string theory. As an example we briefly discuss here bosonic open
string in a flat space in presence of a constant $B$-field and
Dp-branes~\cite{hep-th/9908142}. If we indicate with $r$ the rank
of the matrix $B_{ij}$, we can assume that $r\leq p+1$ since the
component of the field $B$ not along the brane can be gauged away.
Denoting
with $\Sigma$ the stringy worldsheet, the stringy action reads:%
\[
S=\frac{1}{4\pi\alpha^{\prime}}\int_{\Sigma}\left\{  g_{ij}\partial_{a}%
x^{i}\partial^{a}x^{j}-2\pi
i\alpha^{\prime}B_{ij}\varepsilon^{ab}\partial
_{a}x^{i}\partial_{b}x^{j}\right\}  ,
\]
that can be also written as%
\begin{equation}
S=\frac{1}{4\pi\alpha^{\prime}}\int_{\Sigma}g_{ij}\partial_{a}x^{i}%
\partial^{a}x^{j}-\frac{i}{2}\int_{\partial\Sigma}B_{ij}x^{i}\partial_{t}%
x^{j}, \label{SA}%
\end{equation}
where $\partial_{t}$ is the tangent derivative along the string
worldsheet boundary $\partial\Sigma.$ The equations of motion for
$i$ along the Dp branes
are:%
\begin{equation}
\left.  g_{ij}\partial_{n}x^{j}+2\pi
i\alpha^{\prime}B_{ij}\varepsilon ^{ab}\partial_{t}x^{j}\right|
_{\partial\Sigma}=0,
\end{equation}
where $\partial_{n}$ is a normal derivative to the worldsheet
boundary $\partial\Sigma.$

In the strong field limit the propagator becomes%
\begin{equation}
\left\langle x^{i}(\tau)x^{j}(\tau)\right\rangle =-\alpha^{^{\prime}}%
G_{ij}\log(\tau-\tau^{^{\prime}})^{2}+\frac{i}{2}\theta^{ij}\epsilon(\tau
-\tau^{^{\prime}})
\end{equation}

This equation is incompatible with classical (commutative)
coordinates and one
can show~\cite{hep-th/9908142} that this implies%
\begin{equation}
\lbrack x^{i}(\tau),x^{j}(\tau)]=i\theta^{ij}.
\end{equation}

In the strong field limit the first term in (\ref{SA}) becomes
negligible so
that the action takes the form%
\begin{equation}
S=-\frac{i}{2}\int_{\partial\Sigma}B_{ij}x^{i}\partial_{t}x^{j},
\end{equation}
which is purely topological. The boundary degrees of freedom
become the dominant ones.

This is a second example of canonical noncommutativity. We observe
that, though in a different context, the mechanism which generates
this type of noncommutativity is always the same. A strong field
renders the action topological (i.e. dependent only on a first
derivative) and through the equation of motion the original
commutative space transforms in an effective noncommutative
phase-space.

\subsection{Breakup of Lorentz symmetry in canonical spacetimes}

\qquad We now want to comment briefly on the fate of the classical
(Poincar\'{e}) symmetries in canonical noncommutative spacetime
$\left[ x_{\mu},x_{\nu}\right]  =i\theta_{\mu\nu}.$ As already
emphasized canonical noncommutativity is the weakest form of
noncommutativity, in the sense that it involves coordinate
independent commutation relations, and can be viewed as the
zeroth-order approximation of the more general form $\left[
x_{\mu },x_{\nu}\right]  =i\theta_{\mu\nu}(x)$. Here we suppose
that the infinitesimal action of the Poincar\'{e} group on the
coordinates is not changed by noncommutativity (i.e.
$M_{j}\triangleright x_{\mu}=[M_{j},x_{\mu }],$
$N_{j}\triangleright x_{\mu}=[N_{j},x_{\mu}],$
$P_{\alpha}\triangleright
x_{\mu}=[P_{\alpha},x_{\mu}]$). It can be easily verified that%
\begin{align}
M_{j}\triangleright\left[  x_{\mu},x_{\nu}\right]   &  \neq M_{j}%
\triangleright\left(  i\theta_{\mu\nu}\right)  ,\\
N_{j}\triangleright\left[  x_{\mu},x_{\nu}\right]   &  \neq N_{j}%
\triangleright\left(  i\theta_{\mu\nu}\right)  ,\\
P_{j}\triangleright\left[  x_{\mu},x_{\nu}\right]   &
=P_{j}\triangleright \left(  i\theta_{\mu\nu}\right)  .
\end{align}

These expressions show that canonical spacetime is not, in
general, covariant under rotation and boosts while it is covariant
under translations. Physically this means that there is not
equivalence between different inertial observers (in the usual
sense) since there is a preferred class of reference frames.
Moreover in each frame of this class there is a preferred
direction (which breaks rotations) but not a preferred origin
(translations symmetry is preserved). It is also worth noticing
that if noncommutativity does not involve time (i.e.
$\theta_{\mu0}=0$) and one only considers infinitesimal boosts
then
\begin{equation}
N_{j}\triangleright\left[  x_{\mu},x_{\nu}\right]
=N_{j}\triangleright\left( i\theta_{\mu\nu}\right)  .
\end{equation}

Actually in this case we have also covariance under finite
non-relativistic boosts.

These mathematical results are in good agreement with the physical
picture of canonical noncommutativity that comes from the
description of charged particles in strong magnetic field. The
covariance under boosts is lost since we are not considering
transformation of the magnetic field ($\theta_{\mu\nu}$ in the
noncommutative analogy), the rotational covariance is lost because
the magnetic field selects a preferred direction while, the
uniformity of the magnetic field (spacetime independent
$\theta_{\mu\nu}$) ensures translational covariance.

\section{Example of Lie-algebra noncommutative spacetime: $\kappa$-Minkowski spacetime}

\qquad In this section we discuss a much-studied example of
Lie-algebra noncommutative spacetime which turns out to play a
role in attempts to describe Planck length $L_{p}$ as a minimum
wave-length $\lambda$. We first show how to implement such
minimum-wavelength condition in a way that is compatible with the
relativity postulates. Then we discuss the quantum group that can
provide the corresponding mathematical framework and show that
$\kappa$-Minkowski Lie-algebra spacetime is dual to this quantum
group.

\subsection{Planck length as a minimum wave-length}

\qquad We have already discussed how there exist many different
approaches to the unification of General Relativity with Quantum
Mechanics which share the prediction of a minimum length. The
existence of a minimum length is not by itself in contrast with
the Poincar\'{e} symmetry, as the quantization of the angular
momentum is not in contrast with the invariance under continuous
rotations. However if one wants to promote the Planck length to an
invariant length, in the same sense in which the speed of light is
an invariant velocity, one finds immediately inconsistence with
Special Relativity. As the introduction of an invariant velocity
scale necessarily induces some modifications in the Galilei group,
so the introduction of a invariant length scale necessarily leads
to modifications of the Lorentz group. Not all the elements of the
Lorentz group will require modifications. For example, as
intuition suggests, the idea of an invariant length scale is not
in contrast with symmetry under spatial rotations therefore we
should expect an unmodified rotation's sector. Deep modifications
instead will be unavoidable in the boost sector, where the
invariant length scale must appear as a deformation scale. In fact
if the Planck scale is treated not as a (rescaled) coupling
constant but as an observer-independent scale then one becomes in
contrast with the Lorentz-Fitzgerald length contraction that
prohibits observer-independent lengths.

Relativistic theories that admit two invariant scales, a velocity
scale $c$ and a length scale $L_{p}$ were first introduced in the
papers~\cite{lnrt,hep-th/9312153} and have been extensively
studied in the recent literature (see e.g.~\cite{hep-th/9405107,
hep-th/9907110, hep-th/0107054, hep-th/0012238, gr-qc/0106004,
hep-th/0203040}). These theories, commonly called ``Doubly Special
Relativity'' (DSR) theories, can be formulated just in the same
way in which Special Relativity is formulated: by introducing some
corresponding postulates. One example of DSR postulates whose
logical consistency has been analyzed in details
in~\cite{gr-qc/0012051} is the following

\begin{itemize}
\item  The laws of physics involve a fundamental velocity $c$ and
a fundamental length scale $L_{p}$\emph{.}

\item  The value of the fundamental velocity scale $c$ can be
measured by each inertial observer as the speed of light with
wavelength $\lambda$ much larger
than $L_{p}$ (more rigorously, $c$ is obtained as the $\lambda/L_{p}%
\rightarrow\infty$ limit of the speed of light)

\item  Each inertial observer can establish the value of $L_{p}$
(same value for all inertial observers) by determining the
dispersion relations for photons which takes the form
$E^{2}-c^{2}p^{2}+f(E,p,L_{p})=0,$ where the function $f$ is the
same for all the inertial observers and in particular all
inertial observers agree on the leading $L_{p}$ dependence of $f(E,p,L_{p}%
)\simeq cEp^{2}L_{p},$ i.e.%
\begin{equation}
E^{2}-c^{2}p^{2}-cEp^{2}L_{p}=0.\label{dr}%
\end{equation}
\end{itemize}

It is worth noticing that the notion of relativity in DSR theories
is conceptually very similar to the notion of relativity in
Galilean Relativity and in Special Relativity. The laws of physics
are the same for all inertial observers. Inertial observers will
not necessarily agree on the measured value of a given quantity
but those relations among measurement results, which we call laws
of physics, will hold for all inertial observers. For example, two
inertial observers do not, in general, agree on the value of the
momentum of an electron, but if they measure energy and momentum
of the electron, both observers will find that the measurements
results satisfy the same dispersion relation. Among the laws of
physics an important role is played by those which identify
relativistic invariants quantities whose measurement gives the
same result in all inertial frames.

In Galilean Relativity, for example, all inertial observers agree
on the dispersion relations $E-p^{2}/(2m)=0.$ This relation does
not involve any invariant scale other than the mass of the
particle. In Special Relativity all
inertial observers agree on the dispersion relation $E^{2}-c^{2}p^{2}%
-m^{2}c^{4}=0$, which involve only one invariant scale ($c$),
other than the mass of the particle. In DSR theories all inertial
observers agree on the
relation of the type (\ref{dr}), which involves two invariant scale ($c,L_{p}%
$), other than the mass of the particle. All these theories are
relativistic.

As already discussed the introduction of such postulates does not
require to modify the rotation sector, but boosts do need to be
modified. At the first
order in $L_{p}$ one can adopt the ansatz%
\begin{equation}
N_{i}=i\left[  cp_{i}+L_{p}\Delta_{1i}(\vec{p}^{2},E)\right]
\frac{\partial }{\partial E}+i\left[
\frac{E}{c}+L_{p}\Delta_{2i}(\vec{p}^{2},E)\right]
\frac{\partial}{\partial p_{i}},
\end{equation}
and it is easy to verify that one has consistency with the
dispersion relation (\ref{dr}) if $\Delta_{1i}$ and $\Delta_{2i}$
are such that the above expression takes the form
\begin{equation}
N_{i}=icp_{i}\frac{\partial}{\partial E}+i\left[
\frac{E}{c}+L_{p}\left( \frac{E}{c}\right)
^{2}+\frac{L_{p}\vec{p}^{2}}{2}\right]  \frac{\partial
}{\partial p_{i}}. \label{boostlo}%
\end{equation}

Directly from (\ref{dr}) follows that $p^{2}=E^{2}/(1+cEL_{p})$.
This deviation from the Special Relativistic dispersion relation
$p=E$ implies that when $E\gtrsim1/L_{p}$ the dependence of
momentum on the energy change in the softest
$p\simeq\sqrt{E/(cL_{p})}.$ This is a evidence of the momentum
saturation that we will discuss more in detail when an all-orders
form of $f$ will have been obtained.

For a massless particle, if we retain the usual definition of
speed\footnote{We will discuss in the next chapter how this
definition, beyond to be the more natural in the momentum sector,
also is the one which comes from a proper analysis of the
wave-packet motion in spacetime.}, in the
energy-momentum sector we get $v_{\gamma}=\frac{dE}{dp}$ $=c(1+\frac{L_{p}}%
{c}E)$ which predicts for sufficiently-energetic photons the
possibility of $v_{\gamma}\gtrsim c$. Moreover the usual
energy-momentum conservation rule needs to be modified in such way
to be covariant under the new transformations rules. We will
discuss more in details these point in the next section in the
framework of $\kappa$-Poincar\'{e} quantum group.

\subsection{$\kappa$-Poincar\'{e} Hopf algebras}

\qquad It was observed in
Refs.~\cite{hep-th/0102098,hep-th/0107039} that the generators of
(modified) boost and (unmodified) rotations constructed from DSR
postulates correspond at the leading order in $L_{p}$ to the
Lorentz-sector generators of a well known quantum group: the
bicrossproduct-basis $\kappa $-Poincar\'{e} Hopf algebra. The
dispersion relation (\ref{dr}) if then the approximation at the
leading order in $L_{p}$ of the Casimir of this Hopf algebra. It
is useful for us to give here a brief review of the structure of
$\kappa$-Poincar\'{e} Hopf algebra in the ``bicrossproduct
basis''~\cite{hep-th/9405107,hep-th/9312153,hep-th/9907110}
showing in particular the connection with $\kappa$-Minkowski
noncommutative spacetime.

$\kappa$-Poincar\'{e} algebras were introduced in~\cite{lnrt,B293}
as one of the softest possible deformations of the usual
Poincar\'{e} group as an Hopf
algebra. It is defined in the algebraic sector by the commutation relations%
\begin{align}
\lbrack M_{i},M_{j}]  &  =i\epsilon_{ijk}M_{k}\text{ \ \ \ \ \ \ \ \ \ \ }%
[M_{i},P_{j}]=i\epsilon_{ijk}P_{k},\label{l1}\\
\lbrack M_{i},N_{j}]  &  =i\epsilon_{ijk}N_{k},\text{ \ \ \ \ \ \ \ \ \ \ }%
[M_{i},P_{0}]=0,\label{l2}\\
\lbrack N_{i},N_{j}]  &  =-i\epsilon_{ijk}M_{k},\text{ \ \ \ \ \ \ \ \ }%
[P_{\mu},P_{\nu}]=0,\label{l4}\\
\lbrack N_{j},P_{0}]  &  =iP_{j},\text{ \ \ \ \ \ \ }[N_{j},P_{k}%
]=i\delta_{jk}\left\{  \frac{1-e^{-2\lambda
P_{0}}}{2\lambda}+\frac{\lambda
}{2}\vec{P}^{2}\right\}  -i\lambda P_{j}P_{k}. \label{l5}%
\end{align}
where $P_{\mu}=(P_{0},P_{i})$ are the four-momentum generators,
$M_{k}$ are the spatial rotation generators and $N_{i}$ are the
boost generators.

The algebraic relations (\ref{l1}-\ref{l5}) are accompanied by
coalgebraic structures of the coproducts
\begin{align}
&  \Delta(p_{0})=p_{0}\otimes1+1\otimes p_{0},\qquad\quad\Delta(p_{j}%
)=p_{j}\otimes1+e^{-\lambda p_{0}}\otimes p_{j},\label{copro}\\
&  \Delta(M_{j})=M_{j}\otimes1+1\otimes M_{j},\quad\quad\Delta(N_{j}%
)=N_{j}\otimes1+e^{-\lambda p_{0}}\otimes N_{j}+\lambda\epsilon_{jkl}%
p_{k}\otimes M_{l},\nonumber
\end{align}
and by the antipodes
\begin{align}
&  S(N_{j})=-e^{\lambda p_{0}}N_{j}+\lambda e^{\lambda
p_{0}}\epsilon
_{jkl}p_{k}M_{l},\quad\quad S(M_{j})=-M_{j},\label{anti}\\
&  S(p_{j})=-e^{\lambda p_{0}}p_{j},\qquad\qquad \ \ \ \ \ \ \ \ \
\ \ \ \ \ \ \ \ \ \ \ \ S(p_{0})=-p_{0}~.\nonumber
\end{align}

A general description of Hopf algebras and in particular of their
co-algebraic properties is given in Ref~\cite{majidBOOK} (see also
Appendix). Actually for the purposes of this thesis it suffices to
notice that while the algebra describes products of generators the
coalgebra essentially describes sums of generators. Nontrivial
coproducts are related to non-Abelian addition law for the energy
and momentum.

Now we analyze more closely the action of the Lorentz sector of
this (quantum) Hopf algebra on the momentum sector. In a
Lie-algebra context the action is obtained directly by the
commutators (infinitesimal transformation) and by their
exponentiation (finite transformation). In an Hopf algebra the
action of a subalgebra on another subalgebra is generalized by the
concept of action and coaction. From the definition of covariant
(left)-adjoint action and from the expressions of coproducts
(\ref{copro}) and of antipodes (\ref{anti}) we have
that%
\begin{align*}
M_{i}\triangleright M_{j}  &  =M_{i}^{(1)}M_{j}S(M_{i}^{(2)})=[M_{i}%
,M_{j}]=i\epsilon_{ijk}M_{k},\\
M_{i}\triangleright N_{j}  &  =M_{i}^{(1)}N_{j}S(M_{i}^{(2)})=[M_{i}%
,N_{j}]=i\epsilon_{ijk}N_{k},\\
M_{i}\triangleright P_{\mu}  &  =M_{i}^{(1)}P_{\mu}S(M_{i}^{(2)}%
)=[M_{i},P_{\mu}]=i\epsilon_{i\mu k}P_{k},\\
N_{i}\triangleright P_{\mu}  &  =N_{i}^{(1)}P_{\mu}S(N_{j}^{(2)}%
)=[N_{i},P_{\mu}]=i\left[  P_{i}\delta_{\mu0}+\left(
\frac{1-e^{-2\lambda
P_{0}}}{2\lambda}+\frac{\lambda}{2}P^{2}\right)
\delta_{i\mu}-\lambda
P_{i}P_{k}\delta_{k\mu}\right]  ,\\
N_{j}\triangleright M_{k}  &  =N_{j}^{(1)}M_{k}S(N_{j}^{(2)})=[N_{j}%
,M_{k}]+i\delta_{jk}\vec{P}\cdot\vec{M}-iP_{j}M_{k},\\
N_{j}\triangleright N_{k}  &  =N_{j}^{(1)}N_{k}S(N_{j}^{(2)})=[N_{j}%
,N_{k}]-i\lambda P_{k}N_{j}+\frac{i}{2}\epsilon_{ijk}\left(
1-e^{2\lambda P_{0}}+\lambda^{2}P^{2}\right)  M_{l}.
\end{align*}

We observe that

\begin{itemize}
\item  The action of rotations is undeformed.

\item  The action of boosts on the translation generators is
deformed but it is still formulated through commutators.

\item  The action of boosts on boost/rotation generators is
deformed and cannot be formulated through commutators.
\end{itemize}

From the above expression one can also easily write down the
infinitesimal
actions on a generic function, for example, of the momenta%
\begin{align}
P_{\mu}\triangleright G(P)  &  =0,\nonumber\\
M_{i}\triangleright F(P)  &  =M_{i}^{(1)}F(P)S(M_{i}^{(2)})=[M_{i}%
,F(P)]=-i\epsilon_{ijk}P_{k}\frac{\partial}{\partial P_{l}}F(P),\label{un}\\
N_{i}\triangleright F(P)  &  =N_{i}^{(1)}F(P)S(N_{i}^{(2)})=[N_{i}%
,F(P)]=\nonumber\\
&  =i\left[  P_{i}\frac{\partial}{\partial P_{0}}+\left(
\frac{1-e^{-2\lambda
P_{0}}}{2\lambda}+\frac{\lambda}{2}P^{2}\right)
\frac{\partial}{\partial P_{i}}-\lambda
P_{i}P_{k}\frac{\partial}{\partial P_{k}}\right]  F(P).
\label{due}%
\end{align}

These last expressions indicate that the action of the Lorentz
sector $so_{1,3}$ on generic functions of the translation
generators is still
described by commutators. Also the usual Leibniz rule is satisfied%
\begin{equation}
N_{i}\triangleright\lbrack F(P)G(P)]=[N_{i}\triangleright
F(P)]G(P)+F(P)[N\triangleright G(P)].\label{LR1}%
\end{equation}

Using the Leibniz rule (\ref{LR1}) and (\ref{un}-\ref{due}) it is
easy to calculate that for a generic finite transformation
$e^{i(\vec{\theta}\cdot
\vec{M}+\vec{\xi}\cdot\vec{N})}$ the action is%
\begin{equation}
e^{i(\vec{\theta}\cdot\vec{M}+\vec{\xi}\cdot\vec{N})}\triangleright
F(P)=e^{i(\vec{\theta}\cdot\vec{M}+\vec{\xi}\cdot\vec{N})}F(P)e^{-i(\vec
{\theta}\cdot\vec{M}+\vec{\xi}\cdot\vec{N})}.\label{fa}%
\end{equation}

The action (\ref{fa}) on functions of momenta is the usual one in
spite of the non-trivial coalgebraic structure
(\ref{copro}-\ref{anti}). This is due to the fact that the action
of the Lorentz sector on the momentum sector is still obtained
through commutators. The ``mass Casimir'' of this Hopf algebra is
\begin{equation}
\mathcal{C}_{\kappa}(p)=\Big(\frac{2}{\lambda}\sinh\frac{\lambda p_{0}}{2}%
\Big)^{2}-\vec{p}^{2}e^{\lambda p_{0}}. \label{casimir}%
\end{equation}

Since the action of Lorentz sector generator is still through
commutators the Casimir $\mathcal{C}_{\kappa}$ preserves the
property of being invariant under the covariant left adjoint
action (\ref{un}) and (\ref{due}). Moreover the mass Casimir
allows an unique, all-order determination of the function $f$
introduced in \textbf{(}\ref{dr}\textbf{), }whereas DSR postulates
fix it only to the lowest order in $L_{p}$. The same is true for
the boost action (\ref{due}) that is an all-order generalization
of (\ref{boostlo}). Therefore we see that the DSR proposal is
naturally realized in this Hopf-algebra quantum scheme and that
all-order results in the deformation parameter can be obtained in
this scheme. All-order expressions for the energy-momentum
transformation rules between different inertial observers have
been obtained in~\cite{hep-th/0107039} and an all-order analysis
of the scattering processes is reported in~\cite{hep-th/0105120}.

\subsection{Phenomenology of $\kappa$-Poincar\'{e}}

\qquad It is perhaps appropriate to pause for a few considerations
regarding the phenomenological implications of
$\kappa$-Poincar\'{e} kinematics. The mathematical formalism
described so far already implies some profound modifications of
the conventional special-relativistic kinematics framework. In
particular from (\ref{casimir}) one gets
immediately~\cite{hep-th/0102098} the important general
conclusions that

\begin{itemize}
\item $E\rightarrow\infty$ when $\left|  \vec{p}\right|
\rightarrow 1/\lambda,$which means that there is a maximum
momentum ($\left|  \vec {p}\right|  =1/\lambda$).

\item $v_{\gamma}(p)=\frac{dE}{dp}=\left(  1-p/\lambda\right)  ^{-1}%
=\exp(E/\lambda),$ which means that the speed of light tends to
infinity when the momentum tends to the maximum allowed momentum.
\end{itemize}

It should be noticed that infinite velocities are allowed only
when $E\rightarrow\infty$ so that an infinite amount of energy is
needed to obtain a photon with infinite speed. Real photons have
finite energy and therefore finite speed. Of course
$v_{\gamma}\neq1$ is a striking characteristic of this framework
but for our ``low-energy'' particles ($E\ll1/\lambda$) the effects
are negligibly small~\cite{hep-th/9312153,hep-th/9907110}. In
practice this new framework is indistinguishable from the usual
one at the presently-accessible energies. Therefore this striking
prediction of the $\kappa$-Minkowski framework does not have
troublesome phenomenological implications, but there is of course
still an intense debate concerning the logical consistency of this
scenario for $v_{\gamma}(p)$. In particular, it appears necessary
to develop a new concept of causality. Actually, even before a
full development of this new concept of causality, especially in
cosmology there has been interest in the $\kappa$-Minkowski
motivated idea of a light cone that effectively becomes wider as
the energies available increase. Let us consider, for example, the
horizon problem that is one of the main motivations for inflation
theory. Horizon problem consists in the fact that zones of the sky
which are angularly separated by more than a few degrees should be
causally separated, whereas we observe significant large-scale
homogeneity. The analysis of Cosmic Microwave Background
Radiation, for instance, indicates that the temperature is the
same in all directions, with a precision of one part in 10$^{5}.$
The hypothesis, predicted by DSR theories, of an energy-dependent
speed of light might provide a simple explanation of this
paradox~\cite{astro-ph/0006250}, without recurring to the
inflation theory. The greatly-energetic photons of the early
stages of the universe in fact are predicted by DSR theories to
have speeds high enough that they could causally connect zones of
the universe that would otherwise, according to Special
Relativity, be causally disconnected.

There are also other physical applications of
$\kappa$-Poincar\'{e} kinematics that allow to explain certain
paradoxes of physics based on the Special Relativity. A much
studied example is the one in which one gives a $\kappa
$-Poincar\'{e}/DSR description of the cosmic-ray paradox.
According to the classical Poincar\'{e} symmetry of classical
Minkowski spacetime ultra-high-energy-cosmic rays should loose
energy interacting with the Cosmic Microwave Background Radiation
by producing pions ($p+\gamma\rightarrow p+\pi $). Considering the
typical energies ($E_{\gamma_{CMB}}$) of the CMRB photons and the
typical distances from the Earth of the cosmic-ray sources, the
assumption of validity of the Special Relativistic kinematic rules
should lead to an upper (GZK) limit $E<5\cdot10^{19}eV$ on the
energy of the observed cosmic rays~\cite{gzk1,gzk2}. Instead
detection of several cosmic rays above the GZK limit has been
reported~\cite{astro-ph/9807193}. One can calculate the thresholds
for photopion production in the DSR scheme. Using the dispersion
relation (\ref{dr}) and the energy-momentum conservation laws one finds%
\begin{equation}
E>\frac{2m_{p}m_{\pi}+m_{\pi}^{2}}{4E_{\gamma_{CMB}}}+\lambda\frac
{(2m_{p}+m_{\pi})^{3}m_{\pi}^{3}}{256E_{\gamma_{CMB}}^{4}}\left(
1-\frac{m_{p}^{2}+m_{\pi}^{2}}{(m_{p}+m_{\pi})^{2}}\right)  , \label{mt}%
\end{equation}
where $m_{p}$ $(m_{\pi})$ is the proton (pion) mass. Of course in
the $\lambda\rightarrow0$ limit one recovers the usual
photopion-production threshold. We observe that in the correction
term the smallness of $\lambda$ (which, as mentioned, we expect to
be of order $L_{p}\sim10^{-33}cm$) is compensated by the huge
ratios $m_{p}/E_{\gamma_{CMB}},m_{\pi}/E_{\gamma _{CMB}}.$ The
result is that~\cite{astro-ph/0008107,gr-qc/0107086} according to
(\ref{mt}) even the observation of E$\sim3\cdot10^{20}eV$ protons
becomes possible providing an explanation to the observed cosmic
rays.

Conceptually similar to the GZK paradox is the Markarian-501
paradox. High energy photons emitted by Markarian 501 with
energies higher than 10TeV should collide with the Far Infrared
Background Radiation (FIRBR) producing electron-positron pairs$.$
Instead photons from Markarian 501 with energies higher than 20
TeV have been detected~\cite{astro-ph/9903159}. Markarian 501
paradox can be explained in a way analogous to the GZK paradox.
The relevant process in this case is $\gamma+\gamma\rightarrow
e^{-}+e^{+}$. The DSR-threshold for this process obtained from the
dispersion relation
(\ref{dr}) and energy-momentum conservation law is%
\begin{equation}
E>\frac{m_{e}^{2}}{E_{\gamma_{FIRB}}}+\lambda\frac{m_{\pi}^{6}}{8E_{\gamma
_{FIRB}}^{4}}.
\end{equation}

Given the value involved in the energy $E_{\gamma_{FIRB}}$ of the
FIRBR, the threshold is shifted up to $E\simeq20TeV$ explaining
the observations.

There are other physical situations, besides the ones mentioned so
far in which $\kappa$-Poincar\'{e} kinematics might lead to
detectable effects. Particularly promising in this sense appear to
be the so-called time-of-flight studies in astrophysics. The
velocity formula $v_{\gamma}(p)$ obtained in the
$\kappa$-Poincar\'{e} framework predicts a difference in the times
of arrival on Earth for two simultaneously-emitted photons
\begin{equation}
\Delta t_{d}\simeq\frac{\lambda\Delta EL}{c}, \label{td}%
\end{equation}
where $L$ is the source-Earth distance and $\Delta E$ is the
difference between the energies of the two photons. Certain
astrophysical objects produce very energetic photons that travel
very large distances. From searches of a time-of-arrival
difference of the type (\ref{td}) it is possible to obtain bounds
on the deformation parameter $\lambda$. Presently, the best bound
is $\lambda\lesssim500L_{p}$ obtained by~\cite{gr-qc/9810044}.
Future planned experiments, such as AMS~\cite{AMS} and
GLAST~\cite{astro-ph/0009271,GLAST,astro-ph/9912136}, are expected
to move this bound to the Planck scale ($L_{p}$) and beyond. In
summary the phenomenology in the $\kappa$-Poincar\'{e}/DSR
framework is rather rich: the new effects are small enough to
agree with all available robust data, but large enough to provide
candidate solutions for emerging experimental paradoxes and for
testing in planned experiments.

\subsection{$\kappa$-Minkowski Spacetime from $\kappa$-Poincar\'{e} duality}

\qquad Having reassured the reader that the phenomenology of
$\kappa $-Poincar\'{e} kinematics is acceptable and interesting,
we go back to the analysis of the mathematical structure in order
to identify a spacetime on which $\kappa$-Poincar\'{e} acts
covariantly. Given the enveloping algebra\textbf{ }of translation
$T=(P_{0},\vec{P})$, it is rather natural to take for the
spacetime coordinate space its dual $T^{\ast}$ that will also be
an algebra on which $T$ necessarily acts in a covariant way. The
structure of the coordinate space $T^{\ast}$ is univocally
determined by the axioms of the
Hopf algebra duality:%
\begin{align}
&  <t,xy>=<t_{(1)},x><t_{(2)},y>,\text{ \ \ \ }\label{ad1}\\
&  <ts,x>=<t,x_{(1)}><s,x_{(2)}>\text{ \ \ \ }\forall t,s\in
T,\text{ }\forall
x,y\in T^{\ast}\label{ad2}\\
&  <p_{\mu},x_{\nu}>=-ig_{\mu\nu}, \label{ad3}%
\end{align}
where $g_{\mu\nu}=\mathrm{diag}(1,-1,-1,-1)$ and $t_{(1)}$ and
$t_{(2)}$ are
introduced in the sense of the standard notation for the coproduct%
\begin{equation}
\Delta t=\sum t_{(1)}\otimes t_{(2)}.
\end{equation}

We see immediately from (\ref{ad2}) that commutation rules in the
momentum
sector correspond to co-commutation rules in the coordinate sector%
\begin{equation}
\Delta x_{\mu}=\mathbb{I}\otimes x_{\mu}+x_{\mu}\otimes\mathbb{I}.
\end{equation}

The equation (\ref{ad1}) and (\ref{ad3}) can be used to define
spacetime coordinates. One gets that
\begin{align}
&  <p_{i},x_{0}x_{j}>=-\frac{i}{\kappa}\delta_{ij},\label{pa1}\\
&  <p_{i},x_{j}x_{0}>=0, \label{pa2}%
\end{align}
from which it follows that
$<p_{i},[x_{0},x_{j}]>=-\frac{1}{\kappa}\delta
_{ij},$ and comparing with (\ref{ad3}) one finds%
\begin{align}
\lbrack x_{0},x_{i}]  &  =-\frac{i}{\kappa}x_{i},\\
\lbrack x_{i},x_{j}]  &  =0.
\end{align}

This is the noncommutative spacetime called $\kappa$-Minkowski. We
observe that spacetime noncommutativity in this case directly
follows from the fact that the momentum sector, although being
commutative, is not co-commutative, and from the duality relations
pairing the two algebras (\ref{ad1}-\ref{ad3}).

\subsection{Covariance of $\kappa$-Minkowski Spacetime}

\qquad The duality between $\kappa$-Minkowski and (bicrossproduct)
$\kappa $-Poincar\'{e} is related with the covariance, in the
sense of Hopf algebras, of $\kappa$-Minkowski under
$\kappa$-Poincar\'{e} action. The action of an element of the
Lorentz sector $w,$ on the coordinates is implicitly defined by
the relation
\begin{equation}
\left\langle f(p),w\triangleright:g(x):\right\rangle =\left\langle
S(w)\triangleright f(p),:g(x):\right\rangle ,\label{eq:WsuX}%
\end{equation}
where the action of $S(w)$ on functions of the momenta $f(p)$ is
the (left) adjoint one (\ref{un}-\ref{due}) and :$g(x):$ is
ordered in the form\footnote{We observe that every smooth function
of noncommutative
coordinates can be written in this form.}%
\begin{equation}
g(x)=\sum_{n_{0}n_{1}n_{2}n_{3}}g_{n_{1}n_{2}n_{3}n_{0}}x_{1}^{n_{1}}%
x_{2}^{n_{2}}x_{3}^{n_{3}}x_{0}^{n_{0}}.
\end{equation}

By using (\ref{un}-\ref{due}). and (\ref{eq:WsuX}) one obtains the actions%
\begin{align}
M_{j}\rhd &  :f(x):=:-i\epsilon_{jkl}x_{k}\frac{\partial}{\partial x_{l}%
}f(x):,\label{msfx}\\
N_{j}\rhd &  :f(x):=:\left[  ix_{0}\frac{\partial}{\partial x_{j}}%
+x_{j}\left(  \frac{e^{2i\lambda\partial_{t}}-1}{2\lambda}+\frac{\lambda}%
{2}\bigtriangledown^{2}\right)  -\lambda x_{k}\frac{\partial}{\partial x_{k}%
}\frac{\partial}{\partial x_{j}}\right]  f(x):. \label{nsfx}%
\end{align}

These actions are covariant in the Hopf algebra sense%
\begin{align}
M_{j}\rhd\lbrack &  :f(x)::g(x):]=(M_{j}^{(1)}\rhd:f(x):)(M_{j}^{(2)}%
\rhd:g(x):),\\
N_{j}\rhd\lbrack &  :f(x)::g(x):]=(N_{j}^{(1)}\rhd:f(x):)(N_{j}^{(2)}%
\rhd:g(x):).
\end{align}

We also observe that the action of boosts and rotations on the
coordinates is the same as in Special Relativity
\begin{equation}
M_{j}\triangleright x_{0}=0,\qquad\quad M_{j}\triangleright x_{k}%
=i\epsilon_{jkl}x_{l},\qquad\quad N_{j}\triangleright
x_{0}=ix_{j},\qquad\quad N_{j}\triangleright
x_{k}=i\delta_{jk}x_{0},
\end{equation}
and can be described as an action through commutators.

Differences occur only in higher powers of the coordinates (\ref{msfx}%
-\ref{nsfx}). Now we have to define the action of $p_{\mu}$ on the
coordinate
space. This action is defined by the (left) canonical action%
\begin{equation}
p_{\mu}\overset{can}{\triangleright}:f(x):=:f(x):_{(1)}\left\langle
p_{\mu },:f(x):_{(2)}\right\rangle =:-i\frac{\partial}{\partial
x^{\mu}}f(x):
\end{equation}

so that a finite transformation reads%
\begin{equation}
e^{iap}\overset{can}{\triangleright}:f(x):=:f(x+a):,
\end{equation}

that is a simple translation\footnote{It is worth noticing that
defining the action of $p_{\mu}$ on $f(x)$ through the commutators
of some quantum phase space (see for example (\ref{pskow}) ), this
action can't be a translation.}. This concludes our analysis of
covariance of $\kappa$-Minkowski.

\section{$\kappa$-Minkowski spacetime and canonical spacetime from general
Lie-algebra spacetimes}

\qquad In this section we observe that $\kappa$-Minkowski and
canonical spacetime have the noticeable property that, among all
the possible
Lie-algebra spacetimes of the form%
\begin{equation}
\left[  x_{\mu},x_{\nu}\right]
=i\theta_{\mu\nu}+iC_{\mu\nu}^{\alpha
}x_{\alpha}, \label{last2}%
\end{equation}
they are selected from simple symmetry requirements. In particular
if we
demand invariance under \underline{undeformed} rotations%
\begin{equation}
M_{j}\rhd\left[  x_{\mu},x_{\nu}\right]  =M_{j}\rhd\left(
i\theta_{\mu\nu }+iC_{\mu\nu}^{\alpha}x_{\alpha}\right)  ,
\end{equation}

we obtain that it must be $\theta_{\mu\nu}=0$ and%
\begin{equation}
\left[  x_{i},x_{j}\right]  =i\epsilon_{ijk}x_{k},\text{ \ \ \ \
}\left[
x_{i},x_{0}\right]  =0 \label{fs}%
\end{equation}

or%
\begin{equation}
\left[  x_{i},x_{0}\right]  =i\lambda x_{i},\text{ \ \ \ \ \ \ }[x_{i}%
,x_{j}]=0. \label{km}%
\end{equation}

The first spacetime (\ref{fs}) is known as fuzzy sphere, since $\mathcal{C}%
=x_{1}^{2}+x_{2}^{2}+x_{3}^{2}+x_{0}^{2}$\ is a Casimir of this
algebra. In the second spacetime (\ref{km}) we recognize
$\kappa$-Minkowski spacetime.

On the other hand if we require invariance of (\ref{last2}) under
\underline{undeformed} translations%
\begin{equation}
P_{j}\rhd\left[  x_{\mu},x_{\nu}\right]  =P_{j}\rhd\left(
i\theta_{\mu\nu }+iC_{\mu\nu}^{\alpha}x_{\alpha}\right)
\end{equation}

we select%
\begin{equation}
\left[  x_{\mu},x_{\nu}\right]  =i\theta_{\mu\nu}%
\end{equation}
that is just canonical spacetime. This shows that among all the
possible Lie-algebra spacetime the request of (undeformed)
rotational invariance selects $\kappa$-Minkowski spacetime and the
fuzzy sphere spacetime, whereas the request of (undeformed)
translational invariance uniquely selects canonical spacetime.

To resume, with respect to the classical symmetries, we can
construct the
following table%
\[%
\begin{tabular}
[c]{|c|c|c|c|c|}\hline
Spacetime%
$\backslash$%
Undef.Transf. & Rotations & Translations & Time-Translations &
Boosts\\\hline $\kappa$-Minkowski & yes & no & yes & no\\\hline
Canonical & no & yes & yes & no\\\hline Fuzzy sphere & yes & no &
yes & no\\\hline
\end{tabular}
\]

Both $\kappa$-Minkowski and canonical spacetime are candidate
noncommutative version of Minkowski spacetime\footnote{Fuzzy
sphere is not a good candidate since it represents a nonflat
space.} which in general preserve 4 \underline{undeformed
}classical symmetries (3 rotations + 1 time translation for
$\kappa$-Minkowski, 3 spatial translations + 1 time translation
for canonical spacetime). However, as discussed above, in addition
to the 4 classical/undeformed symmetries, $\kappa$-Minkowski also
enjoys 6 additional quantum-deformed symmetries, and its full
symmetry structure is described by the 10 generators of
$\kappa$\textbf{-}Poincar\'{e} Hopf algebra.

In the case of canonical noncommutativity the four classical
symmetries are all there is (there are no additional deformed
symmetries). This reflects the fact that canonical
noncommutativity requires the support of a preferred class of
inertial observers. In particular we know that canonical
noncommutativity can be described as a commutative geometry in
presence of a background (magnetic) field, and the presence of
field allows the selection of a preferred class of inertial
observer.



\chapter{Waves in noncommutative Spacetimes}

\footnotetext{* In this Chapter we discuss in detail the analysis
reported more briefly in Ref.~\cite{hep-th/0211022}.}*\qquad In
this chapter, after a brief review of the properties of waves in
classical Minkowski spacetime, we analyze the concept of wave in a
noncommutative spacetime. We show that the usual picture of
propagating wave-packets can be implemented in a noncommutative
context as well. We discuss how in the case of canonical
noncommutativity this construction is very close to the
commutative case: the phase velocity and the, physically more
relevant, group velocity are the same as in the Minkowski
spacetime in spite of the appearance of some extra (unobservable)
phases that depend on the noncommutativity parameters. On the
other hand, in $\kappa$-Minkowski spacetime, we find significant
differences with respect to the commutative case. In particular we
find that the description of the group velocity is still governed
by the relation $v_{g}=dE(p)/dp,$ but the dispersion relation
$E(p)$ is significantly modified as discussed in the previous
chapter.

\section{Review of waves in Minkowski spacetime}

\qquad Both in theories in Galilei spacetime and in theories in
Minkowski spacetime the relation between the physical velocity of
signals (the group velocity of a wave packet) and the dispersion
relation is governed by the formula
\begin{equation}
v=\frac{dE}{dp}~, \label{veloxrule}%
\end{equation}
in components
\begin{equation}
v_{j}\equiv\frac{dx_{j}}{dt}=\frac{\partial E}{\partial p_{j}}=\frac{p_{j}}%
{p}\frac{\partial E}{\partial p}~. \label{Vrel}%
\end{equation}

This is basically a result of the fact that our theories in
Galilei and Minkowski spacetime admit Hamiltonian formulation. In
classical mechanics this leads directly to
\begin{equation}
\frac{dx_{j}}{dt}=\frac{\partial H(p)}{\partial p_{j}}~.\label{Vhamil}%
\end{equation}
In ordinary quantum mechanics $\vec{x}$ and $\vec{p}$ are
described in terms
of operators that satisfy the commutation relations $[x_{k},p_{j}%
]=i\delta_{jk}$, and in the Heisenberg picture the time evolution
for the position operator is given by
\[
\frac{dx_{j}(t)}{dt}=i[x_{j}(t),H]
\]
Since $x_{j}\rightarrow\partial/\partial p_{j}$ and, again,
$H\rightarrow E(p)$, also in ordinary quantum mechanics one finds
$v=dE/dp$ (but in quantum mechanics $v_{j}$ is the operator
$dx_{j}/dt$ and the group-velocity relation strictly holds only
for expectation values).

Given a spacetime, the concept of group velocity can be most
naturally investigated in the study of the propagation of waves.
It is useful to review that discussion briefly. For simplicity we
consider a classical 1+1-dimensional Minkowski spacetime. We
denote by $\omega$ the frequency of the wave and by $k(\omega)$
the wave number of the wave. [Of course, $k(\omega)$ is governed
by the dispersion relation, by the mass Casimir of the classical
Poincar\'{e} algebra.] A plane wave is described by the
exponential $e^{i\omega t-ikx}$. A wave packet is the Fourier
transform of a function $a(\omega)$ which is nonvanishing in a
limited region of the spectrum ($\omega_{0}-\Delta$,
$\omega_{0}+\Delta$ ):
\[
\Psi_{(\omega_{0},k_{0})}(t,x)=\int_{\omega_{0}-\Delta}^{\omega_{0}+\Delta
}a(\omega)e^{i\omega t-ikx}d\omega~.
\]
%
%
%
%
%
%
%
%
%
%

The information/energy carried by the wave will travel at a
sharply-specified velocity, the group velocity, only if
$\Delta\ll\omega_{0}$. It is convenient to write the wave-packet
as $\Psi_{(\omega_{0},k_{0})}(t,x)=A(t,x)e^{i\omega
_{0}t-ik_{0}x}$, from which the definition of the wave amplitude
$A(t,x)$ follows:
\begin{equation}
A(t,x)=\int_{\omega_{0}-\Delta}^{\omega_{0}+\Delta}a(\omega)e^{i(\omega
-\omega_{0})t-i(k-k_{0})x}d\omega\approx\int_{\omega_{0}
-\Delta}^{\omega
_{0}+\Delta}a(\omega)e^{i(\omega-\omega_{0})\left(  t - \left[
\frac
{dk}{d\omega}\right]  _{0} x\right)  }d\omega\label{eq:appr}%
\end{equation}

The wave packet is therefore the product of the plane-wave factor
$e^{i\omega_{0}t-ik_{0}x}$ and the wave amplitude $A(t,x)$. One
can introduce a ``phase velocity'', $v_{ph}=\omega_{0}/k_{0}$,
associated with the plane-wave factor $e^{i\omega_{0}t-ik_{0}x}$,
but there is no information/energy that actually travels at this
velocity (this ``velocity'' is a characteristic of a pure phase,
with modulus $1$ everywhere). It is the wave amplitude $A(t,x)$
that describes the time evolution of the energy/information
actually carried by the wave packet. From (\ref{eq:appr})
we see that the wave amplitude stiffly translates at velocity $v_{g}%
=d\omega_{0}/dk_{0}$, the group velocity. In terms of the group
velocity and the phase velocity the wave packet can be written as
\[
\Psi_{(\omega_{0},k_{0})}(t,x)=e^{ik_{0}(v_{ph}t-x)}
\int_{\omega_{0}-\Delta }^{\omega_{0}+\Delta}a(\omega)
e^{i(\omega-\omega_{0})(t-x/v_{g})}d\omega~.
\]

In ordinary Minkowski spacetime the group velocity and the phase
velocity both are 1 (in our units) for photons (light waves)
travelling in vacuum. For massive particles or massless particles
travelling in a medium $v_{ph}\neq v_{g}$. The causality structure
of Minkowski spacetime guarantees that $v_{g}\leq1$, whereas,
since no information actually travels with the phase velocity, it
provides no obstruction for $v_{ph}>1$.
%
%
%
%
%
%
%
%
%
%

The main idea to extend the above construction to a noncommutative
spacetime is that of writing the function of noncommutative
variables as the inverse Fourier transform of a commutative
energy-momentum-space function. Thus in
general one writes%
\begin{equation}
f(x)=\frac{1}{(2\pi)^{4}}\int
dk^{4}\tilde{f}(k):\exp(ik^{\mu}x_{\mu}):,
\label{ftnc}%
\end{equation}
where $\tilde{f}(k)$ is the Fourier transform of $f(x):$%
\begin{equation}
\tilde{f}(k)=\int d\alpha^{4}f(\alpha)\exp(ik^{\mu}\alpha_{\mu}).
\label{aftnc}%
\end{equation}

Here $k$ and $\alpha$ are commuting variables while $x$ are
noncommuting variables. The function $:\exp(ik^{\mu}x_{\mu}):$
must be consistent with the Fourier calculus and must reduce in
the commutative limit to the usual $\exp(ik^{\mu}x_{\mu}).$ The
advantage in using Fourier formulation is that one can do products
of fields once it is known how to do products of the phases
$:\exp(ik^{\mu}x_{\mu}):.$ This last ones are usually evaluated
using the Baker-Campbell-Hausdorff formula.

\section{Proposal of a description of wave-packets in Canonical Spacetime}

\qquad We now try to extend the construction just outlined to
waves in canonical spacetime. As a first step we have to define
the phase
$:\exp(ik^{\mu}x_{\mu}):_{\theta}$. The simplest possible choice is%
\begin{equation}
:\exp(ik^{\mu}x_{\mu}):_{\theta}=\exp(ik^{\mu}x_{\mu}).
\end{equation}

Once that single plane wave has been identified the next step is
to consider
products of waves. From the canonical noncommutativity relations we have that%
\begin{equation}
\exp(ik^{\mu}x_{\mu})\exp(ip^{\mu}x_{\mu})=\exp(ik^{\mu}\theta_{\mu\nu}p^{\nu
})\exp[i(k^{\mu}+p^{\mu})x_{\mu}]. \label{wcnc1}%
\end{equation}

Waves packets are constructed summing plane waves in the usual way%
\[
\Psi_{(\omega_{0},k_{0})}(t,\vec{x})=\int_{\omega_{0}-\Delta}^{\omega
_{0}+\Delta}a(\omega)e^{i\omega t-i\text{
}\vec{k}\cdot\vec{x}}d\omega~.
\]

Following the procedure outlined in the Minkowski case we find%
\[
\Psi_{(\omega_{0},k_{0})}(t,\vec{x})\simeq\int_{\omega_{0}-\Delta}^{\omega
_{0}+\Delta}a(\omega)e^{i\left[  \left(  \omega-\omega_{0}\right)
t-\frac{d\vec{k}}{d\omega}\left(  \omega-\omega_{0}\right)
\vec{x}\right] +i\left[
\omega_{0}t-\vec{k}_{0}\cdot\vec{x}\right]  }d\omega~,
\]
that is formally similar to (\ref{eq:appr}). Using (\ref{wcnc1}),
defining
$k^{\mu}\equiv(\omega-\omega_{0},-\frac{d\vec{k}}{d\omega}\left(
\omega-\omega_{0}\right)  )$ and
$p^{\mu}\equiv(\omega_{0},-\vec{k}_{0})$ we
get%
\[
\Psi_{(\omega_{0},k_{0})}(t,\vec{x})\simeq\left[
\int_{\omega_{0}-\Delta }^{\omega_{0}+\Delta}a(\omega)e^{i\left[
\left(  \omega-\omega_{0}\right) t-\frac{d\vec{k}}{d\omega}\left(
\omega-\omega_{0}\right)  \vec{x}\right] }\right]  d\omega
e^{-ik^{\mu}\theta_{\mu\nu}p^{\nu}}e^{i\left[  \omega
_{0}t-\vec{k}_{0}\cdot\vec{x}\right]  }.
\]

As in the Minkowski case we can individuate an amplitude here%
\begin{equation}
A(t,\vec{x})=\left[  \int_{\omega_{0}-\Delta}^{\omega_{0}+\Delta}%
a(\omega)e^{i\left[  \left(  \omega-\omega_{0}\right)  t-\frac{d\vec{k}%
}{d\omega}\left(  \omega-\omega_{0}\right)  \vec{x}\right]
}\right]  ,
\end{equation}
and write the wave packet as%
\begin{equation}
\Psi_{(\omega_{0},k_{0})}(t,\vec{x})\simeq e^{ik^{\mu}\theta_{\mu\nu}p^{\nu}%
}e^{i\vec{k}_{0}\cdot\left[  \vec{v}_{ph}t-\vec{x}\right]  }\left[
\int_{\omega_{0}-\Delta}^{\omega_{0}+\Delta}a(\omega)e^{i\frac{d\vec{k}%
}{d\omega}\left(  \omega-\omega_{0}\right)  \left[
\vec{v}_{g}t-\vec
{x}\right]  }d\omega\right]  , \label{pccnc}%
\end{equation}
where we have defined $v_{g}=\frac{d\omega}{dk},$ and
$v_{ph}=\frac{\omega _{0}}{k_{0}}.$ At this point some remarks are
in order. A first observation regards the overall phase factor
$e^{ik^{\mu}\theta_{\mu\nu}p^{\nu}}$ which appears in
(\ref{pccnc}). This factor doesn't give any modification of the
group velocity and is just an overall phase velocity. A second
observation regards the fact that we would have reached the same
conclusion by factorizing
the phase contribution to the right:%
\begin{equation}
\Psi_{(\omega_{0},k_{0})}(t,x)\simeq\left[  \int_{\omega_{0}-\Delta}%
^{\omega_{0}+\Delta}a(\omega)e^{i\frac{dk}{d\omega}\left(
\omega-\omega _{0}\right)  \left[  v_{g}t-x\right]
}d\omega\right]  e^{-ik^{\mu}\theta
_{\mu\nu}p^{\nu}}e^{ik_{0}\left[  v_{ph}t-x\right]  }.
\end{equation}

The only difference being now that the overall phase factor
acquires an opposite sign with respect to the previous case.
However the results for $v_{g},v_{ph}$ do not depend on this sign.

These observations seem to suggest that in the case of canonical
noncommutativity there is no observable modification of the in
vacuum propagation of wave packets.

\section{The challenge of Waves in $\kappa$-Minkowski Spacetime}

\qquad In this section we try to construct wave packets in
$\kappa$-Minkowski spacetime following the line of the previous
sections. We use the Fourier calculus in noncommutative spaces
developed in~\cite{hep-th/9402037} for a proper definition of the
wave packet. This assure the right covariance properties of the
packets under $\kappa$-Poincar\'{e} transformations and allow us
to get phase velocity and group velocity.

\subsection{Differential calculus and Fourier calculus in $\kappa$-Minkowski}

\qquad The only consistently-developed differential
calculus~\cite{majidBOOK} on the enveloping algebra of
$\kappa$-Minkowski is
\begin{equation}
\partial_{j}:f(x):=:\frac{\partial f(x)}{\partial x_{j}}: \label{derx}%
\end{equation}%
\begin{equation}
\partial_{0}:f(x):=:\frac{e^{i\lambda{\frac{\partial}{\partial t}}}%
-1}{i\lambda}f(x):=:\frac{f(\vec{x},t+i\lambda)-f(\vec{x},t)}{i\lambda}:
\label{dert}%
\end{equation}
The notation $:f(x):$, conventional in the $\kappa$-Minkowski
literature, is reserved for time-to-the-right-ordered\footnote{In
$\kappa$-Minkowski spacetime (with its commuting space coordinates
and nontrivial commutation relations only when the time coordinate
is involved), it is easy to see that the natural functional
calculus should be introduced in terms of
time-to-the-right-ordered functions or (the equivalent alternative
of) intuitive rules for time-to-the-left-ordered functions. In
other noncommutative spacetimes the choice of ordering may not be
so obvious.} functions of the noncommutative coordinates. The
standard symbolism adopted in Eqs.~(\ref{derx})-(\ref{dert})
describes noncommutative differentials in terms of familiar
actions on commutative functions. The symbols ``$\partial_{j}$''
and ``$\partial_{0}$'' refer to elements of the differential
calculus on $\kappa$-Minkowski, while the symbols
``$\partial/\partial x_{j}$'' and ``$\partial/\partial t$'' act as
ordinary derivatives on a time-to-the-right-ordered function of
the $\kappa$-Minkowski coordinates. For example, Eq.~(\ref{derx})
states that in $\kappa$-Minkowski $\partial
_{x}(xt)=t$ and $\partial_{x}[xt^{2}+2i\lambda xt-\lambda^{2}x+x^{2}%
t]=t^{2}+2i\lambda t-\lambda^{2}+2xt$, \textit{i.e.}
$\partial_{x}$ acts as a familiar $x$-derivative on
time-to-the-right-ordered functions. Of course, the
$\kappa$-Minkowski commutation relations impose that, if
derivatives are standard on time-to-the-right-ordered functions,
derivatives must be accordingly modified for functions which are
not time-to-the-right ordered. For example, since
$\partial_{x}(xt)=t$ and $\partial_{x}(x)=1$ (the functions $xt$
and $x$ are time-to-the-right ordered), also taking into account
the $\kappa$-Minkowski commutation relation $xt=tx-i\lambda x$,
one can obtain the $x$-derivative of the function $tx$, which must
be given by $\partial
_{x}(tx)=t+i\lambda$. Similarly, one finds that $\partial_{x}[t^{2}%
x+x^{2}t]=t^{2}+2i\lambda t-\lambda^{2}+2xt$ (in fact, using the
$\kappa
$-Minkowski commutation relations one finds that $t^{2}x+x^{2}t=xt^{2}%
+2i\lambda xt-\lambda^{2}x+x^{2}t]$).

The time derivative described by Eq.~(\ref{dert}) has analogous
structure, with the only difference that the special role of the
time coordinate in the structure of $\kappa$-Minkowski spacetime
forces~\cite{majidBOOK} one to introduce an element of
discretization in the time direction: the time derivative of
time-to-the-right-ordered functions is indeed standard (just like
the $x$-derivative of time-to-the-right-ordered functions is
standard), but it is a standard $\lambda$-discretized derivative
(whereas the $x$-derivative of time-to-the-right-ordered functions
is a standard continuous derivative).

A central role in the $\kappa$-Minkowski functional calculus is
played by the ordered exponentials:
\begin{equation}
e^{-i \vec{q} \vec{x}} e^{i q_{0} t} ~, \label{eq:app}%
\end{equation}
where $\{q_{j},q_{0}\}$ are four real numbers and $\{x_{j},t\}$
are $\kappa $-Minkowski coordinates. These ordered exponentials
enjoy a simple property with respect to the generators $p_{\mu}$
of translations of the $\kappa $-Minkowski coordinates:
\begin{equation}
\left\langle p_{\mu}, e^{-i \vec{q} \vec{x}} e^{i q_{0} t}
\right\rangle
=q_{\mu} ~. \label{nocite}%
\end{equation}
We also note that, using the $\kappa$-Minkowski commutation
relations, one finds the relation
\begin{equation}
e^{-i \vec{q} \vec{x}} e^{i q_{0} t}=\exp\left(  i q_{0} t- i
\vec{q} \vec{x}
\frac{\lambda q_{0} }{1-e^{-\lambda q_{0} }}\right)  \label{eq:appb}%
\end{equation}
which turns out to be useful in certain applications.

The ordered exponentials $e^{-i\vec{q}\vec{x}}e^{i\omega t}$ also
play the role of plane waves in $\kappa$-Minkowski since on the
mass-shell (\textit{i.e.}
$\mathcal{C}_{\kappa}(q_{0},\vec{q})=M^{2}$) they are
solutions~\cite{hep-th/9907110} of the relevant wave (deformed
Klein-Gordon) equation:
\begin{equation}
(\square-M^{2}) \left[  e^{-i \vec{q} \vec{x}} e^{i q_{0} t}
\right]  =0
\label{waveeq}%
\end{equation}
where $\square=\partial_{\mu}\partial^{\mu}L^{-1}$ is the
$\kappa$-deformed D'Alembert operator, properly
defined~\cite{majidBOOK,hep-th/9907110} in terms of the so-called
``$\kappa$-Minkowski shift operator" $L$
\[
\;L:\!f(\vec{x},t)\!: =e^{-\lambda
p_{0}}\triangleright:\!f(\vec{x},t)\!:\;
=\;:\!f(\vec{x},t+i\lambda)\!:
\]

The ordered exponentials are also the basic ingredient of the
Fourier theory on $\kappa$-Minkowski. This Fourier
theory~\cite{majidBOOK} is constructed in terms of the canonical
element $\sum_{i}{e_{i}\otimes f^{i}}$, where $\{e_{i}\}$ and
$\{f^{j}\}$ are dual bases, which satisfy the relation
$<e_{i},f^{j}>\,\,=\delta_{i}^{j}$. On the basis of
(\ref{ad1}-\ref{ad3}) one finds that
%
%
%
%
%
%
%
%
%
the canonical element is
\begin{equation}
\psi_{(q_{0},\vec{q})}(t,\vec{x})=\sum_{n_{0},n_{1},n_{2},n_{3}}^{0,\infty
}\frac{(-iq_{1}x_{1})^{n_{1}}}{n_{1}!}\frac{(-iq_{2}x_{2})^{n_{2}}}{n_{2}%
!}\frac{(-iq_{3}x_{3})^{n_{3}}}{n_{3}!}\frac{(iq_{0}t)^{n_{0}}}{n_{0}%
!}=e^{-i\vec{k}\vec{x}}e^{i\omega t} \label{eq:scan}%
\end{equation}
The canonical element (\ref{eq:scan}) retains the notable feature
that, if we define the transform $\tilde{f}(q)$ of an ordered
function $:\!f(x)\!:$ through
\[
:\!f(x)\!:\;=\int\tilde{f}(q)\,e^{-i\vec{q}\vec{x}}e^{iq_{0}t}\,\frac
{e^{3\lambda q_{0}}d^{4}q}{(2\pi)^{4}}~,
\]
the choice of the integration measure $e^{3\lambda q_{0}}$ and the
definition (\ref{eq:WsuX}) of the actions of boosts/rotations on
the coordinates guarantee that
\[
w\,\triangleright:\!f(x)\!:\;=\int\left(  S(w)\triangleright\tilde
{f}(q)\right)  e^{-i\vec{q}\vec{x}}e^{iq_{0}t}\,\frac{e^{3\lambda q_{0}}%
d^{4}q}{(2\pi)^{4}}%
\]
for each $w\in U(so_{1,3})$. This is a relevant property because
it implies that under a finite transformation both $:\!f\!:$ and
$\tilde{f}$ change, but they remain connected by the
Fourier-transform relations. The action of a transformation on the
$x$ is equivalent to the inverse transformation on the $q$. This
is exactly what happens in the classical-Minkowski case ($\lambda
=0$), through the simple relation
\[
f(x)\mapsto f_{\Lambda}(x)=\int\!\tilde{f}(\Lambda^{-1}q)\,e^{iqx}%
\,\frac{d^{4}q}{(2\pi)^{4}}.
\]
In $\kappa$-Minkowski the action of boosts does not allow
%
%
%
%
%
%
%
%
%
description in terms of a matrix $\Lambda_{\mu}^{\nu}$, but it is
still true that the action of a transformation on the $x$ is
equivalent to the ``inverse transformation'' on the $q$ (where, of
course, here the ``inverse transformation'' is described through
the antipode).\bigskip

\subsection{Group velocity in $\kappa$-Minkowski}

\qquad The elements of $\kappa$-Minkowski functional analysis we
reviewed in Section~2 allow us to implement a consistent
deformation of the analysis that applies in commutative Minkowski
spacetime, here reviewed in the preceding subsection. In order to
present specific formulas we adopt the $\kappa $-Minkowski
functional analysis based on time-to-the-right-ordered
noncommutative functions, but the careful reader can easily verify
that the same result for the group velocity is obtained adopting
the time-to-the-left ordering prescription.

We are little concerned with the concept of phase velocity (which
is not a physical velocity). In this respect we just observe that
the phase velocity should be a property of the $\kappa$-Minkowski
plane wave
\begin{equation}
\psi_{(\omega,\vec{k})}=e^{-i\vec{k}\vec{x}}e^{i\omega t}~, \label{pva}%
\end{equation}
and, since the $\kappa$-Minkowski calculus is structured in such a
way that the properties of time-to-right-ordered functions are
just the ones of the corresponding commutative function, this
suggests that the relation
\begin{equation}
v_{ph}=\frac{\omega}{k} \label{pvc}%
\end{equation}
should be valid.

But let us focus on the more significant (physically meaningful)
analysis of group velocity. Our starting point is the wave packet
\[
\Psi_{(\omega_{0},\vec{k}_{0})}=\int
e^{-i\vec{k}{\cdot}\vec{x}}e^{i\omega
t}d\mu.\label{kmpack}%
\]
In this equation (\ref{kmpack}) for simplicity we denote with
$d\mu$ an integration measure which includes the spectrum of the
packet. In fact, the precise structure of the wave packet is
irrelevant for the analysis of the group velocity: it suffices to
adopt a packet which is centered at some
$(\omega_{0},\vec{k}_{0})$ (with $(\omega_{0}$ and $\vec{k}_{0})$
related through Eq.~(\ref{casimir}), the dispersion relation, the
mass Casimir, of the classical Poincar\'{e} algebra) and has
support only on a relatively small
neighborhood of $(\omega_{0},\vec{k}_{0})$, \textit{i.e.} $\omega_{0}%
-\Delta\omega\leq\omega\leq\omega_{0}+\Delta\omega$ and $\vec{k}_{0}%
-\Delta\vec{k}\leq\vec{k}\leq\vec{k}_{0}+\Delta\vec{k}$.

Next, in order to proceed just following the same steps of the
familiar commutative-spacetime case, we should factor out of the
integral a ``pure phase'' with frequency and wavelength fixed by
the wave-packet center: $(\omega_{0},\vec{k}_{0})$. Consistently
with the nature of the time-to-the-right-ordered functional
calculus the phase $e^{ik_{0}x}$ will be factored out to the left
and the phase $e^{-i\omega_{0}t}$ will be factored out to the
right:
\begin{equation}
\Psi_{(\omega_{0},\vec{k}_{0})}=e^{-i\vec{k}_{0}{\cdot}\vec{x}}\left[
\int e^{-i\Delta\vec{k}{\cdot}\vec{x}}e^{i\Delta\omega
t}d\mu\right]
e^{i\omega_{0}t}\! \label{eq:pacchetto}%
\end{equation}
This way to extract the phase factor preserves the
time-to-the-right-ordered structure of the wave
$\Psi_{(\omega_{0},\vec{k}_{0})}$, and therefore, also taking into
account the role that time-to-the-right-ordered functions have in
the $\kappa$-Minkowski calculus, should allow an intuitive
analysis of its properties.

From (\ref{eq:pacchetto}) one recognizes the $\kappa$-Minkowski
group velocity as
\begin{equation}
v_{g}=\lim_{\Delta\omega\rightarrow0}\frac{\Delta\omega}{\Delta k}
=\frac{d\omega}{dk}~, \label{gva}%
\end{equation}
just as in Galilei and Minkowski spacetime. Just as one does in
commutative Minkowski spacetime, the integral can be seen as the
amplitude of the wave, the group velocity $v_{g}$ is the velocity
of translation of this wave amplitude, which be meaningfully
introduced only in the limit of narrow packet (small
$\Delta\omega$ and $\Delta\vec{k}$).

Notice that
\begin{equation}
e^{-i\Delta\vec{k}\vec{x}}e^{i\Delta\omega t}
=\exp\Big(i\Delta\omega t
-i\Delta\vec{k}\vec{x} \frac{\lambda\Delta\omega}{1-e^{-\lambda\Delta\omega}%
}\Big) ~, \label{pvf}%
\end{equation}
and
\begin{equation}
\left[  \exp\Big(i\Delta\omega t -i\Delta\vec{k}\vec{x}
\frac{\lambda \Delta\omega}{1-e^{-\lambda\Delta\omega}}\Big)
\right]  _{\Delta \omega\rightarrow0} = \exp\Big(i\Delta\omega t
-i\Delta\vec{k}\vec{x}\Big) ~,
\label{pvfbis}%
\end{equation}
and therefore the evaluation of the velocity of translation of
this wave amplitude turns out to be independent of the way in
which the exponentials are arranged (but this is an accident due
to the fact that for small $\Delta
\omega$ and $\Delta\vec{k}$ one finds that $[e^{-i\Delta\vec{k}\vec{x}%
},e^{i\Delta\omega t}]=0$.

%
%
%
%
%
%
%
%
%

\section{Comparison with previous analyses}

\qquad Because of the mentioned interest in the phenomenological
implications~\cite{astro-ph/9712103,gr-qc/9810044,gr-qc/9910089,astro-ph/0004225,
gr-qc/0204051,astro-ph/9912136,astro-ph/0009271}, the introduction
of group velocity in $\kappa$-Minkowski has been discussed in
several studies. In the large majority of these studies the
concept of group velocity was not introduced constructively (it
was not a {\underline{result}} obtained in a full theoretical
scheme: it was just introduced through an {\underline {\textit{ad
hoc}}} relation). This appeared to be harmless since the
\textit{ad-hoc} assumption relied on the validity of the relation
$v_{g}=dE/dp$, which holds in Galilei spacetime and Minkowski
spacetime (and for which the structure of $\kappa$-Minkowski
appears to pose no obstacle).

Taking as starting point the approach to $\kappa$-Minkowski
proposed in Ref.~\cite{hep-th/9907110}, we have here shown through
a dedicated analysis that the validity of $v_{g} = dE/dp$ indeed
follows automatically from the structure of $\kappa$-Minkowski and
of the associated functional calculus.

At this point it is necessary for us to clarify which erroneous
assumptions led to the claims reported in
Refs.~\cite{hep-th/0107054,gr-qc/0111056,gr-qc/0208002}, which
questioned the validity of $v_{g} = dE(p)/dp$ in
$\kappa$-Minkowski.

\subsection{Tamaki-Harada-Miyamoto-Torii analysis}

\qquad It is rather easy to compare our analysis with the study
reported by
Tamaki, Harada, Miyamoto and Torii in Refs.~\cite{gr-qc/0111056,gr-qc/0208002}%
. In fact, Ref.~\cite{gr-qc/0111056} explicitly adopted the same
approach to $\kappa$-Minkowski calculus that we adopted here, with
Fourier transform and functional calculus that make direct
reference to time-to-the-right-ordered functions. Also the scheme
of analysis is analogous to ours, in that it attempts to derive
the group velocity from the analysis of the time evolution of a
superposition of plane waves. However, the $\kappa$-Minkowski
functional calculus was applied inconsistently in
Ref.~\cite{gr-qc/0111056}: at the stage of the analysis were one
should factor out the phases $e^{-i\vec{k}_{0}{\cdot }\vec{x}}$
and $e^{i\omega_{0}t}$ from the wave amplitude (as we did in
Eq.~(\ref{eq:pacchetto})) Ref.~\cite{gr-qc/0111056} does not
proceed consistently with the time-to-the-right-ordered functional
calculus. Of course, as done here, in order to maintain the
time-to-the-right-ordered form
of the wave packet it is necessary to factor out the phases $e^{-i\vec{k}%
_{0}{\cdot}\vec{x}}$ and $e^{i\omega_{0}t}$ respectively to the
left and to the right, as we did here. Instead in
Ref.~\cite{gr-qc/0111056} both phases are factored out to the left
leading to a form of the wave packet which is not
time-to-the-right ordered. In turn this leads to the erroneous
conclusion that $v_{g}(k)\neq d\omega(k)/dk$, \textit{i.e.}
$v_{g}(p)\neq dE(p)/dp$.

This inconsistency with the ordering conventions is the key factor
that affected Ref.~\cite{gr-qc/0111056} failure to reproduce
$v_{g}(p) \neq dE(p)/dp$, but for completeness we note here also
that Ref.~\cite{gr-qc/0111056} leads readers to the erroneous
impression that in order to introduce the group velocity in
$\kappa$-Minkowski one should adopt the approximation
\begin{equation}
e^{-i\vec{k}\vec{x}} e^{i\omega t} \sim e^{-i\vec{k}\vec{x} +
i\omega t} ~,
\label{jap1}%
\end{equation}
for generic values of $\omega$ and $\vec{k}$. Actually, unless
$\omega$ and $\vec{k}$ are very small, this approximation is very
poor: it only holds in zeroth order in the noncommutativity scale
$\lambda$ and therefore it does not describe reliably the
structure of $\kappa$-Minkowski (since it fails already in leading
order in $\lambda$, it does not even reliably characterize the
main differences between classical Minkowski and
$\kappa$-Minkowski). As we showed here there is no need for the
approximation (\ref{jap1}) in the analysis of the group velocity
of a wave packet in $\kappa$-Minkowski.
%
%
%
%
%
%
%
%
%

\subsection{$\kappa$-Deformed phase space}

\qquad As discussed in the preceding Subsection, it is very easy
to compare our study with the study reported in
Ref.~\cite{gr-qc/0111056}, since both studies adopted the same
approach. We must now provide some guidance for the comparison
with the study reported by Kowalski-Glikman in
Ref.~\cite{hep-th/0107054}. Also this comparison is significant
for us since Ref.~\cite{hep-th/0107054}, like
Ref.~\cite{hep-th/0107054}, questioned the validity of the
relation $v_{g}=dE(p)/dp$, which instead emerged from our
analysis.

Our approach to $\kappa$-Minkowski, which originates from
techniques developed in
Refs.~\cite{hep-th/9405107,hep-th/9907110}, is profoundly
different from the one adopted in Ref.~\cite{hep-th/0107054}. In
fact, the differences start off already at the level of the action
of $\kappa$-Poincar\'{e} generators on $\kappa$-Minkowski
coordinates. The actions we adopted are described in Section~2.
They take a simple form on time-to-the-right ordered functions,
but they do not allow description as a ``commutator action'' on
generic ordering of functions in $\kappa$-Minkowski. Instead in
Ref.~\cite{hep-th/0107054} the action of the $\kappa$-Poincar\'{e}
generators on $\kappa$-Minkowski coordinates was introduced in
fully general terms as a commutator action. This
would allow to introduce a ``phase-space extension'' of $\kappa$%
-Minkowski~\cite{hep-th/0107054}
\begin{equation}
\lbrack x_{0},x_{j}]=i\lambda x_{j},\qquad\lbrack
p_{0},x_{0}]=-i,\qquad
\lbrack p_{k},x_{j}]=i\delta_{jk}e^{-\lambda p_{0}},\qquad\lbrack p_{j}%
,x_{0}]=[p_{0},x_{j}]=0. \label{pskow}%
\end{equation}

Taking this phase space (\ref{pskow}) as starting point,
Kowalski-Glikman then found, after a rather lengthy analysis, that
``massless particles move in spacetime with universal speed of
light'' $c$~\cite{hep-th/0107054}, in conflict with the relation
$v_{g}=dE(p)/dp$ and the structure of the mass Casimir
(\ref{casimir}). Kowalski-Glikman argued that this puzzling
conflict with the structure of the mass Casimir might be due to a
mismatch between the mass-Casimir relation, $E(p,m)$, and the
dispersion relation, $\omega(k,m)$: the puzzle could be
explained~\cite{hep-th/0107054} if the usual identifications
$k\sim p$ and $\omega\sim E$ were to be replaced by $k\sim
pe^{\lambda E}$ and $\omega\sim sinh(\lambda E)/\lambda+\lambda p^{2}%
e^{\lambda E}/2$.

We observe that the correct explanation of the puzzling result
obtained by Kowalski-Glikman is actually much simpler: the
commutator action (\ref{pskow}) adopted in
Ref.~\cite{hep-th/0107054}, in spite of the choice of symbols
$p_{j}$,$x_{k}$, {\underline{cannot}} describe the action of
``momenta" $p_{j}$ on coordinates $x_{k}$. Momenta should generate
translations of the coordinates, which requires that they may be
represented as derivatives of
functions of the coordinates, but the commutator action $[p_{k},x_{j}%
]=i\delta_{jk}e^{-\lambda p_{0}}$ clearly does not allow to
represent $p_{k}$ as a derivative with respect to the $x_{k}$
coordinate, because of the spurious factor $e^{-\lambda p_{0}}$.
Similarly, those commutation relations do not allow to represent
the $x_{k}$ coordinate as a derivative with respect to $p_{k}$,
and therefore in a Hamiltonian theory, with Hamiltonian $H$, one
would find
\begin{equation}
{\dot{x}_{j}} \sim[x_{j},H] \neq{\frac{dH }{dp_{j}}}~, \label{kowHAMILT}%
\end{equation}
and this is basically the reason for the puzzling result $v_{g}(p)
\neq dE(p)/dp$ obtained in Ref.~\cite{hep-th/0107054}.
Kowalski-Glikman finds a function $v_{g}(p)$ but this function
cannot be seen as describing the relation between velocity and
momentum, since the ``$p$" symbol introduced in (\ref{pskow}) does
not generate translations of coordinates, and therefore ``$p$" is
not a momentum.



\chapter{Quantum Field Theories on Canonical Spacetimes}

\footnotetext{* In this Chapter we review the issue of QFT
construction in noncommutative spacetimes.}*\qquad In this chapter
we discuss the construction of quantum field theory in
noncommutative spacetimes. We observe that the analysis of
quantization in $\kappa$-Minkowski spacetime is still at a
preliminary stage. Thus we focus mainly on the quantization in
canonical spacetime which has been extensively analyzed in
literature. The most common strategy to construct a quantum field
theory on noncommutative space is through the Weyl map which
allows to introduce certain structures in noncommutative
spacetimes in analogy with the corresponding structures in
commutative spacetime. The product of functions in the commutative
space is described through a deformed ``*'' or ``Moyal'' product
of corresponding functions. Here we will discuss the main features
of the quantum field theories obtained through this procedure. In
particular we focus on the so-called IR-UV mixing which is one of
the most interesting and distinguishing features of these
theories.

\section{Weyl Quantization in canonical and $\kappa$-Minkowski spacetime}

\subsection{Weyl Quantization in the phase space of ordinary quantum mechanics}

\qquad Weyl quantization is a technique used (see
e.g.\cite{WeylMoyal2,WeylMoyal}) to describe quantum mechanics
using the phase-space of classical mechanics. The general idea
consists in defining a mapping between the algebra of operators on
the quantum phase space and the algebra of functions on the
classical phase space. This map must of course be compatible with
the product in the noncommutative algebra of quantum operators. To
achieve this goal one has to deform the usual product of
commutative functions on the phase space in a new product called
Weyl-Moyal
product. In quantum mechanics the Weyl map is given by:%
\begin{equation}
W:F(x,p)\rightarrow W[F]=\frac{1}{(2\pi)^{4}}\int d\alpha^{4}d\beta^{4}%
\tilde{F}(\alpha,\beta)\exp i(\alpha\hat{x}+\beta\hat{p})
\end{equation}
where $\tilde{F}(\alpha,\beta)$ is the Fourier transform of
$F(x,p).$ If we
consider products of functions of the quantum phase space we obtain%
\begin{equation}
W[F]W[G]=\frac{1}{(2\pi)^{8}}\int d\alpha^{4}d\beta^{4}d\alpha^{\prime4}%
d\beta^{\prime4}\tilde{F}(\alpha,\beta)\tilde{F}(\alpha^{\prime},\beta
^{\prime})\exp i(\alpha\hat{x}+\beta\hat{p})\exp
i(\alpha^{\prime}\hat
{x}+\beta^{\prime}\hat{p}). \label{wpqm}%
\end{equation}

Since it holds that $W[F]W[G]\neq W[FG],$ it follows that, as
introduced, the Weyl map does not preserve the usual product of
commutative functions on the phase space. However the same formula
(\ref{wpqm}) suggests how to modify the product to have
compatibility. We notice that (\ref{wpqm}) can be easily
evaluated using the commutation relation $[\hat{x}_{i},\hat{p}_{j}%
]=i\hslash\delta_{ij}$ and the Baker-Campbell-Hausdorff expansion%
\begin{equation}
\exp i(\alpha\hat{x}+\beta\hat{p})\exp
i(\alpha^{\prime}\hat{x}+\beta^{\prime }\hat{p})=\exp
i[(\alpha+\alpha^{\prime})\hat{x}+(\beta+\beta^{\prime})\hat
{p}]\exp[\frac{i}{2}\hslash(\alpha\beta^{\prime}-\alpha^{\prime}\beta)],
\end{equation}
from which follows that%
\begin{align}
W[F]W[G]  &  =\frac{1}{(2\pi)^{8}}\int
d\alpha^{4}d\beta^{4}d\alpha^{\prime
4}d\beta^{\prime4}\tilde{F}(\alpha,\beta)\tilde{F}(\alpha^{\prime}%
-\alpha,\beta^{\prime}-\beta)e^{i[\alpha\hat{x}+\beta\hat{p}]}e^{\frac{i}%
{2}\hslash\lbrack\alpha\beta^{\prime}-\alpha^{\prime}\beta]}=\\
&  =\frac{1}{(2\pi)^{4}}\int d\alpha^{4}d\beta^{4}\exp i[\alpha\hat{x}%
+\beta\hat{p}]\tilde{A}(\alpha,\beta)
\end{align}
where%
\begin{equation}
\tilde{A}(\alpha,\beta)=\frac{1}{(2\pi)^{4}}\int d\alpha^{\prime4}%
d\beta^{\prime4}\tilde{F}(\alpha,\beta)\tilde{F}(\alpha^{\prime}-\alpha
,\beta^{\prime}-\beta)\exp\frac{i}{2}\hslash\lbrack\alpha\beta^{\prime}%
-\alpha^{\prime}\beta].
\end{equation}

Thus one is led to introduce a new product ``$\ast$'' between
commutative functions on the classical phase space such that
\begin{equation}
F(x,p)\ast G(x,p)=W^{-1}\left(  W[F]W[G]\right)
=\frac{1}{(2\pi)^{4}}\int d\alpha^{4}d\beta^{4}\exp i[\alpha
x+\beta p]\tilde{A}(\alpha,\beta).
\end{equation}

One can easily verify that this $\ast$ product has the following
differential
expression on the phase space%
\begin{equation}
F(x,p)\ast G(x,p)=F(x,p)\exp\frac{i}{2}\hslash(\frac{\overleftarrow{\partial}%
}{\partial x}\frac{\overrightarrow{\partial}}{\partial p}-\frac
{\overrightarrow{\partial}}{\partial x}\frac{\overleftarrow{\partial}%
}{\partial p})G(x,p).
\end{equation}

The Weyl-Moyal $\ast$ product so defined is associative but
noncommutative and
in particular one has that commutation relation is properly mapped%
\begin{equation}
\lbrack x,p]_{\ast}=x\ast p-p\ast x=i\hslash.
\end{equation}

\subsection{Weyl quantization for canonical noncommutativity}

\qquad To deal with the quantization of spacetime one can repeat
for spacetime the procedure just outlined for the phase-space of
quantum mechanics. A first point to notice is the relevance of the
choice of the Weyl map and in particular the choice of the order
of noncommuting coordinates in the expression that defines the
Weyl map. Different choices lead to different Weyl-Moyal products.
A second important point is that this proposal of quantization,
eventually, takes into account only the quantization of spacetime.
Quantization of the phase-space must anyway be implemented
separately. We will assume the conservative hypothesis that it can
be implemented with the usual strategies as, for instance, that of
the path-integral formulation. We start considering Weyl
quantization for canonical spacetime. In this case one can define
the map between functions on
noncommutative spacetime and functions on commutative spacetime as%
\begin{equation}
W_{\theta}:F(x)\rightarrow W_{\theta}[F]=\frac{1}{(2\pi)^{4}}\int
d\alpha ^{4}\tilde{F}(\alpha)\exp(i\alpha_{\mu}\hat{x}^{\mu}),
\end{equation}
where $\tilde{F}(\alpha)$ is the Fourier transform of $F(x).$ Now
let us
consider products of functions of canonical spacetime%
\begin{equation}
W_{\theta}[F]W_{\theta}[G]=\frac{1}{(2\pi)^{8}}\int
d\alpha^{4}d\beta ^{4}\tilde{F}(\alpha)\tilde{G}(\beta)\exp
i(\alpha_{\mu}\hat{x}^{\mu})\exp
i(\beta_{\mu}\hat{x}^{\mu}). \label{wp}%
\end{equation}
Al already discussed it holds that%
\begin{equation}
\exp(i\alpha^{\mu}x_{\mu})\exp(i\beta^{\mu}x_{\mu})=\exp(i\alpha^{\mu}%
\theta_{\mu\nu}\beta^{\nu})\exp[i(\alpha^{\mu}+\beta^{\mu})x_{\mu}],
\end{equation}
thus%
\begin{align}
W_{\theta}[F]W_{\theta}[G]  &  =\frac{1}{(2\pi)^{8}}\int
d\alpha^{4}d\beta
^{4}\tilde{F}(\alpha-\beta)\tilde{G}(\beta)\exp(i\alpha^{\mu}\theta_{\mu\nu
}\beta^{\nu})\exp[i\alpha^{\mu}x_{\mu}]=\\
&  =\frac{1}{(2\pi)^{4}}\int
d\alpha^{4}\exp[i\alpha^{\mu}x_{\mu}]\tilde {A}(\alpha),
\end{align}
where%
\begin{equation}
\tilde{A}(\alpha)=\frac{1}{(2\pi)^{4}}\int
d\beta^{4}\tilde{F}(\alpha
-\beta)\tilde{G}(\beta)\exp(i\alpha^{\mu}\theta_{\mu\nu}\beta^{\nu}).
\end{equation}

Again it is natural to introduce a Moyal $\star$-product
\begin{equation}
F(x)\star G(x)=\frac{1}{(2\pi)^{4}}\int
d\alpha^{4}\exp[i\alpha^{\mu}x_{\mu
}]\tilde{A}(\alpha)=F(x)\exp[\frac{i}{2}\theta_{\mu\nu}\frac{\overleftarrow
{\partial}}{\partial
x^{\mu}}\frac{\overrightarrow{\partial}}{\partial x^{\nu
}}]G(x). \label{mp}%
\end{equation}

In this way we automatically have that
$W_{\theta}[F]W_{\theta}[G]=W_{\theta }[F\star G].$ Also this
Moyal $\star$-product is associative but it is noncommutative. One
can easily verify that the commutation relations are
properly mapped:%
\begin{equation}
\lbrack
x_{\mu},x_{\nu}]_{\star}=x_{\mu}x_{\nu}-x_{\nu}x_{\mu}=i\theta_{\mu
\nu}.
\end{equation}

For the product of several waves one finds%
\begin{equation}
e^{ik_{1}x}\star e^{ik_{2}x}\star...\star e^{ik_{n}x}=e^{ix\sum_{i=1}^{n}%
k_{j}-\frac{i}{2}\sum_{i,j=1,i<j}^{n}k_{i}\theta k_{j}},
\end{equation}
whereas complex coniugation gives%
\begin{equation}
\overline{F\star G}=\overline{F}\star\overline{G}.
\end{equation}

One can also introduce integrals directly in the noncommutative
space by
defining%
\begin{equation}
\int
d\hat{x}^{4}:\exp(ik_{\mu}\hat{x}^{\mu}):_{\theta}\equiv\delta^{4}(k),
\label{ci}%
\end{equation}
where $\hat{x}^{\mu}$ are noncommutative coordinates. Using the
above formula
one can calculate integrals of every noncommutative function as follows%
\begin{equation}
\int d\hat{x}^{4}W[F]=\frac{1}{(2\pi)^{4}}\int d\alpha^{4}\tilde{F}%
(\alpha)\int
d\hat{x}^{4}\exp(i\alpha_{\mu}\hat{x}^{\mu})=\tilde{F}(0)=\int
dx^{4}F(x).
\end{equation}

It also results that%
\begin{equation}
\int d\hat{x}^{4}W[F]W[G]=\int dx^{4}F(x)\star G(x).
\end{equation}

Therefore canonical noncommutative integrals defined by (\ref{ci})
reduce to the $\star$-deformed commutative integrals. Using these
relations and the Fourier transform properties it is easy to
verify the following properties of the star product under
integration:

\begin{enumerate}
\item $\int dx^{4}F(x)\star G(x)=\int dx^{4}F(x)G(x),$

\item $\int dx^{4}F_{1}(x)\star F_{2}(x)\star...\star
F_{n}(x)=\int dx^{4}F_{n}(x)\star F_{1}(x)\star...\star
F_{n-1}(x),$

\item $\int dx^{4}F_{1}(x)\star F_{2}(x)\star...\star
F_{n}(x)=\int
\frac{dp_{1}^{4}...dp_{n}^{4}}{\left(  2\pi\right)  ^{n-1}}\widetilde{F}%
_{1}(p_{1})...\widetilde{F}_{n}(p_{n})\delta^{4}(p_{1}+...+p_{n})e^{-\frac
{i}{2}\sum_{i,j,i<j}^{1...n}k_{i}\theta k_{j}}.$
\end{enumerate}

The first property implies that under integration the
$\star$-product of two functions is the same as the common product
of functions. The second properties says that under integrations
the $\star$-product is invariant under cyclic permutations. The
third property relates (under integrations) $\star $-products of
functions with products of their Fourier transforms.

One can also introduce a differential calculus on canonical
noncommutative
spacetime defining derivatives through the formulas%
\begin{align}
\lbrack\hat{\partial}_{i},x_{j}]  &  =\delta_{ij},\label{der1}\\
\lbrack\hat{\partial}_{i},\hat{\partial}_{j}]  &  =0. \label{der2}%
\end{align}

These formulas are compatible with the spacetime commutation relation%
\begin{equation}
\hat{\partial}_{k}[\hat{x}_{i},\hat{x}_{j}]=\hat{\partial}_{k}\theta_{ij}=0.
\end{equation}

It is also easy to verify that
\begin{equation}
\hat{\partial}_{k}W_{\theta}[F]=W_{\theta}[\partial_{k}F],
\end{equation}
and that%
\begin{equation}
\int d\hat{x}^{4}\hat{\partial}_{k}W_{\theta}[F]=\int
d\hat{x}^{4}W_{\theta }[\partial_{k}F]=\int dx^{4}\partial_{k}F=0.
\end{equation}

With these definitions of integrals and derivatives, it is
possible to define noncommutative versions of the usual
commutative actions. An example that we will analyze in detail in
the following sections is provided by the
noncommutative scalar theory whose action is%
\begin{equation}
S=\int d\hat{x}^{4}\left[  \frac{1}{2}\left(
\hat{\partial}_{\mu}W_{\theta }[\varphi]\right)
^{2}+\frac{m^{2}}{2}W_{\theta}[\varphi]^{2}+\frac{\lambda
}{4!}W_{\theta}[\varphi]^{4}\right]  .
\end{equation}

In terms of commutative functions this action is written as:%
\begin{equation}
S=\int dx^{4}\left[  \frac{1}{2}\left(
\partial_{\mu}\varphi\right)
^{2}+\frac{m^{2}}{2}\varphi^{2}+\frac{\lambda}{4!}\varphi\star\varphi
\star\varphi\star\varphi\right]  .
\end{equation}

\subsection{Weyl quantization for $\kappa$-Minkowski spacetime}

\qquad In the case of $\kappa$-Minkowski spacetime the procedure
is exactly
the same. The first step is to fix the Weyl map%
\begin{equation}
W_{\kappa}:F(x)\rightarrow W_{\kappa}[F]=\frac{1}{(2\pi)^{4}}\int
d\alpha ^{4}\tilde{F}(\alpha):\exp
i(\alpha_{\mu}x^{\mu}):_{\kappa},
\end{equation}
which means to choose the ordering in $:\exp i(\alpha
x):_{\kappa}$. On the basis of the considerations already
discussed in the previous chapter one is
naturally led to adopt the ordering prescription%
\begin{equation}
:\exp i(\alpha_{\mu}x^{\mu}):_{\kappa}=\exp i(\alpha_{i}x^{i})\exp
i(\alpha_{0}x^{0}).
\end{equation}
Now let us consider the products of functions in the quantum spacetime%
\begin{equation}
W_{\kappa}[F]W_{\kappa}[G]=\frac{1}{(2\pi)^{8}}\int
d\alpha^{4}d\beta\tilde {F}(\alpha)\tilde{G}(\beta):\exp
i(\alpha_{\mu}x^{\mu}):_{\kappa}:\exp
i(\beta_{\mu}x^{\mu}):_{\kappa}. \label{qqq}%
\end{equation}

Again we have that $W[F]W[G]\neq W[FG],$ which means that the Weyl
map does not preserve the usual product of commutative function.
Using the commutation relation of $\kappa$-Minkowski and the
Baker-Campbell-Hausdorff expansion one
finds%
\begin{equation}
:\exp i(\alpha_{\mu}x^{\mu}):_{\kappa}:\exp
i(\beta_{\mu}x^{\mu}):_{\kappa
}=:\exp i[(\alpha_{\mu}\dot{+}\beta_{\mu})x^{\mu}]:_{\kappa}, \label{stanco}%
\end{equation}
where $\dot{+}$ is such that
\begin{equation}
\alpha_{\mu}\dot{+}\beta_{\mu}\equiv\delta_{\mu0}(\alpha_{0}+\beta
_{0})+(1-\delta_{\mu0})[\alpha_{\mu}+\exp(\beta_{0}/\kappa)\beta_{\mu}].
\end{equation}

Using the above expression we can rewrite (\ref{qqq}) as%
\begin{equation}
W_{\kappa}[F]W_{\kappa}[G]=\frac{1}{(2\pi)^{8}}\int
d\alpha^{4}d\beta ^{4}\tilde{F}(\alpha)\tilde{G}(\beta):\exp
i[(\alpha_{\mu}\dot{+}\beta_{\mu })x^{\mu}]:_{\kappa}.
\end{equation}
Therefore the Weyl-Moyal product for $\kappa$-Minkowski is
\begin{equation}
F(x)\overset{\kappa}{\ast}G(x)=W_{\kappa}^{-1}(W_{\kappa}[F]W_{\kappa
}[G])=F(x)\exp[ix^{\mu}\sigma_{\mu}\left(  \overleftarrow{\partial}%
_{x},\overrightarrow{\partial}_{x}\right)  ]G(x),
\end{equation}

where%
\begin{align}
\sigma_{0}(\alpha,\beta)  &  =\alpha_{0}+\beta_{0}\\
\sigma_{i}(\alpha,\beta)  &
=(1-\delta_{\mu0})[\alpha_{\mu}+\exp(\beta
_{0}/\kappa)\beta_{\mu}]
\end{align}

The $\overset{\kappa}{\ast}$-product is associative but it is
noncommutative.
And in particular we have that the commutation relations are properly mapped%
\begin{align}
\lbrack x_{0},x_{i}]_{\overset{\kappa}{\ast}}  &
=x_{0}\overset{\kappa}{\ast
}x_{i}-x_{i}\overset{\kappa}{\ast}x_{0}=\lambda x_{i},\\
\lbrack x_{i},x_{j}]_{\overset{\kappa}{\ast}}  &
=x_{i}\overset{\kappa}{\ast
}x_{j}-x_{j}\overset{\kappa}{\ast}x_{i}=0,
\end{align}
and waves combine in agreement with (\ref{stanco})%
\begin{equation}
\exp i(\alpha_{\mu}x^{\mu})\overset{\kappa}{\ast}\exp
i(\beta_{\mu}x^{\mu })=\exp
i(\alpha_{\mu}\dot{+}\beta_{\mu})x^{\mu}.
\end{equation}

Again one can introduce integrals directly in the noncommutative
space by
defining%
\begin{equation}
\int
d\hat{x}^{4}:\exp(ik_{\mu}\hat{x}^{\mu}):_{\kappa}\equiv\delta_{\kappa
}^{4}(k),
\end{equation}
where $\hat{x}^{\mu}$ are the noncommutative coordinates. From the
above formula one can calculate the integrals of every
noncommutative function in
$\kappa$-Minkowski spacetime as follows%
\begin{equation}
\int d\hat{x}^{4}W[F]=\frac{1}{(2\pi)^{4}}\int d\alpha^{4}\tilde{F}%
(\alpha)\int
d\hat{x}^{4}\exp(i\alpha_{\mu}\hat{x}^{\mu})=\tilde{F}(0)=\int
dx^{4}F(x).
\end{equation}

It results that%
\begin{align}
\int d\hat{x}^{4}W[F]W[G]  &  =\int
dx^{4}F(x)\overset{\kappa}{\ast}G(x)=\int
d\alpha^{4}d\beta^{4}F(\alpha)G(\beta)\int dx^{4}\exp
i(\alpha_{\mu}\dot
{+}\beta_{\mu})x^{\mu}=\nonumber\\
&  =\int
d\alpha^{4}d\beta^{4}F(\alpha)G(\beta)\delta_{\kappa}^{4}(\alpha
_{\mu}\dot{+}\beta_{\mu}). \label{a}%
\end{align}

One can also verify the useful formula%
\begin{equation}
\int
d\alpha^{4}F(\alpha)\delta_{\kappa}^{4}(\alpha_{\mu}\dot{+}\beta_{\mu
})=\mu(p_{0})F(\dot{-}\beta), \label{c}%
\end{equation}
where $\mu(p_{0})$, plays the role of a integration measure. We
observe that most of the properties under integration of the
canonical $\star$-product are not shared by the $\kappa$-Minkowski
$\overset{\kappa}{\ast}$-product. Most notably the integral
$\overset{\kappa}{\ast}$ product of functions is not symmetric
under cyclic permutations of the argument functions and in the
integral of $\overset{\kappa}{\ast}$-product of two functions the
dependence on the noncommutative parameter does not disappear.

We also observe that Fourier momenta combine nonlinearly in the
arguments of the $\delta$- function. This will produce profound
implications in the construction of the quantum field theory.

\subsection{Functional formalism in noncommutative space}

\qquad The basic hypothesis of the most popular approach to QFT in
noncommutative spacetime is that the Hilbert space is not modified
by the noncommutativity so that the physicals relevant information
is still encoded
in the Green functions defined as%
\begin{equation}
G^{(n)}(x_{1,...,}x_{n})=\left\langle 0\left|  T\left\{  \phi(x_{1}%
)...\phi(x_{n})\right\}  \right|  0\right\rangle =\int D\phi\text{
}\phi
(x_{1})...\phi(x_{n})\exp iS_{\theta}(\phi). \label{gfnc}%
\end{equation}

All the dependence on the noncommutativity parameters is contained
in the new action $S_{\theta}(\phi)$. We observe that this popular
approach may be also viewed as an assumption of ``minimality'':
the action is modified but the entire procedure that from the
action lead us to the physical predictions is assumed to be
unaffected by noncommutativity. It is not inconceivable that a
meaningful QFT in NCST might also be introduced through a more
profound modification of the conceptual structure adopted in
commutative-spacetime frameworks, but such alternative
formulations remain so far largely unexplored.

One can obtain the Green functions (\ref{gfnc}) from a\thinspace\
generating functional defined as usual
\begin{equation}
Z[J]=\int D\phi\exp i\left\{  S(\phi)+\int
dx^{4}J(x)\phi(x)\right\}  ,
\end{equation}
where we notice that the source $J(x)$ is coupled to the field
$\phi(x)$ in the usual way. In this scheme all the machinery of
the ordinary field theory can be carried on. For example we define
as usual the generating functional
for the connected Green's function $W[J]$%
\begin{equation}
Z[J]=e^{iW[J]},
\end{equation}
and the 1PI-effective action%
\begin{equation}
\Gamma\lbrack\varphi]=W[J]-\int J(x)\varphi(x)dx^{4},
\end{equation}
where $\varphi(x)=\left\langle 0\left|  \phi(x)\right|
0\right\rangle =\frac{\delta W[J]}{\delta J(x)}.$

Also the usual perturbative expansions (weak coupling expansion,
loop expansion ecc.) in this framework hold unchanged.

\section{Scalar $\lambda\varphi^{4}$-theory in canonical spacetime}

\subsection{Action, functional derivatives and equation of motion}

\qquad According to the arguments discussed in the previous
chapter noncommutative scalar theory is simply obtained from the
usual commutative scalar theory with the only prescription of
substituting every product of the fields with a $\star$-product.
As already emphasized the substitutions do not change the
quadratic part of the action. For example in the case of scalar
$\lambda\varphi^{4}$-theory the action can be written as
\begin{equation}
S=\int dx^{4}\left\{
\frac{1}{2}\partial_{\mu}\phi\partial^{\mu}\phi+\frac
{1}{2}m^{2}\phi^{2}+\frac{\lambda}{4!}\phi\star\phi\star\phi\star\phi\right\}
. \label{actionscalar}%
\end{equation}

As usual one obtains the equation of motion from the request of
stationarity
of the action%
\begin{equation}
\frac{\delta S}{\delta\phi}=0,
\end{equation}
where we have adopted the usual definition of functional
derivative. We stress that we are now working in a commutative
space of functions, in which however just the product is deformed.

From the action (\ref{actionscalar}) and the fact that
\begin{align}
\frac{\delta\phi\star\phi\star\phi\star\phi(x)}{\delta\phi(y)}  &
=\delta
^{4}(x-y)\star\phi\star\phi\star\phi+\phi\star\delta^{4}(x-y)\star\phi
\star\phi+\\
&
+\phi\star\phi\star\delta^{4}(x-y)\star\phi+\phi\star\phi\star\phi
\star\delta^{4}(x-y)
\end{align}
one easily obtains the equation of motion%
\begin{equation}
(\square+m^{2})\phi=\frac{\lambda}{6}\phi\star\phi\star\phi.
\end{equation}

It is worth noticing some differences for the solutions of this
equation of motion with respect to the commutative case. For
example in noncommutative $\lambda\varphi^{4}$ one finds solitonic
solutions whereas in classical (4d-commutative)
$\lambda\varphi^{4}$-scalar theory Derrick theorem prohibits the
existence of all finite energy classical solutions. Derrick
theorem is based on the observation that if all lengths are scaled
as $L\rightarrow\alpha L$ both the kinetic and the potential
energies decrease so that no finite-size minimum can exist. It is
not surprising that this argument fails in presence of a
characteristic length scale, such as $\sqrt{\theta}$ in the
canonical noncommutative case. In fact it was shown in
Ref.~\cite{hep-th/0003160} that for sufficiently large $\theta$
stable solitons can exist in the noncommutative theory.
Mathematically this is due to the fact that equations of the type
$\lambda\phi\star\phi+\phi=0,$ which is a typical example of
solitonic equation for the scalar theory, admit non-trivial
solutions (whereas the corresponding equation
$\lambda\phi^{2}+\phi=0$ for the commutative case only has
constant solutions).

\subsection{Feynman diagrams}

\qquad The analysis of Feynman diagrams in theories constructed on
canonical noncommutative spacetime is particularly interesting
since important differences emerge with respect to the commutative
counterpart. We start
considering a generic scalar interaction given by the vertex%
\begin{equation}
S_{int}=\frac{\lambda}{4!}\int dx^{4}\underset{n}{\underbrace{\phi
\star...\star\phi}}=\frac{\lambda}{4!}\int\frac{dp_{1}^{4}...dp_{n}^{4}%
}{\left(  2\pi\right)  ^{n-1}}\tilde{\phi}(p_{1})...\tilde{\phi}(p_{n}%
)\delta^{4}(p_{1}+...+p_{n})\exp(-\frac{i}{2}\sum_{i,j=1,i<j}^{n}p_{i}\theta
p_{j}). \label{sint}%
\end{equation}

A first observation is that the usual $\delta$-function of
energy-momentum conservation is still present so that the usual
energy-momentum conservation rules, at each vertex, hold
unchanged. This is in agreement with what suggested by the
analysis of symmetries of canonical spacetime in Section
1.4\textbf{. }The important differences with respect to the usual
Feynman rules, of the corresponding commutative interaction, is
the appearance of the
phase factor $V(p_{1},...,p_{n})=\exp(-\frac{i}{2}\sum_{i,j=1,i<j}^{n}%
p_{i}\theta p_{j})$. This factor must be taken into account and in
particular one must preserve the order of the lines attached to
each vertex. The order of the lines attached at each vertex is not
in general important, but in noncommutative case it is crucial.

\subsubsection{Planar diagrams and nonplanar diagrams}

\qquad Noncommutative theories have the feature that the total
contribution to Green functions, while still symmetries under
momenta exchange, is obtained summing vertices which are not
themselves symmetric under the exchange of momenta entering the
vertex. This means that in the usual perturbative expansion one
must pay particular care, even in a single-field theory, keeping
track of the order of the momenta attached at each
vertex\footnote{This operation is of course without consequences
in the commutative counterpart thanks to the symmetry of the
vertex.}. Particularly important with respect to this issue is the
distinction between diagrams that are or nonplanar in the
noncommutative-theory sense. This noncommutative theory concept of
planarity is best introduced through an example.

\begin{figure}[h]
\begin{center}
\includegraphics[width=4cm]{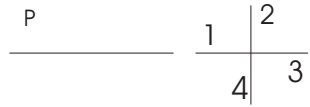}
\end{center}
\caption{External momentum and ordered vertex.}%
\label{figura1}%
\end{figure}

Let us consider the diagrams contributing to the self-energy of
$\lambda \varphi^{4}$ scalar theory. We distinguish the line\
incoming into a vertex by the numbers 1,2,3,4. Given the first
external line (of associate momentum $p$) we can attach the vertex
by one of the lines 1,2,3 or 4. In a vertex without any symmetry
under exchange of the momenta, these different choices correspond
to different contributions. In our case, thanks to the symmetry
under cyclical exchanges, all these choices gives the same
contributions. We have only to multiply the results times the
number of possible choices (four in this case). Now let us
consider the case in which the external momentum $p$ is attached
to the vertex line-1 (Fig.\ref{figura1}).

\begin{figure}[h]
\begin{center}
\includegraphics[width=8cm]{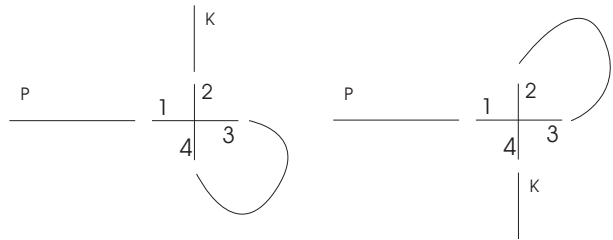}
\end{center}
\caption{Planar connections of the vertex lines.}%
\label{figura2}%
\end{figure}

There are of course three possible ways to attach the second
external line of momentum $k$ to the vertex. If we choose the
vertex line-2 or the vertex line-4 (Fig.\ref{figura2}) we can
connect by a propagator the remaining vertex lines in a way to
obtain planar diagrams

\begin{figure}[h]
\begin{center}
\includegraphics[width=4cm]{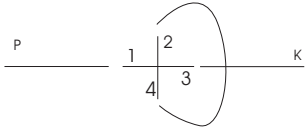}
\end{center}
\caption{Nonplanar connection of the vertex lines.}%
\label{figura3}%
\end{figure}

These diagrams all have will give the same $\star$-product induced
phase contribution ($\exp(-\frac{i}{2}p\theta p)=1$). Instead, if
we attach the second external line to the vertex line-3,
(Fig.\ref{figura3}) the only way we have to connect the remaining
vertex lines 2,4 is through a propagator which intersects at least
one line

This means that this diagrams is nonplanar. Its phase factor is
$\exp (-\frac{i}{2}p\theta k)$. We will discuss the implications
of these phases in the next sections. Finally for the self energy
we have 4 nonplanar diagrams with the phase factor
$\exp(-\frac{i}{2}p\theta k)$ and 8 planar diagrams with phase
factor 1.

More in general nonplanar diagrams are those diagrams than cannot
be, in any way, drawn in a plane without intersecting (at least)
two lines. Planar diagrams are those diagrams that are not
nonplanar. It the usual Feynman diagrammatic planar diagrams,
though always well distinct from the nonplanar ones, give the same
numerical contributions. This is essentially due to the invariance
of the vertices under generic permutation of the momenta. In
theories in which the order of the lines incoming in each vertex
is important, in general, different combinations give different
contributions. However, in the case of the theory we are
considering we have the important properties of the
momenta-conservation at each vertex and the invariance of the
vertex under cyclic permutation of the attached momenta. With this
properties one can prove that the phase factor is the same for all
the possible complex-planar diagrams
and reads%
\begin{equation}
V(p_{1},...,p_{n})=\exp(-\frac{i}{2}\sum_{i<j<n}p_{i}\theta p_{j}%
).\label{plan}%
\end{equation}

It depends only on the order of the external momenta: it is
insensitive to the internal structure of the graph. This implies
that the contribution of a planar graph is precisely the same of
the corresponding diagrams of the corresponding commutative theory
multiplied by $V(p_{1},...,p_{n}).$ The eventual divergencies also
are the same and they can be treated similarly to the commutative
case.

We said that nonplanar diagrams cannot be drawn in a plane in such
way that propagators do not cross each other. It is rather easy to
see that any nonplanar graph, for each crossing of the momenta
$k_{i}$ and $k_{j},$ will
acquire an extra phase%
\begin{equation}
\exp ik_{i}\times k_{j},
\end{equation}
in addition to the phase associated with the ordering of external
momenta.
Therefore one has for the complete phase factor of a nonplanar graph%
\begin{equation}
V(p_{1},...,p_{n})\exp(-\frac{i}{2}\sum_{i<j<n}C_{ij}k_{i}\theta
k_{j}),
\end{equation}
where $V(p_{1},...,p_{n})$ is as in (\ref{plan}) and $C_{ij}$ is
an intersection matrix that counts the number of times that the
i-th line (internal or external)\ crosses the over the j-th line.
Crossing are counted as positive if $p_{i}$ crosses $p_{j}$ with
$p_{j}$ moving to the left. There is not an one to one
correspondence between graphs and $C_{ij}$ matrices since
different way of drawing the graph will lead to different
intersections. However all these matrices give the same Feynman
integral. We see that in the case on nonplanar graph the
$\theta$-dependence cannot be factorized out as in the planar
case. This $\theta$-dependence is the cause of profound
differences in the behavior of the diagrams. In particular one has
that the phase factor improve the UV convergence of the diagrams
and one might expect that, with the exception of divergent planar
subgraphs, all nonplanar graphs to be finite. We will see however
that new IR divergences appear. Moreover as a consequence of the
internal phase factor nonplanar diagrams vanishes in the $\theta
\rightarrow\infty$ limit (strong commutativity limit). In this
limit the theory is the sum of planar diagrams only (planar
limit).

\subsection{One-loop 1PI\ effective action and the IR/UV Mixing.}

\begin{figure}[h]
\begin{center}
\includegraphics[width=4cm]{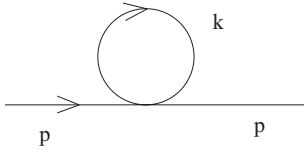}
\end{center}
\caption{Planar tadpole contribution.}%
\label{tp}%
\end{figure}

\begin{figure}[h]
\begin{center}
\includegraphics[width=4cm]{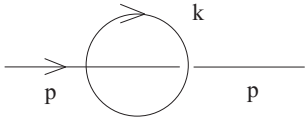}
\end{center}
\caption{Nonplanar tadpole contribution.}%
\label{tnp}%
\end{figure}

\qquad As an explicit example of calculation with this new
diagrammatic we consider the 1PI\ two-point function. At the
lowest order we have that it is
simply the inverse propagator%
\begin{equation}
\Gamma_{0}^{2}(p)=p^{2}+m^{2},
\end{equation}
that is unchanged. At one loop one has to sum the diagrams of
Figs.\ref{tp}-\ref{tnp}

The first (Fig.\ref{tp}) is a planar diagram while the second
(Fig.\ref{tnp}) is a nonplanar diagram. We observe that in the
$\theta\rightarrow0$ limit the two diagrams become the same
(planar) diagrams with the right commutative
combinatorial factor. Their contributions are respectively%
\begin{align}
\Gamma_{\text{pl}}^{2}(p)  &  =\frac{\lambda}{3}\int\frac{dk^{4}}{(2\pi)^{2}%
}\frac{1}{k^{2}+m^{2}},\label{gp}\\
\Gamma_{\text{npl}}^{2}(p)  &  =\frac{\lambda}{6}\int\frac{dk^{4}}{(2\pi)^{2}%
}\frac{\exp[ip\theta k]}{k^{2}+m^{2}}. \label{gnp}%
\end{align}

We observe that in the $\theta\rightarrow0$ limit the integrands
of the two diagrams become equal. The planar contribution
(\ref{gp}) is the same as in the commutative case (up to a
numerical factor) and it is quadratically divergent in the
ultraviolet sector, whereas the nonplanar contribution (\ref{gnp})
is finite thanks to the oscillation produced by the phase in the
integrand. To evaluate explicitly (\ref{gp}-\ref{gnp}) it is
useful to use the
Schwinger\textbf{ }parametrization%
\begin{equation}
\frac{1}{k^{2}+m^{2}}=\int_{0}^{\infty}d\alpha
e^{-\alpha(k^{2}+m^{2})}.
\label{SP}%
\end{equation}
Substituting (\ref{SP}) in (\ref{gp}) and (\ref{gnp}) and
integrating in the
Gaussian variables $k$ one obtains%
\begin{align}
\Gamma_{\text{pl}}^{2}(p)  &
=\frac{\lambda}{48\pi^{2}}\int_{0}^{\infty
}d\alpha\frac{e^{-\alpha m^{2}-\frac{1}{\alpha\Lambda^{2}}}}{\alpha^{2}},\\
\Gamma_{\text{npl}}^{2}(p)  &
=\frac{\lambda}{96\pi^{2}}\int_{0}^{\infty
}d\alpha\frac{e^{-\alpha m^{2}-\frac{p\theta^{2}
p}{\alpha}-\frac{1}{\alpha \Lambda^{2}}}}{\alpha^{2}}.
\end{align}

We have introduced in both diagrams explicitly the cut-off
$\Lambda.$ We observe that the planar contribution behaves as
usual (i.e. it has a leading-quadratic divergence for large
momenta), whereas the nonplanar contribution is finite even after
removing the cutoff. The evaluation of the
above expressions gives%
\begin{align}
\Gamma_{\text{pl}}^{2}(p)  &  =\frac{\lambda}{48\pi^{2}}(\Lambda^{2}-m^{2}%
\ln(\frac{\Lambda^{2}}{m^{2}})+O(1)),\label{ft}\\
\Gamma_{\text{npl}}^{2}(p)  &
=\frac{\lambda}{96\pi^{2}}m^{2}\sqrt
{\frac{\Lambda_{\text{eff}}^{2}}{m^{2}}}K_{1}\left(
\frac{m}{\Lambda
_{\text{eff}}^{2}}\right)  =\frac{\lambda}{96\pi^{2}}(\Lambda_{\text{eff}}%
^{2}-m^{2}\ln(\frac{\Lambda_{\text{eff}}^{2}}{m^{2}})+O(1)),
\end{align}
where%
\begin{equation}
\Lambda_{\text{eff}}^{2}=\frac{1}{1/\Lambda^{2}+p\theta^{2} p}%
\end{equation}
and $K_{1}(x)$ is a modified-Bessel function of the first kind.

We observe that the contributions coming from the nonplanar
diagrams remain finite in the $\Lambda\rightarrow\infty$ limit and
that in the same limit $\Lambda_{\text{eff}}^{2}\rightarrow (
p\theta^{2}p)^{-1} .$ Noncommutativity regularizes the divergences
in the nonplanar diagrams but the planar ones remain divergent as
in the commutative case. Moreover unexpected\footnote{In the usual
quantum picture of spacetime a caracteristic length scale $\alpha$
introduce new features only at lenght scale smaller than $\alpha$.
Instead in the case of canonical noncommutativity, where the new
length scale is $\sqrt{\theta},$ corrections are introduced at
lenght scale larger than 1/$\sqrt{\theta}.$} (IR-)divergences
appear in the limit of vanishing momenta $p\rightarrow0.$
Explicitly up to one-loop the two-point effective action reads%
\begin{align}
\Gamma^{2}(\phi)  &  =\int dp^{4}\phi(p)\phi(-p)\frac{1}{2}\left\{
p^{2}+m_{R}^{2}+\right. \label{e12}\\
&  \left.  -\frac{\lambda}{96\pi^{2}}\left(
\frac{1}{1/\Lambda^{2}+p\theta^{2}
p}-m_{R}^{2}\ln(\frac{1}{m_{R}^{2}(1/\Lambda^{2}+p\theta^{2}
p)})+O(1)\right) +O(\lambda^{2})\right\}  ,
\end{align}
where
\begin{equation}
m_{R}^{2}=m^{2}+\frac{\lambda^{2}}{48\pi^{2}}(\Lambda^{2}-m^{2}\ln
(\frac{\Lambda^{2}}{m^{2}})+O(1))
\end{equation}
is the renormalized mass. It is worth noticing that

\begin{itemize}
\item  In the limit $p\theta^{2}p\gg1/\Lambda^{2}$ one recovers
$\Lambda _{\text{eff}}^{2}\simeq\tfrac{1}{p\theta^{2}p}$ and the
effective action
(\ref{e12}) becomes%
\begin{equation}
\Gamma^{2}(\phi)=\int dp^{4}\phi(p)\phi(-p)\frac{1}{2}\left\{  p^{2}+m_{R}%
^{2}-\frac{\lambda^{2}}{96\pi^{2}}\left(  \frac{1}{p\theta^{2} p}-m_{R}^{2}%
\ln(\frac{1}{m_{R}^{2}p\theta^{2} p})+O(1)\right)
+O(\lambda^{2})\right\}
\label{gnc2}%
\end{equation}

\item  Instead if $p\theta^{2} p\ll1/\Lambda^{2}$ one obtains
$\Lambda _{\text{eff}}\simeq\Lambda$ and the commutative
expression
\begin{equation}
\Gamma^{2}(\phi)=\int dp^{4}\phi(p)\phi(-p)\frac{1}{2}\left\{  p^{2}%
+\widetilde{m}_{R}^{2}+O(\lambda^{4})\right\}
\end{equation}

is recovered, where $\widetilde{m}_{R}^{2}=m^{2}+\frac{\lambda^{2}}{96\pi^{2}%
}(\Lambda^{2}-m^{2}\ln(\frac{\Lambda^{2}}{m^{2}})+O(1))$ is the
commutative renormalized mass.
\end{itemize}

We see that in (\ref{gnc2}) there are singularities in the
infrared ($p\rightarrow0$) limit, involving quadratic and
logarithmic poles. Surprisingly the same poles persist even in the
limit $\theta\rightarrow0$. This means that after the
renormalization of the theory (i.e. after the removal of the
cutoff $\Lambda$) the limit $\theta\rightarrow0$ does not give
back the commutative theory. If we instead work at fixed cutoff
$\Lambda$ the limit $\theta\rightarrow0$ can always be taken and
it always gives back the commutative theory. The fact that the
ultra-violet (UV) limit $\Lambda \rightarrow\infty$ and the
infra-red (IR) limit $p\rightarrow0$ do not commute is a
manifestation of a mixing of the ultraviolet degrees of freedom
with the infrared ones. In literature this mixing is known as
IR/UV mixing \cite{hep-th/9912072}.

\subsubsection{On the validity of perturbative expansion}

\qquad We want to report some observations regarding the validity
of the perturbative expansion in these theories
\cite{hep-th/9912072}. The point is that even if the 1-loop
contributions are all of order $\lambda$, with respect to the tree
level, they diverge in the $p\rightarrow0$ limit. This might
motivate some skepticism toward the validity of the perturbative
expansion.
The one-loop contribution becomes greater than the tree level one when%
\begin{equation}
p^{2}+m_{R}^{2}\lesssim\frac{\lambda}{p\theta^{2} p}.
\end{equation}

At the $n$-th order the divergent dependence in $p$ of the
nonplanar diagram may be read from the dependence on $\Lambda$ of
the planar diagrams. Thus one
expects at the $n$-th order the leading singularities in $p$ are of the type%
\begin{equation}
\Gamma_{n}^{2}(p)\approx\frac{\lambda^{n}}{p\theta^{2} p}\left[  \ln(m_{R}%
^{2}p\theta^{2} p)\right]  ^{n-1}.
\end{equation}

This higher order contributions are as large as the first order when%
\begin{equation}
\frac{\lambda^{n}}{p\theta^{2} p}\left[  \ln(m_{R}^{2}p\theta^{2}
p)\right] ^{n-1}\approx\frac{\lambda}{p\theta^{2} p},
\end{equation}
which yields%
\begin{equation}
m_{R}^{2}p\theta^{2} p<e^{-\frac{c}{\lambda^{2}}},
\end{equation}
where $c$ is a dimensionless constant. Therefore the one loop
approximation is
valid for%
\begin{equation}
O(e^{-\frac{c}{\lambda}})<m_{R}^{2}p\theta^{2} p<O(\lambda),
\end{equation}
which means that the range of momenta in which the loop expansion
is meaningful is exponentially small in terms of the inverse of
the coupling constant.

\subsubsection{One-loop vertex function}

\qquad The tree-level vertex is easily obtained from (\ref{sint})
by
functional derivatives. It reads%
\begin{equation}
\Gamma^{4}(p,q,r,s)=\frac{\lambda}{4!}\delta^{4}(p+q+r+s)V_{s}(p,q,r,s),
\end{equation}
where%
\begin{equation}
V_{s}(p,q,r,s)=\frac{1}{3}\left[  \cos\left(  \frac{p\theta^{2} s-q\theta^{2} r}%
{2}\right)  +\cos\left(  \frac{p\theta^{2} r+q\theta^{2}
s}{2}\right) +\cos\left( \frac{p\theta^{2} q-r\theta^{2}
s}{2}\right)  \right] .
\end{equation}

We have already observed how the usual rules of energy-momentum
conservation still hold. The new observation here is that we have
recovered the symmetry under any exchange of external momenta in
spite of the fact that the phase associated to each vertex is only
cyclically symmetric.

Now let us consider one-loop corrections. We have already
discussed the emergence of infrared singularities in the two-point
function connected to the IR/UV mixing. Here we want to
investigate if the IR/UV mixing has similar implications for the
4-point vertex function. The relevant one-loop diagrams have the
same structure of the corresponding commutative ones but as usual
an appropriate $\theta$-dependent phase factor is present in each
vertex. The analysis proceeds using the same techniques already
used in eqs.
(\ref{gp})-(\ref{ft}) and the final results is \cite{hep-th/9912072}%
\begin{align}
\Gamma^{4}(p,q,r,s)  &
=-\frac{\delta^{4}(p+q+r+s)V_{s}(p,q,r,s)}{3\cdot
2^{5}\pi^{2}}\lambda\left\{
2\ln(\frac{\Lambda^{2}}{m_{R}^{2}})+\ln(\frac
{1}{m_{R}^{2}p\theta^{2} p})+\ln(\frac{1}{m_{R}^{2}q\theta^{2} q})+\right. \nonumber\\
&  +\ln(\frac{1}{m_{R}^{2}r\theta^{2} r})+\ln(\frac{1}{m_{R}^{2}s\theta^{2} s}%
)+\ln(\frac{1}{m_{R}^{2}(q+r)\theta^{2}(q+r)})+\\
&  +\ln(\frac{1}{m_{R}^{2}(q+s)\theta^{2}(q+s)})+\left.  \ln(\frac{1}{m_{R}%
^{2}(s+r)\theta^{2}(s+r)})+...\right\}
\end{align}
that is again divergent for vanishing external momenta (or
vanishing noncommutative parameter $\theta$). As one could guess
from simple arguments, whereas for the two point function the
divergences were quadratic, here we find only logarithmic infrared
divergences. However again we observe the effect of the IR/UV
mixing: the UV logarithmic divergences of the commutative theory
become the IR divergences of the corresponding nonplanar diagrams.
It is also worth noticing that in spite of these new infrared
divergences the theory has been shown to be renormalizable up to
two loops \cite{hep-th/0008057}.

\section{Unsolved problems for QFT in $\kappa$-Minkowski spacetime:
$\lambda\varphi^{4}$ example}

\qquad The issue of quantization of noncommutative theories on
$\kappa $-Minkowski spacetime has not yet been extensively studied
in literature (see however
\cite{hep-th/0105120,hep-th/9902037,hep-th/0205047,hep-th/0112252,
hep-th/0103127}). The obstacles to the implementation of these
theories are closely connected to the properties of the
$\overset{\kappa}{\ast}$ products. Here we want to briefly
consider one example of these theories, $\lambda \varphi^{4}$
scalar theory, and discuss the basic differences with respect to
the canonical counterparts. Construction of scalar field theory in
functional formalism has been discussed in
\cite{hep-th/0105120,hep-th/0205047}. The
starting point is the generating functional%
\begin{equation}
Z[J]=\int D\varphi\exp\left(  i\int dx^{4}\frac{1}{2}\partial_{\mu}%
\phi\overset{\kappa}{\ast}\partial^{\mu}\phi-\frac{1}{2}m^{2}\phi
\overset{\kappa}{\ast}\phi+\lambda\phi\overset{\kappa}{\ast}\phi
\overset{\kappa}{\ast}\phi\overset{\kappa}{\ast}\phi+\frac{1}{2}%
J\overset{\kappa}{\ast}\phi+\frac{1}{2}\phi\overset{\kappa}{\ast}J\right)
.
\label{fgkm}%
\end{equation}

A first point to notice is that in the $\kappa$-Minkowski case
some ambiguities arise already in the introduction of the sources.
In this case in fact expressions like $\int
dx^{4}J\overset{\kappa}{\ast}\phi$ and $\int
dx^{4}\phi\overset{\kappa}{\ast}J$ do not give the same
contributions. This is different from the canonical case where the
corresponding terms give the same contributions thank to the
properties of the canonical $\star$-product. The ambiguity between
a $\int dx^{4}J\overset{\kappa}{\ast}\phi$ and a $\int
dx^{4}\phi\overset{\kappa}{\ast}J$ source term has been
tentatively approached, as shown in \cite{hep-th/0105120}, by
introducing both terms but some of the pathologies of QFT in
$\kappa$-Minkowski might be even due to this initial assumption
(which one may have to modify eventually).

Using relations (\ref{a}-\ref{c}) one can write the above
expression in the
momentum space as%
\begin{equation}
Z[J]=\int D\varphi\exp i(S_{0}+S_{int}+\frac{1}{2}\int
dk^{4}\mu(k_{0})\left[
J(k)\phi(\dot{-}k)+J(\dot{-}k)\phi(k)\right]  ),
\end{equation}
where%
\begin{align}
S_{0}  &  =\frac{1}{2}\int dk_{1}^{4}dk_{2}^{4}\delta^{4}(k_{1}\dot{+}%
k_{2})\phi(k_{1})\left(  \mathcal{C}_{\kappa}(k_{2})-m^{2}\right)
\phi
(k_{2})=\nonumber\\
&  =\frac{1}{2}\int dk^{4}\mu(k_{0})\phi(\dot{-}k)\left(
\mathcal{C}_{\kappa
}(k)-m^{2}\right)  \phi(k),\label{s1}\\
S_{int}  &  =\lambda\int dk_{1}^{4}...dk_{4}^{4}\phi(k_{1})\phi(k_{2}%
)\phi(k_{3})\phi(k_{4})\delta^{4}(k_{1}\dot{+}k_{2}\dot{+}k_{3}\dot{+}k_{4}).
\label{sas2}%
\end{align}

One can also perform the Gaussian integration of (\ref{fgkm})
obtaining
\begin{equation}
Z_{0}[J]\equiv\exp\left(  -\frac{i}{2}\int
dk^{4}\mu(k_{0})\frac{J(k)J(\dot
{-}k)}{\mathcal{C}_{\kappa}(k)-m^{2}}\right)  . \label{partik}%
\end{equation}

To obtain Green functions from the above expression one first
needs a generalization of the functional derivative. A proper
generalization results to be
\begin{align}
\frac{\delta F[J]}{\delta J(k)}  &  \equiv\lim_{\varepsilon\rightarrow0}%
\frac{F[J(p)+\varepsilon\delta^{4}(p\dot{+}(\dot{-}k))]-F[J(p)]}{\varepsilon
},\\
\frac{\delta F[J]}{\delta J(\dot{-}k)}  &  \equiv\lim_{\varepsilon
\rightarrow0}\frac{F[J(p)+\varepsilon\delta^{4}(p\dot{+}k)]-F[J(p)]}%
{\varepsilon}.
\end{align}

Using these new functional derivative, expression
(\ref{a}-\ref{c}) and the fact that
$\mathcal{C}_{\kappa}(k)=\mathcal{C}_{\kappa}(\dot{-}k),$ from
(\ref{partik}) one can find the two-point function at the tree level%
\begin{equation}
G(k,\dot{-}k^{\prime})=\frac{i}{2}\mu(k_{0})\mu(-k_{0})\frac{\delta^{4}%
(p\dot{+}(\dot{-}p^{\prime}))+\delta^{4}((\dot{-}p^{\prime})\dot{+}%
p)}{\mathcal{C}_{\kappa}(k)-m^{2}}. \label{2km}%
\end{equation}

An important point to notice is that being
$\delta^{4}(p\dot{+}(\dot
{-}p^{\prime}))=\delta^{4}((\dot{-}p^{\prime})\dot{+}p)e^{-3p_{0}/\kappa
}=\delta^{4}(p-p^{\prime})e^{-3p_{0}/\kappa},$ expression
(\ref{2km}) predicts the usual rule of energy momentum
conservation in spite of the nontrivial coproducts governing the
algebra of $\kappa$-Poincar\'{e} symmetries. So far, at the level
of having considered only the tree level propagator, the theory
still looks healthy. However serious pathologies are encountered
already in the analysis of the one loop contribution to propagator
and tree level vertex.

One-loop formulas for the propagator may be obtained
\cite{hep-th/0105120,hep-th/0205047} with the usual procedure from
(\ref{sas2}) and (\ref{partik}). A distinction between planar and
nonplanar diagrams, in analogy with the canonical case, results to
be useful. For planar diagrams no problems arise and the
energy-momentum conservations rules: they are the same as in the
tree level formulas. Instead for planar diagrams nontrivial
problems are encountered, mainly due to the fact that a
modification of the momentum-conservation rule occurs that cannot
even be described as a modified conservation law. The terms
involving loop momenta in fact do not cancel each other in the
argument of delta functions. With respect to the vertex function,
functional formalism provide a way to overcome the problem of the
ordering in the vertex, that has been considered one of the main
obstacles towards the construction of a field theory. However
other urgent problems in the vertex function occur already at the
tree level. The most relevant problems is represented by the
lacking of covariance under $\kappa$-Poincar\'{e} transformations
of the argument of the $\delta $\footnote{More properly the
argument of the $\delta$ is covariant in form but when it vanishes
in a given inertial system it does not vanish in all other
inertial systems. This has for example the illogical consequence
that different inertial observers may not agree on the creation of
a particle in a given process.} function which represents the
energy-momentum-conservation law. These fundamental problems
render non-reliable the construction of QFT on $\kappa$-Minkowski
spacetime and however no general consensus is present in
literature on the strategy adopted for quantization. These
consideration led us to stop here our analysis of QFT\ theories in
$\kappa$-Minkowski spacetime. From here on we will focus
exclusively on QFT in canonical spacetime.

\section{Gauge theories in canonical spacetime}

\qquad In this section we want to sketch the construction of gauge
theories in the case of canonical noncommutativity. Commutative
non-abelian-gauge theories contain at most logarithmic
divergences. So one might conjecture that noncommutative gauge
theory to be free from quadratic and linear poles. We will see
that this naive expectation is not fulfilled.

Noncommutative gauge theories are constructed as in the
commutative case \cite{hep-th/0104153}, starting from a Lie
algebra whose generators satisfy
the commutation rules%
\begin{equation}
\lbrack T^{a},T^{b}]=if^{abc}T^{c}, \label{struc}%
\end{equation}
where $f^{abc}$ are the structure constants of the algebra. The
transformations of the fields can be defined as
\begin{align}
\delta_{\alpha}\psi(x)  &  =i\alpha(x)\star\psi(x),\label{gpsi}\\
\delta_{\alpha}\bar{\psi}(x)  &  =-i\psi(x)\star\alpha(x),\\
\delta_{\alpha}A_{\mu}  &
=\partial_{\mu}\alpha(x)+i[\alpha(x),A_{\mu }]_{\star},\nonumber
\end{align}
where $[a,b]_{\star}=a\star b-b\star a$ has been already
introduced in Section 3.1.2. and
$\alpha(x)\equiv\alpha_{a}(x)T^{a}$.

One can also define the field strength as%
\[
F_{\mu\nu}=\partial_{\mu}A_{\nu}-\partial_{\nu}A_{\mu}+i[A_{\nu},A_{\mu
}]_{\star}.
\]
This field strength has the noteworthy property of transforming
under infinitesimal gauge transformations according to the adjoint
representation of
the gauge group%
\begin{equation}
\delta_{\alpha}F_{\mu\nu}=i[\alpha(x),F_{\mu\nu}]_{\star}. \label{tgf}%
\end{equation}
We observe also that gauge symmetry is realized on the fields%
\begin{equation}
(\delta_{\alpha}\delta_{\beta}-\delta_{\beta}\delta_{\alpha})\psi
(x)=\delta_{\lbrack\alpha,\beta]}\psi(x),
\end{equation}%
\begin{equation}
(\delta_{\alpha}\delta_{\beta}-\delta_{\beta}\delta_{\alpha})F_{\mu\nu}%
=\delta_{\lbrack\alpha,\beta]}F_{\mu\nu}.
\end{equation}

The usual Yang-Mills Lagrangian density can be modified in
\begin{equation}
L=-\frac{1}{4\pi g^{2}}Tr[F_{\mu\nu}\star F^{\mu\nu}], \label{lqa}%
\end{equation}
where $Tr$ acts, as usual, on the gauge indices. We notice that
(\ref{lqa}) is not invariant under gauge transformations. In fact
using (\ref{tgf}) is easy
to check that%
\begin{align}
\delta_{\alpha}L  &  =-\frac{1}{4\pi
g^{2}}Tr[\delta_{\alpha}F_{\mu\nu}\star
F^{\mu\nu}+F^{\mu\nu}\star\delta_{\alpha}F_{\mu\nu}]=\\
&  =-\frac{i}{4\pi g^{2}}Tr[\alpha(x)\star F_{\mu\nu}\star F^{\mu\nu}%
-\alpha(x)\star F_{\mu\nu}\star F^{\mu\nu}]\neq0.
\end{align}

However when we consider the action%
\begin{equation}
S=-\frac{1}{4\pi g^{2}}\int dx^{4}Tr[F_{\mu\nu}\star F^{\mu\nu}],\label{ncym}%
\end{equation}
we recover the gauge invariance ($\delta_{\alpha}S=0$). We observe
that to recover gauge invariance the cyclicity property of the
$\star$-product under integration is crucial\footnote{In the
$\kappa$-Minkowski case where the $\overset{\kappa}{\ast}$ product
is not cyclically invariant, attempts of construction of gauge
theories have been so far unsuccessful.}. The action (\ref{ncym})
has all the required features to represent a generalization of the
$U(N)$ Yang-Mills action in a canonical noncommutative framework.
We observe also that the procedure just outlined is of rather wide
applicability in dealing with noncommutative spacetimes. The only
property one needs is the mentioned cyclical invariance of the
star product under integration. It is also worth noticing that,
differently from the commutative case, (\ref{ncym}) is written in
terms of the U(N)\ gauge bosons in such a way that one cannot
separate out a SU(N) sector from a residual U(1)
sector\footnote{The interaction terms couple these two sectors
since in general $\det(A\star B)\neq\det(A)\star\det(B).$}.

One can also define covariant derivatives as%
\begin{equation}
D_{\mu}\psi\equiv\partial_{\mu}\psi-iA_{\mu}\star\psi. \label{cd}%
\end{equation}

It is easy to check that they have the right gauge transformation properties%
\begin{equation}
\delta_{\alpha}(D_{\mu}\psi)=i\alpha(x)\star D_{\mu}\psi.
\end{equation}

With the notion of covariant derivative one can construct an
action for the
spinor fields%
\begin{equation}
S=\int dx^{4}\bar{\psi}\star(\gamma^{\mu}D_{\mu}-m)\psi, \label{actio}%
\end{equation}
whose gauge invariance is again easily verified though again its
Lagrangian density is not gauge invariant.

To analyze the properties of noncommutative Yang-Mills theory, one
needs to adapt the standard Faddeev-Popov technique to the
noncommutative case. Here we will not discuss this technical point
(see however \cite{hep-th/0005208,hep-th/9903187,hep-th/9903077})
since we are mainly interested in the phenomenon of the IR/UV
connection, which is largely independent from this issue.

Feynman rules for canonical noncommutative gauge theory can be
obtained from the actions (\ref{ncym}) and (\ref{actio}). In
particular explicit calculations of the gauge bosons self energy
\cite{hep-th/0002075} lead to the
result%
\begin{equation}
\Pi^{\mu\nu}(p)=8g^{2}(N_{s}+2-N_{f})\alpha\frac{\tilde{p}^{\mu}\tilde{p}%
^{\nu}}{\tilde{p}^{4}},
\end{equation}
where $N_{s}$ is the number of scalar degrees of freedom and
$N_{f}$ is the number of fermionic degree of freedoms in the
theory.

We will comment on the phenomenological implication of this result
in the next chapter. Here we observe only that again the quadratic
pole arises in the limit $p\rightarrow0$ and in limit
$\theta\rightarrow0$ (again for $\theta\rightarrow0$ one does not
recover the commutative theory). We also observe that the
coefficient of the quadratic divergent term is proportional to the
number of bosonic degrees of freedom minus the number of fermionic
degrees of freedom of the theory and that the gauge bosons
contributes with two degrees of freedom (the contribution 2 in
$N_{s}+2-N_{f}$). In particular if one has the same number of
bosonic and fermionic degrees of freedom this coefficient
vanishes. This occurs in supersymmetric theories and also in
softly-broken SUSY theories. We will show an example in the next
chapter. Moreover it is worth noticing that while
renormalizability of gauge theories has been largely studied (see
e.g.\cite{hep-th/9903077,hep-th/0001182}) an all-order prove of
renormalizability is still lacking.

\section{Supersymmetric theories in canonical spacetime}

\qquad There are two main motivations that render noncommutative
supersymmetric theories interesting
\cite{hep-th/0002084,hep-th/0002119,hep-th/0007050}. The first is
that, as discussed in the case of the noncommutative gauge
theories, an equal number of fermionic degrees of freedom and
bosonic degrees of freedom may improve the infrared behavior of
noncommutative theories. The second is the expectation that, as in
the commutative case, SUSY noncommutative theories might manifest
a more regular ultraviolet behavior. It will be not surprising to
discover that actually both ultraviolet and infrared properties of
noncommutative theories are improved by SUSY.

A first point to consider in the construction of a supersymmetric
noncommutative theory is the compatibility of the commutation
relations $[x_{\mu},x_{\nu}]=i\theta_{\mu\nu}$ whit the
supersymmetric algebra. Here we will consider only $\mathcal{N}$=1
supersymmetric theories although the general construction does not
depend on the number of supersymmetries nor on the number of
spacetime dimensions.

A superspace formulation of supersymmetry was given in
\cite{hep-th/0012009} where instead of investigating the
noncommutative superspace formalism it was considered the usual
superspace and superfields. The result is that given the
commutative supersymmetric action written in terms of superfields
it is possible to obtain the noncommutative supersymmetric action
by the only prescription of replacing the ordinary product between
superfields with the $\star$-product.

We use the standard notation of \cite{WeBa}, (except for the
spacetime indices), which are denote here by $\mu,\nu,$... We
start considering the chiral superfields which satisfy
$\bar{D}_{\dot{\alpha}}\Phi=0$. Using the coordinates
$y^{m}=x^{m}+i\theta\sigma^{m}\bar{\theta}$, chiral superfields
can be written as $\Phi(y,\theta,\bar{\theta})=A(y)+\sqrt{2}\theta
\psi(y)+\theta\theta F(y)$. The supersymmetry transformations are
identical to the ones of the commutative counterparts simply
because we are considering the
ordinary superfields%
\begin{align}
\delta_{\xi}A  &  =\sqrt{2}\xi\psi,\nonumber\\
\delta_{\xi}\psi &
=i\sqrt{2}\sigma^{m}\bar{\xi}\partial_{m}A+\sqrt{2}\xi
F,\nonumber\\
\delta_{\xi}F  &
=i\sqrt{2}\bar{\xi}\bar{\sigma}^{m}\partial_{m}\psi.
\label{st}%
\end{align}

The most generic action which can be constructed from the chiral
superfields
${\Phi^{i}}$ takes the form\footnote{We use the notation $(\prod_{i=1}%
^{n}f_{i})_{\star}=f_{1}\star f_{2}\star\cdots\star f_{n}.$ Also
notice that standard notation $\theta$ for both noncommutative
parameters and SUSY Grassmann variables. From the context it
should be clear when we refer to one or to the other.}
\begin{equation}
S=\int d^{4}x\left(  \int d^{2}\theta d^{2}\bar{\theta}K(\Phi^{i}%
,\Phi^{\dagger j})_{\star}+\left[  \int d^{2}\theta
W(\Phi^{i})_{\star }+h.c.\right]  \right)  ,
\end{equation}
where $\int d^{2}\theta\,\theta^{2}=1$ and $\int d^{2}\bar{\theta}%
\,\bar{\theta}^{2}=1$. This is invariant under
$K(\Phi^{i},\Phi^{+j})_{\star }\rightarrow
K(\Phi^{i},\Phi^{+j})_{\star}+F(\Phi)_{\star}+F(\Phi^{+})_{\star
}^{+}$. The action can be written in terms of the component fields
straightforwardly, but rather than doing this in full generality
we focus on a couple of specific examples. First we consider the
action with $K=\Phi
^{+}\star\Phi+a\Phi\star\Phi\star(\Phi^{+})+a^{\ast}\Phi\star(\Phi^{+}%
)\star(\Phi^{+})$ and $W=0$, where $a$ is some numerical
coefficient. Note that in this case the part of the action which
depends on $F$ becomes
\begin{equation}
S|_{F}=\int d^{4}x\left(  F^{+}F+(aF(F\star A)+aF(A\star
F^{+})+aF(F\star A^{+})+h.c)\right)  .
\end{equation}

This action clearly contains the derivatives of the auxiliary
field $F.$ Thus $F$ may become a propagating field if the
noncommutative parameter $\theta^{0\mu}\neq0$ for some $\mu$.
However in the case which are relevant in our future discussions
the canonical K\"{a}hler potential is of the type
$K=\sum_{i}\Phi_{i}^{+}\star\Phi_{i}$ and in this case the action
with non vanishing superpotential does not involve derivatives of
$F$ and then $F$ plays the role of an auxiliary field, which can
be eliminated as in the commutative case.

A model on which we want focusing our attention (and also the
simplest supersymmetric model) is the Wess-Zumino model. Its
action in terms of
superfields reads%
\begin{equation}
S_{WZ}=\!\int d^{4}x\left(  \int d^{2}\theta d^{2}\bar{\theta}\,\Phi_{i}%
^{+}\!\star\!\Phi_{i}+\left[  \int d^{2}\theta\left(  {\frac{1}{2}}m_{ij}%
\Phi_{i}\!\star\!\Phi_{j}+\frac{1}{3}g_{ijk}\Phi_{i}\!\star\!\Phi_{j}%
\!\star\!\Phi_{k}+g_{i}\Phi_{i}\right)  +h.c.\right]  \right)  ,
\end{equation}
where the mass matrix $m_{ij}$ is symmetric in its indices but the
coupling $g_{ijk}$ is not necessarily symmetric. One easily finds
that
\begin{align}
S_{WZ}\!\!\!\!  &  =\!\!\!\int d^{4}x\left(  -\partial_{\mu}A_{i}^{+}%
\partial^{\mu}A_{i}+i\partial_{\mu}\psi_{i}^{+}\bar{\sigma}^{\mu}\psi
_{i}+F_{i}^{+}F_{i}\right) \nonumber\\
&  \!\!\!\!\!\!\!\!+\!\!\int d^{4}x\left[
\frac{1}{3}g_{ijk}\left(
F_{i}\,A_{j}\!\star\!A_{k}+F_{j}\,A_{k}\!\star
A_{i}+F_{k}\,A_{i}\!\star
A_{j}\!\!-A_{i}\,\psi_{j}\!\star\psi_{k}\!-\!A_{j}\,\psi_{k}\!\star\psi
_{i}\!-\!A_{k}\,\psi_{i}\!\star\psi_{j}\right)  \right. \nonumber\\
&  \left.  +g_{i}F_{i}+m_{ij}\left(
A_{i}F_{j}-{\frac{1}{2}}\psi_{i}\psi
_{j}\right)  +h.c\right]  . \label{WZ}%
\end{align}
The equation of motions of $F_{i}$ is
\begin{equation}
F_{i}^{+}=g_{i}+m_{ij}A_{j}+\frac{1}{3}\left(
g_{ijk}+g_{kij}+g_{jki}\right) A_{j}\!\star\!A_{k},
\end{equation}
and the supersymmetry transformation becomes (\ref{st}) with this
$F_{i}$. We note that the typical scalar potential has the form
$A^{+}\!\star \!A^{+}\!\star\!A\!\star\!A$ and the notion of
holomorphy is still valid at $\theta\neq0$. One can also consider
vector superfield $V=V^{+}$ and construct supersymmetric gauge
theories \cite{hep-th/0002084}. Here we only state the main
results. A first result is that the Wess-Zumino model is
renormalizable to all orders of perturbation theory
\cite{hep-th/0005272} and no signs of infrared poles are found in
the two point effective action.

Instead $\mathcal{N}$=1 and $\mathcal{N}$=2 theories with generic
$U(N)$ gauge group were found
\cite{hep-th/0011218,hep-th/0009043,hep-th/0009174,hep-th/0203141,hep-th/0102007}
to be divergent, at one loop, only in the two point function.
However no quadratic divergences were found. Only logarithmic
divergences appear. UV divergences in the planar sector and IR
divergences in the nonplanar sector have been found. They signal
that UV/IR mixing is present in these theories
though it has less strong effects. Supersymmetric noncommutative $\mathcal{N}%
$=4 theory was studied in
\cite{hep-th/0009196,hep-th/0010275,hep-th/9907166}. In
\cite{hep-th/9907166} it was shown to be free from infrared poles
and, more remarkably, it was also argued to be finite (like its
commutative counterpart).

\section{Causality and unitarity in canonical spacetimes}

\qquad In this section we briefly describe the issues of unitarity
and causality in canonical noncommutative field theories.
Perturbative unitarity was for the first time discussed in this
context in \cite{hep-th/0005129}. It was noticed that unitarity is
lost if noncommutativity involves the time coordinate. If $M_{ab}$
is the transition matrix element between the state $a$
and the state $b$, for on-shell matrix elements unitarity implies that%
\begin{equation}
2\operatorname{Im}M_{ab}=\sum_{n}M_{an}M_{nb} \label{1}%
\end{equation}

\begin{figure}[h]
\begin{center}
\includegraphics[width=8cm]{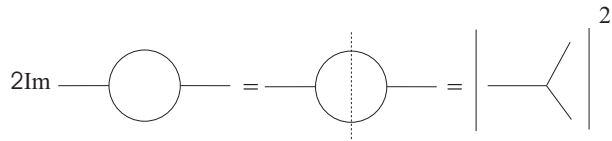}
\end{center}
\caption{Cutting rule in the 2-point function for the $\varphi^{3}$ theory.}%
\label{cutrule}%
\end{figure}

The sum over intermediate states is intended in the right hand
side of the above expression. The rule (\ref{1}) can be expressed
in terms of Feynman graphs. This produces the so-called
generalized-unitarity relations or cutting rules\footnote{Actually
cutting rules are more restrictive than (\ref{1}) since they
involve off-shell conditions. Unitarity of the S-matrix (\ref{1})
follows from the cutting rules.}. Cutting rules state that the
imaginary part of a Feynman diagram can be obtained as follows:
first one must cut the diagrams by a line through virtual lines,
then one must place that virtual
particle on-shell by replacing the propagator with a delta function%
\begin{equation}
\frac{1}{p^{2}-m^{2}+i\varepsilon}\rightarrow-2\pi
i\delta(p^{2}-m^{2}),
\end{equation}
wherever the cut intersects the virtual line, and finally the sum
over all cuts is the imaginary part of the Feynman diagram. For
example in the case of the two-point function in the
noncommutative $\varphi^{3}$ theory one has that unitarity implies
what reported in Fig.\ref{cutrule}\ \ $\ $and explicit calculations shows
\cite{hep-th/0005129,hep-th/0201222}\textbf{ }that this relation
is only satisfied for the case $\theta^{0i}=0.$ The same result
has been proved for other graphs of the $\varphi^{4}$ scalar
theory\footnote{However for other studies that seem to disagree
with these results see \cite{hep-th/0206011}.}.

The analysis of causality leads to similar conclusions. It was
noticed in \cite{hep-th/0005015} that nonlocal effects arising in
NC field theories may lead to a violation of causality if the time
coordinate is involved in noncommutativity. The example treated in
\cite{hep-th/0005015} is that of scattering of wave packets. An
outgoing signal appears before the incoming particles reach each
other. Acausal effects are not found in the case of space-space
noncommutativity.

In general it seems that time-space noncommutativity leads to
violation of both unitarity and causality whereas space-space
noncommutativity preserves both of them\footnote{There are however
some recent attempts to circumvent these problems (see for example
\cite{hep-th/0201222,hep-th/0209253}).}.

\section{Open problems related to the IR/UV mixing in canonical spacetime}

\qquad So far we have discussed how low-energy poles appear in the
Green functions of different NCQFT. We have also stressed how
these poles originate from the ultraviolet sector of loop
integrals so that they can be viewed as manifestations of the
mixing between the UV and IR degrees of freedoms. It is also worth
noticing that infrared poles are only one of the manifestations of
mixing and theories that are free from these singularities must
not be considered free from IR/UV mixing\footnote{This point is
often misunderstood in literature where absence of poles in the
propagator if often identified with absence of the IR/UV mixing.}.
Moreover this correlation between short distances\ (IR) and large
distances (UV) is not accidental in canonical noncommutative
theories but, on the contrary, it is in a sense to be expected.
This expectation comes directly from the commutation relations,
which imply
that%
\begin{equation}
\Delta x_{\mu}\Delta x_{\nu}\gtrsim\theta.
\end{equation}

This uncertainty relations implies that if a coordinate (say $x$)
is known with an uncertainty $\Delta x\lesssim1/\Lambda_{0}$ the
other coordinate (say $y$) must be determined with an uncertainty
$\Delta y\gtrsim\theta\Lambda _{0}.$ This implies correlation
between energies larger than $\Lambda_{0}$ with energies smaller
than $1/\theta\Lambda_{0}.$ Correlations of this type are rather
unusual in the common language of physics where one usually
observes the decoupling of the energy scales. Now we want to
analyze more in detail the wide implications of this IR/UV mixing.

\subsection{IR/UV mixing and renormalization group flow}

\qquad We start our analysis of the implications of the IR/UV
mixing from a discussion on the fate of the Wilsonian picture in
QFT in a canonical spacetime. According to the usual Wilson
picture to QFT, every theory, unless seen as fundamental, is
defined with some cut-off which indicates our ignorance of the
correct theory beyond the cut-off energy scale \cite{wilson,
polchi} . Thus the theory is predictive at least below the
cut-off, say $\Lambda_{0}$, and is understood as embedded in an
unknown more fundamental theory beyond it. One typically studies
how the theory appears at an observer which tests the theory at
scales much lower than the cut-off. High-energy modes of the
theory only generate the low-energy couplings of the effective
theory. In the usual Wilsonian picture the high-energy cut-off,
the bare couplings and the high-energy degrees of freedom may be
entirely encoded in a definition of the parameters of the
low-energy theory. Thus whatever a theory is at its natural
high-energy scale its predictions for the low-energy regime depend
only on a finite number of parameters. In this way different
energy scales may be decoupled and one can perform low-energy
experiments (i.e. ``studying chemistry'') and make predictions,
knowing very little on the details of the high-energy interactions
(i.e. the fundamental interactions). All the necessary information
are encoded in a definition of the couplings (i.e. the fine
structure constant).

Wilson-Polchinski renormalization group tells us how these
couplings change upon varying the scale. Given an action
$S_{\Lambda_{0}}$ that describes quantum field theory up to the
scale $\Lambda_{0}$ one can obtain the action $S_{eff}$ which
describes physics up to the scale $\Lambda<\Lambda_{0}$ by
integrating out the degrees of freedom between $\Lambda$ and $\Lambda_{0}$%
\begin{align}
Z_{\Lambda_{0}}[J]  &  =\int D\phi_{\Lambda_{0}}\exp\left\{  -S_{\Lambda_{0}%
}(\phi)\right\}  =\\
&  =\int D\phi_{\Lambda}\int D\phi_{\Lambda\Lambda_{0}}\exp\left\{
-S_{\Lambda_{0}}(\phi)\right\}  =\int D\phi_{\Lambda}\exp\left\{
-S_{eff}(\phi;\Lambda,\Lambda_{0})\right\}  ,
\end{align}
where%
\begin{equation}
\exp\left\{  -S_{eff}(\phi;\Lambda,\Lambda_{0})\right\}  =\int
D\phi _{\Lambda\Lambda_{0}}\exp\left\{  -S(\phi)\right\}  .
\end{equation}

If one is interested in processes that take place at energy
$E\ll\Lambda$ one can equivalently use $S_{\Lambda_{0}}$ or
$S_{eff}(\phi;\Lambda,\Lambda_{0}).$ The difference between the
two possible choices lies in the fact that while using
$S_{\Lambda_{0}}$ one has to integrate in a huge range in the loop
momenta (the range $0-\Lambda_{0}$), in
$S_{eff}(\phi;\Lambda,\Lambda_{0})$ the integration in the range
$\Lambda$-$\Lambda_{0}$ has already been
performed and its effects are encoded in the coefficients of $S_{eff}%
(\phi;\Lambda,\Lambda_{0})$. However the price to pay in this
second case is that in general $S_{eff}(\phi;\Lambda,\Lambda_{0})$
involves infinite interactions. From the fact that it must be
\begin{equation}
\partial_{\Lambda}Z_{\Lambda_{0}}\left[  J\right]  =0,\label{eqz}%
\end{equation}
one can also write the following integral-differential-flux
equation (known as
Polchinski equation) for the effective action%
\begin{equation}
\Lambda\partial_{\Lambda}S_{eff}=\frac{1}{2}\int dp^{4}\left(
2\pi\right) ^{8}\Lambda\partial_{\Lambda}(D_{\Lambda})\left\{
{\displaystyle{\frac{\delta
S_{eff}}{\delta\phi(p)}}}{\displaystyle{\frac{\delta S_{eff}}{\delta\phi(-p)}%
}}-{\displaystyle{\frac{\delta^{2}S_{eff}}{\delta\phi(p)\delta\phi(-p)}}%
}\right\}  \label{polchi}%
\end{equation}
with the initial condition
\begin{equation}
S_{eff}\left[  \phi,\Lambda_{0}\right]  =S\left[
\phi,\Lambda_{0}\right]  ,
\end{equation}
where $D_{\Lambda}$ is a cut-off function which is equal to one
below $\Lambda$ and rapidly vanishes above $\Lambda$ .As for the
action one might rewrite the same equation for the connected
action $W[J]$ or the 1PI-effective action $\Gamma$\footnote{It is
worth observing that the equation for $W[J]$ can be regarded as an
infinite-dimensional heat equation.}. Although in principle
(\ref{polchi}) might be valid at non-perturbative level (see i.e.
\cite{hep-th/9607188,hep-th/0012111}), one may use it for a
perturbative
calculation by performing a vertex expansion of the action%
\begin{equation}
S_{eff}\left[  \varphi,\Lambda,\Lambda_{0}\right]
=\mathop{\displaystyle\sum
}_{n=1}^{\infty}\frac{1}{(2n!)}\int_{p1...p_{2n}}C_{2n}(p_{1},...,p_{2n}%
,\Lambda,\Lambda_{0})\left(  2\pi\right)  ^{4}\delta^{4}(p_{1}+...+p_{2n}%
)\varphi(p_{1})...\varphi(p_{2n})\text{ \ ,}%
\end{equation}
where vertices $C_{2n}$ may be evaluated by loop-expansion. The
same may be done for the 1PI-effective action
\begin{equation}
\Gamma_{\Lambda,\Lambda_{0}}\left[  \varphi\right]
=\mathop{\displaystyle
\sum}_{n=1}^{\infty}\frac{1}{(2n!)}\int_{p1...p_{2n}}\Gamma_{\Lambda
,\Lambda_{0}}^{2n}(p_{1},...,p_{2n})\left(  2\pi\right)  ^{4}\delta^{4}%
(p_{1}+...+p_{2n})\varphi(p_{1})...\varphi(p_{2n}).
\end{equation}

One obtains for the vertices an equation of the form
\begin{equation}
\Lambda\partial_{\Lambda}\Gamma_{\Lambda,\Lambda_{0}}^{2n}=\mathcal{F[}%
\Gamma_{\Lambda,\Lambda_{0}}^{2},...,\Gamma_{\Lambda,\Lambda_{0}}%
^{2n+2}\mathcal{]} \label{e1piv}%
\end{equation}
with an appropriate function $\ \mathcal{F}$. Then one can isolate
in each of $\Gamma_{\Lambda,\Lambda_{0}}^{2n}$ the contributions
of the relevant operators and of the irrelevant operators.
Relevant operators are those operators whose couplings increase
along the renormalization-group flow whereas irrelevant operators
are those whose coupling are suppressed in the flow. The couplings
of the relevant operators are all one needs for a low-energy
theory (the irrelevant-operator couplings vanish in the infrared).
For example one can write%
\begin{align}
\Gamma_{\Lambda,\Lambda_{0}}^{2}(p)  &  =\left.  \Gamma_{\Lambda,\Lambda_{0}%
}^{2}(p)\right|  _{p=p_{0}}+\left.  \frac{\partial\Gamma_{\Lambda,\Lambda_{0}%
}^{2}(p)}{\partial p^{2}}\right|
_{p=p_{0}}(p^{2}-p_{0}^{2})+\Gamma
_{\Lambda,\Lambda_{0}}^{2\text{ \ }irr}(p)\label{flusso}\\
\Gamma_{\Lambda,\Lambda_{0}}^{4}(p)  &  =\left.  \Gamma_{\Lambda,\Lambda_{0}%
}^{4}(p)\right|  _{p=p_{0}}+\Gamma_{\Lambda,\Lambda_{0}}^{4\text{
\ }irr}(p),
\end{align}
where $p_{0}$ is the renormalization point (e.g. the scale at
which renormalized coupling are fixed) and
$\Gamma_{\Lambda,\Lambda_{0}}^{2\text{ \ }irr}(p),$
$\Gamma_{\Lambda,\Lambda_{0}}^{4\text{ \ }irr}(p)$ represent the
irrelevant contributions to the two-point and four-point effective
actions.

The same formal techniques can be applied to the
canonical-noncommutative framework but some important differences
emerge. In the commutative case, at any order of perturbation
theory, it can be shown that irrelevant operators
$\Gamma_{\Lambda,\Lambda_{0}}^{n,irr}(p)$ are suppressed by
positive powers of $\Lambda/\Lambda_{0}$. This means that the
low-energy theory $\Lambda <<\Lambda_{0}$ depends on the
high-energy theory only through the renormalized mass $m_{R}$ and
the renormalized coupling $\lambda_{R}$ (up to corrections of
order $\Lambda/\Lambda_{0}$)$.$ This is a finite number of
parameters. The theory is renormalizable and the energy-scale
decoupling mechanism works$.$

In the noncommutative case one finds \cite{hep-th/0104217} that as
long as $p\ll{\Lambda_{0}}$ and
$p_{0},p\gg{\displaystyle1/\theta\Lambda_{0}}$,
$\Gamma_{\Lambda,\Lambda_{0}}^{2\text{ \ }irr}$, for example,
depends on $\Lambda_{0}$ in a exponentially suppressed way so that
Wilsonian picture still holds: there is negligible influence on
the physics at momentum scales $p$ from the high-energy
($\Lambda_{0}$) sector of the theory. In these regime,
energy-scale decoupling still works. Instead if one considers
external momenta $p$ less than
${\displaystyle1/\theta\Lambda_{0}{,}}$one finds
\begin{equation}
\Gamma_{\Lambda,\Lambda_{0}}^{2\text{ \
}irr}(p)\simeq\frac{\lambda}{96\pi ^{2}}\Lambda_{0}^{2}+...,
\end{equation}
which means that the Wilsonian picture of energy-scale decoupling
is spoiled and that low-energy prediction under the scale
${1/\theta\Lambda_{0}}$ are highly sensitive to the (unknown)
details of the ultraviolet sector of the theory\footnote{It is
important to notice that this does not imply the theory \ is
nonrenormalizable. One can still formally consider the infinite
cut-off limit and obtain predictions in terms of a finite number
of parameters (this is what we mean with renormalizability).
However physically, the infinite cut-off limit is only justified
by the mechanism of energy-scale decoupling and therefore it is
unmotivated in these theories in canonical noncommutative
spacetime. In the next chapter we consider a theory with a large
mass scale and find that, after removal of the cut-off scale, this
large mass scale still affects significantly the low-energy \
sector of the theory.}. This spoils the usefulness of the concept
of effective low-energy action useless and affects the procedure
usually adopted to test a physical model by comparing the
predictions of the model with the low-energy data. We will
investigate this important point in the next chapter.

\subsection{IR/UV mixing and the subtraction point}

\qquad Other effects related to the IR/UV mixing are the different
scaling laws of the Green functions at different momenta and
certain problems with the choice of the subtraction point. Let us
consider as an example the case of the scalar theory already
analyzed. For the two-point function we have for large
$\Lambda$ the scaling%
\begin{equation}
\Gamma^{2}(\mu)\simeq\frac{\lambda}{48\pi^{2}}\Lambda^{2},
\end{equation}
if $\mu\neq0$. Instead, if $\mu=0,$ we have the scaling%
\begin{equation}
\Gamma^{2}(\mu)\simeq\frac{\lambda}{32\pi^{2}}\Lambda^{2}.
\end{equation}

Closely related to this difference in the scaling of the Green
function is the problem of the choice of the renormalization
point. If we set renormalization conditions at a momentum scale
$\mu\neq0$ we find for the one-loop renormalized parameters
\begin{align}
m_{R}^{2}  &  =m^{2}+\frac{\lambda}{8\pi^{2}}\left[
\Lambda^{2}-m^{2}\ln
\frac{\Lambda^{2}}{m^{2}}\right]  ,\\
\lambda_{R}  &
=\lambda-\frac{\lambda}{8\pi^{2}}\ln\frac{\Lambda^{2}}{m^{2}}.
\end{align}

Instead if we choose a subtraction point $\mu=0$ the renormalized
parameters
are the same as in the commutative case and they read%
\begin{align}
m_{R}^{2}  &  =m^{2}+\frac{3\lambda}{16\pi^{2}}\left[  \Lambda^{2}-m^{2}%
\ln\frac{\Lambda^{2}}{m^{2}}\right]  ,\nonumber\\
\lambda_{R}  &  =\lambda-\frac{9\lambda}{16\pi^{2}}\ln\frac{\Lambda^{2}}%
{m^{2}}. \label{mr}%
\end{align}

This implies that if we choose $\mu=0$ as renormalization point,
(subtracted) Green functions for general external momenta are not
finite in the $\Lambda\rightarrow\infty$ limit and the theory will
appear to be non renormalizable. If we choose a subtraction point
at $\mu\neq0$, (subtracted) Green functions are finite but poles,
of the type of the ones already described, appear for
$\mu\rightarrow0.$

Another way to obtain the renormalized parameters (\ref{mr}) is
the one of considering the effective potential which is the
generator of 1PI-Green
functions at zero momentum%
\begin{equation}
V_{eff}=\sum_{i=1}^{\infty}\frac{1}{n!}\Gamma^{n}(0,...,0)\varphi^{n}.
\end{equation}

At one-loop level the effective potential is the same as in the
noncommutative case since at zero momentum nonplanar diagrams give
the same contributions of
the planar ones:%
\begin{equation}
V_{eff}=\frac{1}{2}m^{2}\varphi^{2}+\frac{\lambda}{4}\varphi^{4}+\frac{1}%
{2}\int^{\Lambda}\frac{dk^{4}}{\left(  2\pi\right)
^{4}}\ln(1+\frac {3\lambda\varphi^{2}}{k^{2}+m^{2}}).
\end{equation}

The renormalized parameters are now obtained from the relations%
\begin{align}
\left.  \frac{d^{2}V_{eff}}{d\varphi^{2}}\right|  _{\varphi=0}  &  =m_{R}%
^{2},\\
\left.  \frac{d^{4}V_{eff}}{d\varphi^{4}}\right|  _{\varphi=0}  &
=6\lambda_{R}.
\end{align}
\qquad\qquad

These relations will lead to a couple of equations identical to
(\ref{mr}) which correspond to nonrenormalizable Green functions.

We also notice that the considerations we are doing hold rather in
general and are not restricted to the scalar-theory case. For
example similar problems manifest in the analysis of the
noncommutative Gross-Neveu model \cite{hep-th/0103199}. In general
if the renormalization conditions are set at zero external
momentum the theory does not renormalize whereas if the
renormalization conditions are set at a nonzero external momentum
the theory renormalizes. The origin of this behaviors is in a sort
of non analytic structure that canonical noncommutativity induces
at zero momentum. Since the noncommutativity parameters appear in
diagrams only through factor of the type $\exp(ip\theta k),$ for
$p=0$ the diagrams reproduce the commutative spacetime limit,
whereas as soon as $p\neq0$ the $\theta$ parameter cannot be
ignored and induce large contribution at low energy. As we shall
show in Chapter 5 this peculiarities of the zero-momentum limit
have profound implications for nonperturbative estimates of the
effective potential.

\subsection{IR/UV Mixing and the Goldstone theorem.}

\qquad In this section we discuss the implication of the UV/IR
mixing for the validity of the Goldstone theorem which is at the
basis of the mechanism of mass generation (Higgs mechanism). The
statement of the Goldstone theorem roughly is that for every
spontaneously broken symmetry (i.e. a symmetry of the action that
is not a symmetry of the ground state of the theory) there must be
a massless particle. For example in the case of the linear sigma
model with $O(N)$ symmetry broken to $O(N-1),$ the numbers of
symmetries changes from $N(N-1)/2$ to $(N-1)(N-2)/2$ so that $N-1$
symmetries are broken and $N-1$ massless particles (Goldstone
bosons) appear. At the quantum level instead of considering the
action $S$ one has to consider the effective action $\Gamma$ which
besides having has the same symmetries of the classical
theory\footnote{If regularization does not break these
symmetries.}

The problem of the validity of the Goldstone theorem in
noncommutative theories was first addressed in
\cite{hep-th/0003137}. We consider scalar
$O(N)\ $theory whose action is%
\begin{equation}
S=\frac{1}{2}\left(  \partial_{\mu}\phi^{i}\right)
^{2}+\frac{1}{2}\mu
^{2}\left(  \phi^{i}\right)  ^{2}-\frac{\lambda}{4}\phi^{i}\star\phi^{i}%
\star\phi^{i}\star\phi^{i}, \label{aon}%
\end{equation}
where the sum over the $i$ index is omitted.

This model enjoys the symmetry $\phi^{i}\rightarrow
R^{ij}\phi^{j}$ where $R$ is a spacetime constant, $N\times N$
orthogonal matrix \ $RR^{T}=I.$ If
$\mu^{2}>0$ the classical potential%
\begin{equation}
V_{cl}(\phi)=\frac{1}{2}\mu^{2}\left(  \phi^{i}\right)  ^{2}-\frac{\lambda}%
{4}\left[  \left(  \phi^{i}\right)  ^{2}\right]  ^{2},
\end{equation}
has a minimum for the constant field configuration%
\begin{equation}
(\phi_{0}^{i})^{2}=\frac{\mu^{2}}{\lambda}. \label{vacua}%
\end{equation}

Therefore the $O(N)$ symmetry of the action is no more a symmetry
of the vacuum, which is to say that the $O(N)\ $symmetry is
broken. The relation (\ref{vacua}) identifies the $(N-1)$
dimensional manifold (actually, in this case, a sphere) on which
the classical potential assumes its minimum. We can choose one
point on this manifold to identify with the vacuum of the broken
phase. We choose the configuration%
\begin{equation}
\phi_{0}=\left(  0,...,0,v\right)  ,
\end{equation}
where $v\equiv\dfrac{\mu}{\sqrt{\lambda}}.$ Hence we define the new fields%
\begin{equation}
\left\{
\begin{array}
[c]{c}%
\pi_{i}\equiv\phi^{i}\text{ \ \ \ \ }\\
\sigma\equiv\phi^{n}-v
\end{array}
\right.
\end{equation}
in such a way that $\left\langle \sigma\right\rangle =0.$ The
action
(\ref{aon}) in terms of the new fields reads%
\begin{equation}
S=\frac{1}{2}\left(  \partial_{\mu}\pi^{k}\right)
^{2}+\frac{1}{2}\left(
\partial_{\mu}\sigma\right)  ^{2}+\frac{1}{2}(2\mu^{2})\sigma^{2}%
+\frac{\lambda}{2}\left[  \left(  \pi^{k}\right)  ^{2}\right]  ^{2}%
+\frac{\lambda}{2}\sigma^{4}+\lambda v\sigma\left(  \pi^{k}\right)  ^{2}%
+\frac{\lambda}{2}\sigma^{2}\left(  \pi^{k}\right)  ^{2}+\lambda
v\sigma^{3}.
\label{sigmam}%
\end{equation}

The absence of the terms quadratic in the pion filed $\pi^{k}$ is
a prove of the Goldstone theorem at the tree level in canonical
noncommutative theories. Also the breaking of the symmetry from
$O(N)$ down to $O(N-1)$ is manifest. To check if the Goldstone
theorem also holds at the quantum level one must verify that pion
fields remain massless at the quantum level as well. This is the
case if $\Gamma_{\pi\pi}^{2}(p)$ vanish at $p=0.$ Explicit
one-loop calculation of $\Gamma_{\pi\pi}^{2}(p)$ have been carried
out in \cite{hep-th/0003137}. The result is that if one considers
first the $\Lambda\rightarrow\infty$ limit and then the
$p\rightarrow0$ limit $\Gamma_{\pi\pi}^{2}(p)$ does not vanish,
actually it diverges. Instead if one first imposes $p=0$ and then
considers the $\Lambda\rightarrow\infty$ limit one recovers
$\Gamma_{\pi\pi}^{2}(0)=0$ and the validity of the Goldstone
theorem\footnote{It is worth observing that absence of violations
of the Goldstone theorem has been proved at one loop in
noncommutative-scalar U(N) theories in \cite{hep-th/0102022} or in
other particular cases \cite{hep-th/0202011}.}. This is another
manifestation of the failure of the commutation of the
zero-momentum limit ($p\rightarrow0$ ) with large cut-off limit
($\Lambda\rightarrow\infty$), again a consequence of the IR/UV
Mixing.

\subsection{IR/UV Mixing and the scalar-theory phase diagram}

\qquad We have discussed various problems connected to the
zero-momentum limit of these theories. For example we have seen
that Green functions at $p\neq0$ are not renormalized by
renormalization conditions fixed at zero-momentum, whereas, if the
renormalization conditions are set away from $p=0,$ the Green
functions exhibit a pole in the zero-momentum limit after removal
of the cut- off ($\Lambda\rightarrow\infty)$. The stiffness of the
zero-momentum modes
($\Gamma^{2}(p)\overset{p\rightarrow0}{\rightarrow}\infty$) has
direct, and deep, implications also for the analysis of phase
transitions. We know in fact that phase transitions are related to
the condensations of some momentum modes and, in particular,
phase-transition to translation-invariant vacuum are related to
the condensation of the zero-momentum modes. We want to observe
how in canonical noncommutative theories just because
zero-momentum modes are stiff, transition to translation-invariant
ordered phases are not trivial. As an example we consider the Ward
identities for the scalar theory with global
$O(2)$ symmetry \cite{hep-th/0104106}. The action of this theory is%
\begin{equation}
S=-\frac{1}{2}\left(  \partial_{\mu}\phi^{i}\right)
^{2}+\frac{1}{2}\mu
^{2}\left(  \phi^{i}\right)  ^{2}+\frac{\lambda}{4}\phi^{i}\star\phi^{i}%
\star\phi^{i}\star\phi^{i}.
\end{equation}

In the symmetric phase the Ward identities are%
\begin{align}
\left.  \frac{\delta^{2}\Gamma}{\delta\phi_{1}^{2}}\right|  _{\phi_{1}%
=\phi_{2}=0}  &  =\left.
\frac{\delta^{2}\Gamma}{\delta\phi_{2}^{2}}\right|
_{\phi_{1}=\phi_{2}=0},\nonumber\\
\left.  \frac{\delta^{4}\Gamma}{\delta\phi_{1}^{4}}\right|  _{\phi_{1}%
=\phi_{2}=0}  &  =3\left.  \frac{\delta^{4}\Gamma}{\delta\phi_{1}^{2}%
\delta\phi_{2}^{2}}\right|  _{\phi_{1}=\phi_{2}=0}. \label{wardsymm}%
\end{align}

In the broken phase the symmetric vacuum becomes unstable and one,
as usual,
finds the new vacuum by a shift of the fields%
\begin{align}
\phi_{1}  &  =\sigma+v,\nonumber\\
\phi_{2}  &  =\pi.
\end{align}

Ward identities in the broken phase now read%
\begin{equation}
\left.  v\frac{\delta^{2}\Gamma}{\delta\pi^{2}}\right|
_{\sigma=\pi =0}=\left.  \frac{\delta\Gamma}{\delta\sigma}\right|
_{\sigma=\pi=0}.
\label{WardBS}%
\end{equation}

We know that in the corresponding theory in commutative spacetime
both (\ref{wardsymm}) and (\ref{WardBS}) hold true. In the
noncommutative theory explicit one-loop calculations
\cite{hep-th/0104106} show that the identities of the symmetric
case (\ref{wardsymm}) are still satisfied but the identities of
the translational-invariant broken phase (\ref{WardBS}) are
violated. The point is that the shift $\phi_{1}=\sigma+v$
implicitly assumes a translational-invariant vacuum. If one
considers transitions to a vacuum $v(x)$ which is not
translational invariant one obtains the following Ward
identities%
\begin{equation}
\int\frac{dp^{4}}{\left(  2\pi\right)  ^{4}}v(-p)\left.
\frac{\delta ^{2}\Gamma}{\delta\pi(p_{1})\pi(p)}\right|
_{\sigma=\pi=0}=\left.
\frac{\delta\Gamma}{\delta\sigma(p_{1})}\right|  _{\sigma=\pi=0},
\end{equation}
which have been shown to be verified at one loop
\cite{hep-th/0104106,hep-th/0202171}. This argument strongly
favors the idea of stable nonuniform phases.

Phase transitions in scalar theories have been more carefully
analyzed in Ref.\cite{hep-th/0006119} using a self-consistent
one-loop analysis. The authors find some evidence of condensation
of nonzero modes corresponding to an ordered phase which breaks
translational invariance. Also relying on the natural assumption
that at fixed cut-off $\Lambda,$ in the $\theta \rightarrow0$
limit one must recover the ordered translational-invariant phase
of the commutative theory, Ref.\cite{hep-th/0006119} the phase
diagrams here reported in for the scalar $\lambda\phi^{4}$
theory\begin{figure}[h]
\begin{center}
\includegraphics[width=8cm]{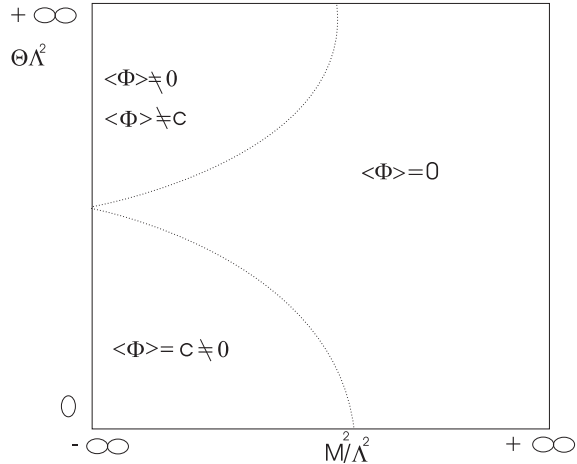}
\end{center}
\caption{Phase diagram of the noncommutative
$\lambda\varphi^{4}$-theory in
the ($m^{2}/\Lambda^{2}$, $\theta\Lambda^{2}$) plane.}%
\label{phased}%
\end{figure}

Essentially three phases can be distinguished. The first phase is
the ordered phase, characterized by the zero vacuum expectation
value $\left\langle \phi\right\rangle =0.$ This phase is dominant
for large positive values of the ratio $m^{2}/\Lambda^{2}$. For
sufficiently large but negative values of the ratio
$m^{2}/\Lambda^{2}$ one encounters the ordered phase $\left\langle
\phi\right\rangle \neq0$. For small value of $\theta\Lambda^{2}$
one has a translational-invariant ordered phase characterized by
$\left\langle \phi\right\rangle =c$ similar to the one found in
the commutative limit. For large value of $\theta\Lambda^{2}$
(i.e. in the limit in which only planar diagrams contribute) one
finds an ordered translational-non-invariant phase which has been
argued \cite{hep-th/0006119} to be a stripe phase, characterized
by a vev of the type $\left\langle \phi\right\rangle
=A\cos(p_{c}x).$ Where $p_{c}$ is the condensating momentum. In
Chapter 5 \ we explore these issues concerning phase transitions
in canonical noncommutative spacetimes using the
Corwall-Jackiw-Tomboulis approach, a powerful nonperturbative
technique of evaluation of the effective potential. This will also
allow us to investigate the implications for nonperturbative
renormalizability of the peculiar structure of the zero-momentum
limit of theories in canonical noncommutative spacetime which was
here briefly introduced in Subsection 3.7.



\chapter{Critical analysis of the phenomenology in CNC spacetimes}

\footnotetext{* \ In this Chapter we discuss in detail the
analysis reported more briefly in
Ref.\cite{hep-th/0209254}.}*\qquad In this chapter {we investigate
the implications of the IR/UV mixing for the derivation of
experimental limits on the parameters of canonical noncommutative
spacetimes. By analyzing a simple Wess-Zumino model in canonical
noncommutative spacetime with soft supersymmetry breaking we
explore the implications of ultraviolet supersymmetry on
low-energy phenomenology. The fact that new physics in the
ultraviolet can modify low-energy predictions affects
significantly the derivation of limits on the noncommutativity
parameters based on low-energy data. These are, in an appropriate
sense here discussed, ``conditional limits''. We also find that
some standard techniques for an effective low-energy description
of theories with non-locality at short distance scales are only
applicable in a regime where theories in canonical noncommutative
spacetime lack any predictivity, because of the strong sensitivity
to unknown UV physics. }

\section{IR/UV Mixing and Phenomenology in canonical spacetime}

\qquad We have discussed how a key characteristic of field
theories on canonical spacetimes, which originates from the
commutation rules, is nonlocality. At least in the case of
space/space noncommutativity ($\theta_{0i}=0$), to which we limit
our analysis for simplicity, this nonlocality is still tractable
although it induces a characteristic mixing of the ultraviolet and
infrared sectors of the theory. This IR/UV mixing has wide
implications, including the possible emergence of infrared
(zero-momentum) poles in the one-loop two-point functions. In
particular one finds a quadratic pole for some integer-spin
particles in non-SUSY theories~ \cite{hep-th/9912072}, while in
SUSY theories the poles, if at all present, are
logarithmic~\cite{hep-th/0002075,hep-th/0005272,hep-th/0011218}.
It is noteworthy that these infrared singularities are introduced
by loop corrections and originate from the ultraviolet part of the
loop integration: at tree level the two-point functions are
unmodified, but loop corrections involve the interaction vertices,
which are modified already at tree level.

There has been considerable work attempting to set limits on the
noncommutativity parameters $\theta$ by exploiting the
modifications of the interaction
vertices~\cite{hep-th/0105082,hep-ph/0010354,hep-ph/0106356,hep-ph/0205040}
and the modifications of the dressed/full
propagators~\cite{hep-th/0109191}. Most of these analyses rely on
our readily available low-energy data. The comparison between
theoretical predictions and experimental data is usually done
using a standard strategy (the methods of analysis which have
served us well in the study of conventional theories in
commutative spacetime). We are here mainly interested in
understanding whether one should take into account some of the
implications of the IR/UV mixing also at the level of the
techniques by which one compares theoretical predictions with
data. In Ref.~\cite{hep-th/0109191} it was argued that the way in
which low-energy data can be used to constrain the
noncommutativity parameters is affected by the IR/UV mixing. These
limits on the entries of the $\theta$ matrix might not have the
usual interpretation: they could be seen only as ``conditional
limits'', conditioned by the assumption that no contributions
relevant for the analysis are induced by the ultraviolet. The
study we report here is relevant for this delicate issue. By
analyzing a simple noncommutative Wess-Zumino-type model, with
soft supersymmetry breaking, we explore the implications of
ultraviolet supersymmetry on low-energy phenomenology. Based on
this analysis, and on the intuition it provides about other
possible features of ultraviolet physics, we provide a
characterization of low-energy limits on the noncommutativity
parameters. Our analysis provides additional encouragement for
combining, as proposed in Ref.~\cite{hep-th/0109191}, high-energy
data, from astrophysics, with the more readily available
low-energy data.

\section{Effects of UV SUSY on IR physics}

\qquad In this section we analyze a mass deformed Wess-Zumino
model in canonical noncommutative spacetime. We emphasize the role
that the UV scale of SUSY restoration plays in the IR sector of
the model, and we also provide some more general remarks on the
IR/UV mixing. This analysis will provide material for one of the
points we raise in the later part of the paper, which concerns the
nature of the bounds that can be set on the noncommutativity
parameters using low-energy data.

\subsection{A model with SUSY restoration in the UV}

For definiteness, we present our observations, which have rather
wide applicability, in the specific context of a mass deformed
Wess-Zumino model,
with action%

\begin{align}
S_{dwz}  &  =S_{0}+S_{m}+S_{g},\\
S_{0}  &  =\int dx^{4}\left\{  \frac{1}{2}\partial_{\mu}\varphi_{1}%
\partial^{\mu}\varphi_{1}+\frac{1}{2}\partial_{\mu}\varphi_{2}\partial^{\mu
}\varphi_{2}+\frac{1}{2}\overline{\psi}i\setbox0=\hbox{$p$}\dimen0=\wd
0\setbox1=\hbox{/}\dimen1=\wd1\ifdim\dimen0>\dimen1\rlap{\hbox to
\dimen0{\hfil/\hfil}}\partial\else\rlap{\hbox
to \dimen1{\hfil$p$\hfil}}/\fi\psi\right\}  ,\\
S_{m}  &  =\int dx^{4}\left\{  \frac{1}{2}F^{2}+\frac{1}{2}G^{2}+m_{s}%
F\varphi_{1}+m_{s}G\varphi_{2}-\frac{1}{2}m_{f}\overline{\psi}\psi\right\}
,\\
S_{g}  &  =\int dx^{4}g\left\{  F\star\varphi_{1}\star\varphi_{1}%
-F\star\varphi_{2}\star\varphi_{2}+G\star\varphi_{1}\star\varphi_{2}+\right.
\\
&  \left.  \text{ \ \ \ \ \ \ \ \ \ \
}+G\star\varphi_{2}\star\varphi
_{1}-\overline{\psi}\star\psi\star\varphi_{1}-\overline{\psi}\star
i\gamma ^{5}\psi\star\varphi_{2}\right\}  ~.\nonumber
\end{align}

$\varphi_{1}$ and $\varphi_{2}$ are bosonic/scalar degrees of
freedom, while $\psi$ denotes fermionic spin-$1/2$ degrees of
freedom. $F$ and $G$ are auxiliary fields. The model is exactly
supersymmetric (SUSY) if $m_{s}=m_{f}$. We consider the case
$m_{s}<m_{f}$ in which supersymmetry is only ``restored'' in the
ultraviolet (UV), where both $m_{s}$ and $m_{f}$ are negligible
with respect to the high momenta involved.

The free propagators are not modified by canonical
noncommutativity:
\begin{align}
\Delta_{m_{s}}(p)  &
\equiv\Delta_{\varphi_{1}\varphi_{1}}(p)=\Delta
_{\varphi_{2}\varphi_{2}}(p)=\frac{i}{p^{2}-m_{s}^{2}+i\varepsilon
}~,~~~~~~~~~~~\Delta_{FF}(p)=\Delta_{GG}(p)=p^{2}\Delta_{\varphi_{1}%
\varphi_{1}}(p),\\
\Delta_{F\varphi_{1}}(p)  &  =\Delta_{\varphi_{1}F}(p)=\Delta_{\varphi_{2}%
G}(p)=\Delta_{G\varphi_{2}}(p)=-m_{s}\Delta_{\varphi_{1}\varphi_{1}%
}(p)~,~~~~~~~~~~~S(p)=\frac{i}{\setbox0=\hbox{$p$}\dimen0=\wd0\setbox
1=\hbox{/}\dimen1=\wd1\ifdim\dimen0>\dimen1\rlap{\hbox to
\dimen0{\hfil/\hfil }}p\else\rlap{\hbox to
\dimen1{\hfil$p$\hfil}}/\fi-m_{f}}~.\nonumber
\end{align}

The vertices acquire the familiar $\theta$-dependent phases:
\begin{align}
V_{[\overline{\psi}\psi\varphi_{1}]}=-ig\cos(p_{1}\widetilde{p}_{2})
&
~,~~~V_{[\overline{\psi}\psi\varphi_{2}]}=-i\gamma^{5}g\cos(p_{1}\widetilde
{p}_{2})~,~~~\\
V_{[F\varphi_{1}\varphi_{1}]}=ig\cos(p_{1}\widetilde{p}_{2})~,~~~V_{[F\varphi
_{2}\varphi_{1}]}=  &
-ig\cos(p_{1}\widetilde{p}_{2})~,~~~V_{[G\varphi
_{1}\varphi_{2}]}=2ig\cos(p_{1}\widetilde{p}_{2})~.\nonumber
\end{align}

[Notice that, taking into account momentum conservation at
vertices, the momenta $p_{1}$and $p_{2}$ can be attributed
equivalently to any of the three particles involved in each of the
vertices.]

\subsection{Self-energies and IR singularities}

Self-energies will play a key role in our observations. Using the
NC Feynman rules the self-energies for fermions and scalars can be
evaluated straightforwardly. The one loop self-energy of the
scalar field receives contributions from five Feynman diagrams,
leading to the result
\begin{align}
-i\Sigma_{1loop}(p)  &  =-g^{2}\int\frac{d^{4}k}{\left(  2\pi\right)  ^{4}%
}\left\{  \left(  8k^{2}+8m_{s}^{2}\right)  \Delta_{m_{s}}(p)\Delta_{m_{s}%
}(p+k)+\right. \\
&  \left.  -\left(  8k^{2}+8m_{f}^{2}+8p{{\cdot} }k\right)  \Delta_{m_{f}%
}(p)\Delta_{m_{f}}(p+k)\right\}  \cos^{2}(k\widetilde{p}).
\end{align}

This expression can be seen as the sum of three terms, and each of
these terms is the sum of a planar and of a nonplanar part:
$-i\Sigma_{1loop}(p) =
I_{1}^{P}(p) +I_{1}^{NP}(p) +I_{2}^{P}(p)+I_{2}^{NP}(p) +I_{3}^{P}%
(p)+I_{3}^{NP}(p)$ with

\medskip\noindent$I_{1}^{P}(p)+I_{1}^{NP}(p) \equiv\frac{1}{2}g^{2}\int
\frac{dk^{4}}{\left(  2\pi\right)
^{4}}\frac{8k^{2}+8m_{s}^{2}}{\left(
k^{2}-m_{s}^{2}\right)  \left(  (k+p)^{2}-m_{s}^{2}\right)  }+\frac{1}{2}%
g^{2}\int\frac{dk^{4}}{\left(  2\pi\right)  ^{4}}\cos(2p\widetilde{k}%
)\frac{8k^{2}+8m_{s}^{2}}{\left(  k^{2}-m_{s}^{2}\right)  \left(
(k+p)^{2}-m_{s}^{2}\right)  };$

\medskip\noindent$I_{2}^{P}(p)+I_{2}^{NP}(p) \equiv-\frac{1}{2}g^{2}\int
\frac{dk^{4}}{\left(  2\pi\right)
^{4}}\frac{8k^{2}+8m_{f}^{2}}{\left(
k^{2}-m_{f}^{2}\right)  \left(  (k+p)^{2}-m_{f}^{2}\right)  }-\frac{1}{2}%
g^{2}\int\frac{dk^{4}}{\left(  2\pi\right)  ^{4}}\cos(2p\widetilde{k}%
)\frac{8k^{2}+8m_{f}^{2}}{\left(  k^{2}-m_{f}^{2}\right)  \left(
(k+p)^{2}-m_{f}^{2}\right)  };$

\medskip\noindent$I_{3}^{P}(p)+I_{3}^{NP}(p)\equiv-\frac{1}{2}g^{2}\int
\frac{dk^{4}}{\left(  2\pi\right)  ^{4}}\frac{8p{{\cdot}}
k}{\left(
k^{2}-m_{f}^{2}\right)  \left(  (k+p)^{2}-m_{f}^{2}\right)  }.-\frac{1}%
{2}g^{2}\int\frac{dk^{4}}{\left(  2\pi\right)  ^{4}}\cos(2p\widetilde{k}%
)\frac{8p{{\cdot}} k}{\left(  k^{2}-m_{f}^{2}\right)  \left(  (k+p)^{2}%
-m_{f}^{2}\right)  };$

The planar terms involve integrations which are already done
ordinarily in field theory in commutative spacetime. Their
contributions lead, as in the commutative case, to logarithmic
mass and wavefunction renormalization. We are here mainly
interested in $\Sigma(p)_{1loop}^{NP(E)}$, the sum of the
nonplanar contributions, which we study in the euclidean region.
One easily finds$\footnote{$K_{0}(x)$ and $K_{1}(x)$ are modified
Bessel functions of the second kind.}$
\begin{equation}
\Sigma_{1loop}^{NP(E)}(p)=I_{1E}^{NP}(p)+I_{2E}^{NP}(p)+I_{3E}^{NP}(p)~,
\label{revgamma}%
\end{equation}
where
\begin{align}
I_{1E}^{NP}(p)  &  =\frac{g^{2}}{2\left(  2\pi\right)  ^{2}}\int_{0}%
^{1}da\left\{  \left[  8m_{s}^{2}+4p^{2}(1-a)(2a-1)\right]
K_{0}(2\left|
\widetilde{p}\right|  \sqrt{m_{s}^{2}+p^{2}a(1-a)})+\right. \nonumber\\
&  \left.  -\frac{4}{\left|  \widetilde{p}\right|  }\sqrt{m_{s}^{2}%
+p^{2}a(1-a)}K_{1}(2\left|  \widetilde{p}\right|  \sqrt{m_{s}^{2}+p^{2}%
a(1-a)})\right\}  ,
\end{align}
\bigskip%
\begin{equation}
I_{2E}^{NP}(p)=-\left[  I_{1E}^{NP}(p)\right]  _{m_{s}\rightarrow
m_{f}},
\end{equation}
\bigskip%
\begin{equation}
I_{3E}^{NP}(p)=-\frac{4}{\left(  2\pi\right)  ^{2}}p^{2}\frac{g^{2}}{2}%
\int_{0}^{1}dbbK_{0}(2\left|  \widetilde{p}\right|  \sqrt{m_{f}^{2}%
+p^{2}b(1-b)})~. \label{cont3}%
\end{equation}

In the case of exact SUSY, $m_{s}=m_{f}$, the contributions
$I_{1E}^{NP}$ and $I_{2E}^{NP}$ cancel each other, so that
$\Sigma_{1loop}^{NP(E)}=I_{3E}^{NP}$ and there are no IR
divergencies~\cite{hep-th/0005272,hep-th/0002075}.

In the general case, $m_{s}\neq m_{f}$, IR divergencies are
present. Their structure depends on the relative magnitude of the
SUSY-restoration scale $\Lambda_{SUSY}\simeq m_{f}$ and the
noncommutativity scale $M_{nc}=\tfrac {1}{\sqrt{\left|
\theta\right|  }}$ (where $\left|  \theta\right|  $ denotes
generically a characteristic size of the elements of the matrix
$\theta _{\mu\nu}$).

If $M_{nc} < m_{f}$ and $p\ll\frac{M_{nc}^{2}}{m_{f}}$ the
non-planar part of the self energy is well approximated by
\begin{align}
\Sigma_{1loop}^{NP(E)}(p)  &  \simeq\frac{g^{2}}{\left(  2\pi\right)  ^{2}%
}\int_{0}^{1}da\left\{  6m_{f}^{2}\ln\left(  2\left|
\widetilde{p}\right|
\sqrt{m_{f}^{2}+p^{2}a(1-a)}\right)  +\right. \nonumber\\
&  -6m_{s}^{2}\ln\left(  2\left|  \widetilde{p}\right|  \sqrt{m_{s}^{2}%
+p^{2}a(1-a)}\right)  +\nonumber\\
&  +2p^{2}(1-a)(3a-1)\left[  \ln\left(
\sqrt{m_{f}^{2}+p^{2}a(1-a)}\right)
-\ln\left(  \sqrt{m_{s}^{2}+p^{2}a(1-a)}\right)  \right] \nonumber\\
&  +\left(  m_{s}^{2}-m_{f}^{2}\right)  [6\ln2-6\gamma+1]+\nonumber\\
&  \left.  +2p^{2}a\left[  \ln\left(  2\left|
\widetilde{p}\right| \sqrt{m_{f}^{2}+p^{2}a(1-a)}\right)
-(\ln2-\gamma)\right]  \right\}  ~.
\label{GIANLUCA1}%
\end{align}
[This approximation is also valid for all $p < M_{nc}$ if $M_{nc}
> m_{f}$, but we are mainly interested here in the case $M_{nc} <
m_{f}$ which allows us to explore the implications for low-energy
phenomena of SUSY restoration above $M_{nc}$.]

If $M_{nc} < m_{f}$ and $\frac{M_{nc}^{2}}{mf}\ll p\ll M_{nc}$ the
non-planar part of the self energy is well approximated by
\begin{align}
\Sigma_{1loop}^{NP(E)}(p)  &  \simeq\frac{g^{2}}{\left(  2\pi\right)  ^{2}%
}\int_{0}^{1}da\left\{  -\frac{1}{\left|  \widetilde{p}\right|
^{2}}+\right.
\nonumber\\
&  -\ln\left(  2\left|  \widetilde{p}\right|
\sqrt{m_{f}^{2}+p^{2}a(1-a)}
\right)  \left[  6m_{s}^{2}+2p^{2}(1-a)(3a-1)\right]  +\nonumber\\
&  \left.  +m_{s}^{2}[6\ln2-6\gamma+1]+2p^{2}(1-a)[a(3\ln2-3\gamma+\frac{1}%
{2})-(\ln2-\gamma)]\right\}  . \label{GIANLUCA2}%
\end{align}

As a result of contributions coming from the UV portion of loop
integrals, we are finding that (for $m_{s}\neq m_{f}$) the model
is affected by logarithmic IR singularities (\ref{GIANLUCA1}) if
$\frac{M_{nc}^{2}}{mf}\gg p$, but as soon as momenta are greater
than $\frac{M_{nc}^{2}}{m_{f}}$ the dependence of
the self-energy on momentum turns into an inverse-square law (\ref{GIANLUCA2}%
). In the limit $m_{f}\rightarrow\infty$, the case in which there
is absolutely no SUSY (not even in the UV), the inverse-square law
takes over immediately and the theory is affected by quadratic IR
singularities. The case of exact SUSY $m_{f}=m_{s}$ is free from
IR singularities, but of no interest for physics (Nature clearly
does not enjoy exact SUSY).

The IR/UV mixing manifests in two (obviously connected) ways which
is worth distinguishing: (1) The UV portion of loop integrals is
responsible for some IR singularities of the self-energies, (2)
the low-energy structure of the model can depend on $m_{f}$ even
when $m_{f}$ is much higher than the energy scales being probed.
There is no IR/UV decoupling.

\subsection{Further effects on the low-energy sector from UV physics}

The implications of supersymmetry for the IR sector of canonical
noncommutative spacetimes are very profound. In our illustrative
model one finds that exact SUSY leads to absence of IR
divergences, if SUSY is only present in the UV (UV restoration of
SUSY) one finds soft, logarithmic, IR divergences, and total
absence of SUSY ($m_{f}\rightarrow\infty$) leads to quadratic IR
divergences. While the presence of SUSY in the UV is clearly an
example of UV physics with particularly significant implications
for the IR sector of canonical noncommutative spacetimes, from
this example we must deduce that in general the loss of decoupling
between UV and IR sectors can be very severe. Other features of
the UV sector, which perhaps have not even yet been contemplated
in the literature, might have similarly pervasive implications for
the IR sector.

A particularly interesting scenario is the one in which
supersymmetry is restored at some high scale (which in our
illustrative model is $m_{f}$) and then at some even higher scale,
possibly identified with the so-called ``quantum-gravity scale'',
the theory predicts additional structures, which in turn, again,
would affect the infrared. The example of quantum gravity is
particularly significant since we have no robust (experimentally
supported) information on this realm of physics, so it represents
an example of UV physics for which our intuition might easily
fail, and as a consequence our intuition for its implications for
the IR sector of a field theory in canonical noncommutative
spacetime might also easily fail.

As a way to emphasize the sensitivity of the IR sector to such
unknown UV physics, it is worth noting here some formulas that
describe features of our illustrative model from the perspective
of a theory with fixed cutoff scale $\Lambda$. For renormalizable
field theories in commutative spacetime the presence of such a
cutoff would be basically irrelevant: if the cutoff is much higher
than all scales of interest it will negligibly affect all
predictions and it can be uneventfully removed through the limit
$\Lambda\rightarrow \infty$. Importantly, in a renormalizable
field theory in commutative spacetime the limit
$\Lambda\rightarrow\infty$ is uneventful independently of whether
or not we have introduced in the theory all the correct UV degrees
of freedom hosted by Nature: the low-energy physics is anyway
independent of (decoupled from) the UV sector.

For field theories in canonical noncommutative spacetime the limit
$\Lambda\rightarrow\infty$ is not at all trivial, meaning that the
structures/degrees of freedom encountered along the limiting
procedure can in principle affect significantly the low-energy
physics. One can take the $\Lambda\rightarrow\infty$ limit in a
physically meaningful way only under the assumption that one has
complete knowledge of the full theory of Nature (something which
of course we cannot even contemplate).

The sensitivity of the IR sector to unknown UV physics is well
characterized by considering, for fixed cutoff scale $\Lambda$,
the nonplanar contributions to the two point functions. For the
two-point function we already considered previously one finds:
\begin{align}
I_{1E}^{NP}  &  =\frac{g^{2}}{2}\left\{  \frac{1}{\left(  2\pi\right)  ^{2}%
}\int_{0}^{1}da\left[  8m_{s}^{2}+4p^{2}(2a-1)(1-a)\right]  K_{0}%
(2\sqrt{\widetilde{p}^{2}+\frac{1}{\Lambda^{2}}}\sqrt{m_{s}^{2}+p^{2}%
a(1-a)})+\right. \nonumber\\
&  \left.
+\frac{4}{\sqrt{\widetilde{p}^{2}+\frac{1}{\Lambda^{2}}}}\left[
\frac{\widetilde{p}^{2}}{\widetilde{p}^{2}+\frac{1}{\Lambda^{2}}}-2\right]
\sqrt{m_{s}^{2}+p^{2}a(1-a)}K_{1}(2\sqrt{\widetilde{p}^{2}+\frac{1}%
{\Lambda^{2}}}\sqrt{m_{s}^{2}+p^{2}a(1-a)})\right\}
\end{align}%
\begin{equation}
I_{2E}^{NP}=-I_{1E}^{NP}(m_{s}\rightarrow m_{f})
\end{equation}%
\begin{equation}
I_{3E}^{NP}=-\frac{4}{\left(  2\pi\right)
^{2}}p^{2}\frac{g^{2}}{2}\int
_{0}^{1}dbbK_{0}(2\sqrt{\widetilde{p}^{2}+\frac{1}{\Lambda^{2}}}\sqrt
{m_{f}^{2}+p^{2}b(1-b)})
\end{equation}
Note that nonplanar diagrams are cutoff by $\Lambda_{eff}=\frac{1}%
{\sqrt{\widetilde{p}^{2}+\frac{1}{\Lambda^{2}}}}$. The self-energy
is insensitive to the value of $\Lambda$ as long as the condition
$\left| \widetilde{p} \right|  \gg\frac{1}{\Lambda}$ is satisfied.
But for $\left| \widetilde{p} \right|  <\frac{1}{\Lambda}$ there
is an explicit dependence\footnote{It is worth noticing that for
fixed cutoff $\Lambda$ and $\left|  \widetilde{p}\right|
<\frac{1}{\Lambda}$ the self-energy is essentially independent of
the noncommutativity parameters. This is due to the fact that
under those conditions the nonplanar contributions are completely
negligible. This might encourage one to contemplate the
possibility of a physical cutoff scale $\Lambda$, but it is
important to notice that such a scale would be observer dependent
since ordinary Lorentz transformations still govern the
transformations between inertial observers in canonical
noncommutative spacetime~\cite{gr-qc/0205125}. (In other
noncommutative spacetimes, where the action of boosts is deformed,
a cutoff scale can be introduced in an observer-independent
way~\cite{hep-th/0012238,gr-qc/0205125}, but this is not the case
of canonical noncommutative spacetimes.) We shall disregard this
possibility; however, in theories that already identify a
preferred class of inertial observers, such as theories in
canonical noncommutative spacetimes, the possibility of an
observer-dependent cutoff scale cannot~ \cite{gr-qc/0205125} be
automatically dismissed.} on $\Lambda$ signaling that the infrared
sector is sensitive to new physics in the UV.

\section{Conditional bounds on noncommutativity parameters from low-energy data}

The main point of our work is that the observations made in the
previous Section have significant implications for the comparison
of low-energy experimental data with a theory in canonical
noncommutative spacetime.

It is useful to note here a brief description of the conventional
technique that allows to use low-energy data to set absolute
(unconditional!) limits on the parameters of theories in
commutative spacetime:

\begin{itemize}
\item \textbf{1C.} Data are taken in experiments involving
particles with energies/momenta from some lower (IR) limit,
$\mathcal{S}_{min}$ (we of course do not have available probes
with wavelength, \textit{e.g.}, larger than the size of the
Universe) up to an upper limit, $\mathcal{S}_{max}$, which
naturally coincides with the highest energy scales attainable in
our laboratory experiments (and, in appropriate cases, the energy
scales involved in certain observations in astrophysics).

\item \textbf{2C.} We then compare these experimental results
obtained at
energy/momentum scales within the range $\{\mathcal{S}_{min},\mathcal{S}%
_{max}\}$ to the corresponding predictions of the theory of
interest. In deriving these predictions we sometimes formally
appear to use the whole structure of the theory, all the way to
infinite energy/momentum; however, in reality, because of the
IR/UV decoupling that holds in (renormalizable) theories in
commutative Minkowski spacetime, the theoretical prediction only
depends on the IR structure of the theory, up to energy/momentum
scales which are not much bigger than $\mathcal{S}_{max}$. (For
example, degrees of freedom with masses of order, say,
$10^{5}\mathcal{S}_{max}$ would anyway not affect the relevant
predictions).

\item \textbf{3C.} If the theoretical predictions obtained in this
way do not
agree with the observations performed in the range $\{\mathcal{S}%
_{min},\mathcal{S}_{max}\}$ we then conclude that the theory in
question is to be abandoned.

\item \textbf{4C.} If the theoretical predictions obtained in this
way agree
with the observations performed in the range $\{\mathcal{S}_{min}%
,\mathcal{S}_{max}\}$ we then conclude that the theory in question
provides a valid description of phenomena up to energy/momentum
scales of order $\mathcal{S}_{max}$. Typically the predictions of
the theory will depend on some free parameters and this parameter
space will be constrained by the requirement of agreeing with the
observations. Values of the parameters that do not belong to this
allowed portion of the parameter space are definitely
(unconditionally) excluded, since nothing that we could introduce
in the ultraviolet could modify the low-energy predictions. In
light of the fact that the structure of the theory above
$\mathcal{S}_{max}$ did not play any true role in the derivation
of the predictions, the successful comparison with
$\{\mathcal{S}_{min},\mathcal{S}_{max}\}$ experiments provides no
particular encouragement for what concerns the validity of the
theory at scales much above $\mathcal{S}_{max}$.

\item \textbf{5C.} With precision measurements in the range $\{\mathcal{S}%
_{min},\mathcal{S}_{max}\}$ we can sometimes put limits on
features of the
theory also slightly (up to a few orders of magnitude) above $\mathcal{S}%
_{max}$. For example, one of the parameters of the theory could be
the mass of a certain particle and the contributions to low-energy
processes due to that particle, while suppressed by its mass, can
be tested in high-precision measurements.
\end{itemize}

For theories in canonical noncommutative spacetime the situation
is quite different, as one infers from the analysis reported in
the previous Section.
The comparison between the theory and data taken in the range $\{\mathcal{S}%
_{min},\mathcal{S}_{max}\}$ is much more delicate:

\begin{itemize}
\item \textbf{2NC.} From the observations made in the previous
Section it follows that in a canonical noncommutative spacetime a
truly reliable
derivation of the predictions for the energy/momentum range $\{\mathcal{S}%
_{min},\mathcal{S}_{max}\}$ requires full knowledge of the theory
at all energy/mo- mentum scales up to
$M_{nc}^{2}/\mathcal{S}_{min}$ (and of course, if
$M_{nc}\gg\mathcal{S}_{max}$, the scale
$M_{nc}^{2}/\mathcal{S}_{min}$ can be much higher than both
$M_{nc}$ and $\mathcal{S}_{max}$). In particular, the IR/UV mixing
is such that degrees of freedom with masses that are much above
$\mathcal{S}_{max}$ still affect significantly the predictions of
the theory in the range $\{\mathcal{S}_{min},\mathcal{S}_{max}\}$.

\item \textbf{3NC.} So the theory can only be taken as a full
description of Nature. It cannot be intended to give the right
predictions only in some low-energy limit. If the predictions of
such a theory are found to be in conflict with observations, it
might still well be that the theory contains the right low-energy
degrees of freedom, and that the disagreement is due to having
adopted the wrong UV sector. So, from our more conventional
perspective (in which we try to identify theories that contain the
right degrees of freedom up to a certain scale) disagreement with
observations does not force us to abandon the theory: it only
invites us to introduce appropriate new physics in the UV sector.

\item \textbf{4NC.} Similarly, if the theoretical predictions are
found to
agree with the observations performed in the range $\{\mathcal{S}%
_{min},\mathcal{S}_{max}\}$ when some free parameters fall within
a certain allowed portion of parameter space, values of the
parameters that do not belong to that region of the parameter
space cannot be conclusively excluded. They are excluded only
\textbf{conditionally}, in the sense that their exclusion is only
tentative, pending further exploration of the UV sector. Think for
example of the illustrative model we considered in the preceding
Section. The $m_{f}\rightarrow\infty$ of that model is a model
without any SUSY (not even in the UV sector). One could propose
such a non-SUSY model and
compare it to data obtained in the range $\{\mathcal{S}_{min},\mathcal{S}%
_{max}\}$. Clearly the need to agree with observations would then
impose a severe (lower) bound on the noncommutativity scale, a key
parameter of the theory, in order to suppress the IR divergences
(\textit{e.g.} effectively relegating those divergences at scales
below $\mathcal{S}_{min}$). However, this bound on the
noncommutativity scale would be only conditional, in the sense
that modifying the theory only in the ultraviolet (\textit{i.e.}
where
we would say it has not been tested with our data in the range $\{\mathcal{S}%
_{min},\mathcal{S}_{max}\}$) may be sufficient to lift the bound.
In fact, SUSY in the ultraviolet sector ($m_{f}$ large but finite)
significantly softens the divergences used to set the bound.
Whereas in commutative spacetime the bounds on parameter space
apply directly to the structure of the theory in the range of
energy/momentum scales that have been probed experimentally, in
canonical noncommutative spacetime the information gained
experimentally in the range
$\{\mathcal{S}_{min},\mathcal{S}_{max}\}$ leaves open two
possibilities: it may still, as in the case of theories in
commutative spacetime, constrain the parameters of the theory in
that same range of energy/momentum scales, but one cannot exclude
the possibility that our low-energy observations are instead
primarily a manifestation of some features of the UV sector
(transferred to the low-energy sector via the IR/UV mixing) and
therefore cannot be used to constrain the low-energy structure of
the theory. If there is disagreement between theory and
experiments in the range $\{\mathcal{S}_{min},\mathcal{S}_{max}\}$
one would normally assume that some aspects (\textit{e.g.} the
field content) of the theory must be changed in that same range of
energy/momentum scales, instead in canonical noncommutative
spacetime that same disagreement could be solved not only by
introducing new features in the
$\{\mathcal{S}_{min},\mathcal{S}_{max}\}$ region but also by
introducing new features in the UV sector of the theory.

\item \textbf{5NC.} Since data taken in the range $\{\mathcal{S}%
_{min},\mathcal{S}_{max}\}$ do not even give definitive
information on the structure of the theory in that same range, it
is of course true that measurements in the range
$\{\mathcal{S}_{min},\mathcal{S}_{max}\}$ cannot be used to put
limits on features of the theory even just slightly above
$\mathcal{S}_{max}$, no matter how precise those measurements are.
However, just because features of the UV sector affect the
low-energy physics, under the assumption that the spacetime is
indeed canonically noncommutative, one can gain insight of the UV
structure of the theory, even just using low-energy data. For
example, some of the observations made in the previous Section
provide an opportunity to discover UV SUSY even just using
low-energy data: if data allowed us to identify an energy/momentum
scale at which the self-energy changed its qualitative dependence
on momentum in the way described by comparison of
Eqs.~(\ref{GIANLUCA1}) and (\ref{GIANLUCA2}), we could then infer
rather robustly the presence of SUSY at high energies and (if the
value of the noncommutativity scale was deduced from some other
observations) we could even deduce the scale of SUSY restoration.
\end{itemize}

In summary we found that the predictions of a canonical
noncommutative theory in the low-energy (i.e. experimentally
accessible) sector of theory depend strongly not only on the
low-energy structure of the theory but also on its high-energy
structure. This is different from the case of commutative
theories, where low-energy predictions are independent of the
high-energy degrees of freedom. The phenomenological implications
of this lack of energy-scale decoupling are of course very
striking. To reliably falsify or accept a theory with low energy
data it is not enough to specify the low-energy sector of the
theory one is considering, but one must also fully specify the
high-energy sector. Two theories with the same low-energy sector
but different high energy sectors may require different parameter
values to fit the data.

\section{Futility of approaches based on expansion in powers of $\theta$}

The observations reported in the preceding section indicate that
some of the standard techniques used in phenomenology require a
prudent implementation in the context of theories in canonical
noncommutative spacetimes. We want to emphasize in this section
that for one of the techniques which served us well in the
analysis of theories in commutative spacetime there are even more
severe limitations to the applicability in the context of theories
in canonical noncommutative spacetimes. This is the technique that
relies on the truncation of a power series in one of the
parameters of the theory: we argue that, at the
quantum-field-theory level, the results obtained by truncating a
power series in $\theta$ do not provide a reliable approximation
of the full theory. This type of truncation, which has been widely
used in the
literature~\cite{hep-th/0104153,hep-th/0205153,hep-ph/0112320,hep-th/0205214,hep-ph/0202223,
hep-ph/0107291,hep-ph/0202121}), is based on the inclusion of only
a few terms in the $\theta$-expansion of the Moyal
$\star$-product. For example up to the second order in $\theta$
one could write
\begin{align}
\varphi_{1}(x)\star\varphi_{2}(x)  &
=\varphi_{1}(x)\varphi_{2}(x)+\frac
{i}{2}\theta^{\mu\nu}\partial_{\mu}\varphi_{1}(x)\partial_{\nu}\varphi
_{2}(x)+\nonumber\\
&  -\frac{1}{8}\theta^{\alpha\beta}\theta^{\mu\nu}\partial_{\alpha}%
\partial_{\mu}\varphi_{1}(x)\partial_{\beta}\partial_{\nu}\varphi
_{2}(x)+O(\theta^{3}) \label{expa}%
\end{align}
The resulting action constructed with the truncated
$\star$-product (\ref{expa}) depends only on a finite number of
derivatives so it is local, unlike the full theory. Moreover,
since $\theta$ has negative mass dimensions, the action will also
certainly be power-counting nonrenormalizable, whereas the full
theory might be renormalizable
\cite{hep-th/0008057,hep-th/9912072,hep-th/0005272,hep-th/0104217,hep-th/9912075}%
.

Even more serious concerns emerge from the realization that the
expansion one is performing is (of course) not truly based on a
power series in the dimensionful quantity $\theta$: it is rather
an expansion in dimensionless quantities of the type $p\theta p$.
Therefore already at tree level the truncated $\theta$-expanded
theory can only give a good approximation of the full theory at
scales $p$ such that $p\theta p\lesssim1$, \textit{i.e.}
$p\lesssim1/\sqrt{\theta}$.

But actually even in that range of momenta the expansion cannot be
used reliably. Its reliability is spoiled by quantum corrections.
The quantum corrections involve the Moyal $\star$-product inserted
in loop diagrams, and the truncation will reliably describe these
loop corrections only for loop momenta such that
$p\lesssim1/(\theta\Lambda)$. In fact, in loop integrals involving
factors of the type $p\theta k$, with $p$ playing the role of
external momentum and $k$ playing the role of integration/loop
momentum, one would like a reliable truncation that is valid over
the whole loop-integration range, which extends at least up to a
cutoff $\Lambda$. In order to have $p\theta k\lesssim1$ even for
$k$ as large as $\Lambda$ it is necessary to assume that indeed
$p\lesssim1/(\theta\Lambda)$. This can also be inferred
straightforwardly in the illustrative example of the
``$\lambda\Phi^{4}$'' scalar-boson field theory: there one finds
that the full theory predicts nonplanar terms giving a leading
contribution of the form
\begin{equation}
\Sigma_{NP}^{1}(p)\simeq\dfrac{g^{2}}{\widetilde{p}^{2}+1/\Lambda^{2}}%
=\Lambda^{2}\dfrac{g^{2}}{\Lambda^{2}\widetilde{p}^{2}+1}.
\end{equation}
whereas the truncated $\theta$-expansion of the $\star$-product
would replace this prediction with
\begin{equation}
\Sigma_{NP}^{1}(p)\simeq g^{2}\Lambda^{2}\left\{
1-\Lambda^{2}\widetilde {p}^{2}+O(\theta^{4})\right\}  .
\end{equation}
Clearly the two expressions are equivalent only if
$\Lambda^{2}\widetilde {p}^{2}\lesssim1$, which indeed corresponds
to $p\lesssim1/(\theta\Lambda)$.

Therefore, when one includes quantum/loop effects, the truncated
$\theta $-expansion could be a good approximation of the full
theory only in the range of momenta $p\lesssim1/(\theta\Lambda)$.
But as we have discussed in the preceding section this is just the
range of momenta in which the theory is maximally sensitive to
ultraviolet physics, which we must assume to be unknown. In other
words the truncated $\theta$-expansion reliably approximates the
full theory only in a regime where the full theory is itself void
of predictive power,
%
%
%
%
%
%
%

because of its sensitivity to unknown physics that might be
present in the ultraviolet. It therefore appears that these
truncated $\theta$-expansions cannot be used for a meaningful
comparison between data and theories in canonical noncommutative
spacetime. In other contexts expansions in powers of $p$ versus
some characteristic momentum scale have been proven to give a
reliable low-energy effective-theory description of the full
theory one intends to study, but in this case of field theories in
canonical noncommutative spacetime the IR/UV mixing provides a
powerful obstruction for any attempt to obtain a meaningful
low-energy effective-theory description.



\chapter{CJT formalism for phase transition on CNC spacetime}

\footnotetext{* In this Chapter we discuss in detail the analysis
reported more briefly in Ref.~\cite{gmgacld}.}*\qquad We have
discussed how the IR/UV mixing, which significantly affects
canonical noncommutative theories, causes strong IR problems. We
have also emphasized that one manifestation of these IR problems
(after the removal of the cut-off) is through zero-momentum poles
in certain Green functions. IR problems of different origin but
similar form, are known to plague also Thermal-Quantum-Field
theories and have been successfully treated using a
nonperturbative technique developed by Cornwall-Jackiw-Tomboulis
(CJT). We apply the CJT formalism to the scalar
$\lambda\varphi^{4}$ theory focusing in the so-called ``bubble
approximation''. Assuming translational invariance of the vacuum
we construct the gap equation and the CJT effective potential. We
discuss the renormalizability of the CJT effective potential both
in ordered and in the disordered phase for general values of the
non-commutativity parameter $\theta$. We comment in particular on
the commutativity limit ($\theta \rightarrow0$) and on the strong
non-commutativity limit ($\theta \rightarrow\infty$). We observe
that while in the disordered phase the hypothesis of translational
invariance leads to a renormalizable effective potential, in the
translational-invariant ordered phase, differently from the
commutative case, the effective potential and the gap equation do
not renormalize. We argue that our result, essentially based on a
selective all-order resummation, appears to confirm the other
(perturbative, one-loop) results, we reported in Chapter 3, that
indicate the incompatibility of a translational-invariant ordered
phase with the infrared structure of the canonical-noncommutative
theories.

\section{CJT formalism}

In this section we briefly review the CJT formalism for the scalar
theory in the commutative case
\cite{CJT,hep-ph/9311324,hep-ph/9211211}. The starting
point is the definition of the partition function in the form%
\[
Z[J,K]=\exp W[J,K]=\int D\phi\exp\left\{  S\left(  \phi\right)
+\int dx^{4}J(x)\phi(x)+\int
dy^{4}dx^{4}\phi(x)K(x,y)\phi(y)\right\}  ,
\]
in which two sources $J(x)$ and $K(x,y)$ are present.

One defines also $\varphi(x)$ and $G(x,y)$ by the relations
$\frac{\delta
W[J(x),K(x,y)]}{\delta J(x)}=\varphi(x)$ and $\frac{\delta W[J(x),K(x,y)]}%
{\delta K(x,y)}=\frac{1}{2}\left\{
\varphi(x)\varphi(y)+G(x,y)\right\}  .$
Then one considers the double Legendre transformation of $W[J,K]$%

\begin{align}
\Gamma\lbrack\varphi(x),G(x,y)]  &  =W[J(x),K(x,y)]-\int
dx^{4}\varphi
(x)J(x)-\frac{1}{2}\int dx^{4}dy^{4}\varphi(x)K(x,y)\varphi(y)+\nonumber\\
&  -\frac{1}{2}\int dx^{4}dy^{4}G(x,y)K(x,y) \label{gamma}%
\end{align}

which satisfies the relation%
\begin{align*}
\frac{\delta\Gamma\lbrack\varphi,G]}{\delta\varphi(x)}  &
=-J(x)+\int
dy^{4}K(x,y)\varphi(y),\\
\frac{\delta\Gamma\lbrack\varphi,G]}{\delta G(x,y)}  &
=-\frac{1}{2}K(x,y).
\end{align*}

The physical point corresponds to vanishing sources $K(x,y)=0,$
$J(x)=0$, so
that $\varphi(x)$ and $G(x,y)$ are solutions of the stationarity equations:%
\begin{align}
\frac{\delta\Gamma\lbrack\varphi,G]}{\delta\varphi(x)}  &  =0\label{peq}\\
\frac{\delta\Gamma\lbrack\varphi,G]}{\delta G(x,y)}  &  =0 \label{gap}%
\end{align}

It can be shown \cite{CJT}\ that $\Gamma\lbrack\varphi,G]$ so
defined is the generating functional for the two-particle
irreducible(2PI) Green's functions, with propagator given by
$G(x,y)$ and vertices given by $S_{int}(\varphi ;\phi)$, where
$S_{int}(\varphi;\phi)$ is obtained from $S(\varphi)$ by retaining
only cubic and higher $\varphi$ -terms in the expression of
$S(\varphi+\phi)$.

\begin{figure}[h]
\begin{center}
\includegraphics[width=8cm]{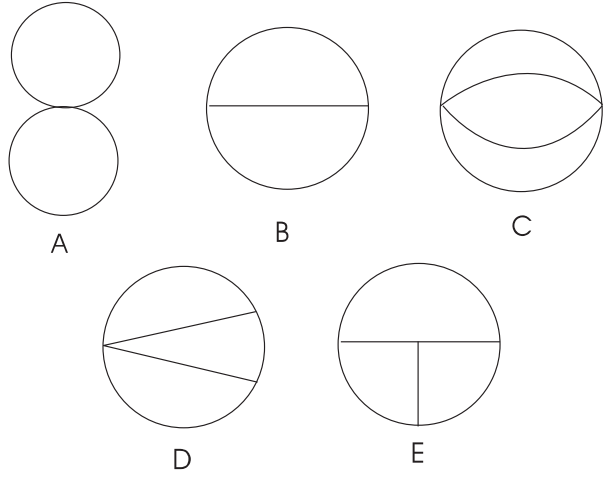}
\end{center}
\caption{Two-particle irreducible graphs contributing to $\Gamma^{2}%
(\varphi,G)$ up to the three loop level.}%
\label{fig:vvdiagrams}%
\end{figure}

One can expand (\ref{gamma}) to obtain%

\begin{equation}
\Gamma(\varphi,G)=S_{cl}(\varphi)-\frac{1}{2}TrLnD_{0}^{-1}G+\frac{1}%
{2}Tr\left\{  D^{-1}G-1\right\}  +\Gamma^{2}(\varphi,G) \label{gammae}%
\end{equation}

where%
\begin{align*}
D^{-1}(x,y)  &  =\frac{\delta^{2}S}{\delta\varphi(x)\delta\varphi(y)},\\
D_{0}^{-1}(x,y)  &  =D^{-1}|_{S_{free}},
\end{align*}

and $\Gamma^{2}(\varphi,G)$ is the sum of vacuum
diagrams(Fig.\ref{fig:vvdiagrams}) with vertices given by $S_{int}%
(\varphi;\phi)$ and propagators given by
$G(x,y).$

\begin{figure}[h]
\begin{center}
\includegraphics[width=8cm]{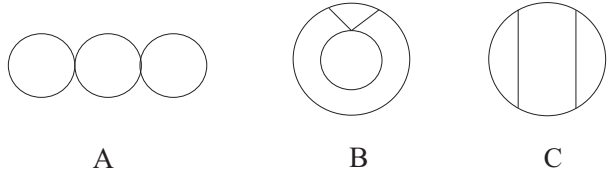}
\end{center}
\caption{Examples of two-particle reducible graphs.}%
\label{fig:vvdiagramspq}%
\end{figure}

Example of two particle reducible graphs which do not contribute
to $\Gamma^{2}(\varphi,G)$ are in
Fig.\ref{fig:vvdiagramspq}\begin{figure}[hh]
\begin{center}
\includegraphics[width=8cm]{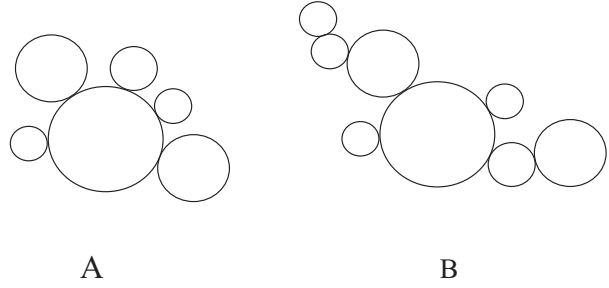}
\end{center}
\caption{Vacuum to vacuum bubble diagrams: daisy (a) and super-daisy (b).}%
\label{bd}%
\end{figure}

Using (\ref{gammae}) the gap equation (\ref{gap}) may be rewritten in the form%
\begin{equation}
G^{-1}(x,y)=D^{-1}(x,y)+2\frac{\delta\Gamma^{2}(\varphi,G)}{\delta
G(x,y)}.
\label{gap2}%
\end{equation}

One can also recover the usual 1PI-effective action
$\Gamma^{1PI}(\varphi)$
simply evaluating $\Gamma\lbrack\varphi,G]$ for vanishing $K(x,y)$:%
\begin{equation}
\Gamma_{1PI}(\varphi)=\Gamma_{2PI}[\varphi,G_{0}],
\end{equation}
where $G_{0}$ is solution of the gap equation%
\begin{equation}
\frac{\delta\Gamma_{2PI}[\varphi,G]}{\delta G(x,y)}=0.
\end{equation}

This 2PI formalism which at a first sight might appear more
involved than the standard 1PI formalism turns out to be very
useful in certain calculations. This is the case for example of
the so-called ``bubble resummation'', which means taking into
account for all the diagrams generate by the vacuum to vacuum
diagrams of the type of Fig.(\ref{bd}). In the case of the
standard 1PI formalism the bubble resummation requires the
evaluation of an infinite number of diagrams. In the 2PI-CJT
formalism one obtains the whole ``bubble
resummation''(\cite{hep-ph/9311324}) simply considering the
``eight''-diagram (A in Fig.\ref{fig:vvdiagrams}) contribution to
$\Gamma^{2}(\varphi,G),$ and the corresponding gap equation. This
``bubble resummation'' turns out to be useful in theories in which
the insertion of a tadpole is not effectively suppressed by the
coupling constant as in the thermal field theories or
canonical-noncommutative theories\footnote{In Thermal Field
Theories\ the insertion of a tadpole which for example
distingushes diagrams Fig.5.1 A and Fig.5.2 A costs a factor
$\lambda\frac{T^{2}}{m^{2}},$ which for large temperatures can be
even large than 1. In noncommutative Thermal Field
Theories the same insertion roughly comes with a factor $\lambda\frac{1}%
{p^{2}\theta^{2}m^{2}}$ which, depending on the momentum entering
the inserted tadpole, can be large.}.

We will use this approximation in the following sections\textbf{.}

\section{CJT formalism in canonical-noncommutative spacetime}

\qquad We recall that once the Moyal $\star$-product (\ref{mp}) is
introduced the scalar $\lambda\varphi^{4}$-theory in canonical
noncommutative spacetime takes the form of a commutative theory
with a deformed interaction given by substituting the products of
fields with the $\star$-products. This implies that the CJT
formalism should be applicable; in fact no specific assumptions
are made in the CJT procedure about the form of the interaction.
In particular Eq.(\ref{gammae}) is still valid in our
noncommutative context.

It is easy to see that $\lambda\varphi^{4}$-theory in canonical
noncommutative
spacetime%
\begin{align}
D^{-1}(x,y)  &  =\frac{\delta^{2}S}{\delta\varphi(x)\delta\varphi
(y)}=\nonumber\\
&  =-\left[  \square+m^{2}\right]  _{x}\delta^{4}(x-y)+\nonumber\\
&  -\frac{\lambda}{3!}\left\{
\delta^{4}(x-y)\star\varphi\star\varphi
+\varphi\star\delta^{4}(x-y)\star\varphi+\varphi\star\varphi\star\delta
^{4}(x-y)\right\}  ,\label{invD}\\
D_{0}^{-1}(x,y)  &  =D^{-1}|_{S_{free}}=-\left[
\square+m^{2}\right]
_{x}\delta^{4}(x-y) \label{invD0}%
\end{align}

As expected $D_{0}^{-1}(x,y)$ is not modified by noncommutativity
since the integrals of terms quadratic in the fields are not
modified by the $\star $-product (\ref{mp}), while $D^{-1}(x,y)$
acquires the $\theta$-dependence. One can easily calculate the
$\star$-products which appear in (\ref{invD})
obtaining, in the momentum space,%
\begin{align*}
\delta^{4}(x-z)\star\varphi(x)\star\varphi(x)  &  =\int dp_{1}^{4}dp_{2}%
^{4}dp_{3}^{4}\exp
F_{1}(\underline{p})\exp[ix(p_{1}+p_{2}+p_{3})]\exp
[-ip_{3}z]\widetilde{\varphi}(p_{1})\widetilde{\varphi}(p_{2})\\
\varphi(x)\star\delta^{4}(x-z)\star\varphi(x)  &  =\int dp_{1}^{4}dp_{2}%
^{4}dp_{3}^{4}\exp
F_{2}(\underline{p})\exp[ix(p_{1}+p_{2}+p_{3})]\exp
[-ip_{3}z]\widetilde{\varphi}(p_{1})\widetilde{\varphi}(p_{2})\\
\varphi(x)\star\varphi(x)\star\delta^{4}(x-z)  &  =\int dp_{1}^{4}dp_{2}%
^{4}dp_{3}^{4}\exp
F_{3}(\underline{p})\exp[ix(p_{1}+p_{2}+p_{3})]\exp
[-ip_{3}z]\widetilde{\varphi}(p_{1})\widetilde{\varphi}(p_{2})
\end{align*}

where%
\begin{align*}
F_{1}(\underline{p})  &  =-\frac{i}{2}\theta_{\mu\nu}\left\{  p_{1}^{\mu}%
p_{2}^{\nu}+p_{3}^{\mu}p_{1}^{\nu}+p_{3}^{\mu}p_{2}^{\nu}\right\} \\
F_{2}(\underline{p})  &  =-\frac{i}{2}\theta_{\mu\nu}\left\{  p_{1}^{\mu}%
p_{3}^{\nu}+p_{1}^{\mu}p_{2}^{\nu}+p_{3}^{\mu}p_{2}^{\nu}\right\} \\
F_{3}(\underline{p})  &  =-\frac{i}{2}\theta\left\{
p_{1}^{\mu}p_{2}^{\nu
}+p_{1}^{\mu}p_{3}^{\nu}+p_{2}^{\mu}p_{3}^{\nu}\right\}  .
\end{align*}

Now we have to calculate the $S_{int}$ action of the 2PI
formalism. One can proceed in two steps. The first step is the
translation of the action
$S(\phi)\rightarrow S(\phi+\varphi)$%
\begin{align}
S(\phi+\varphi)  &  =\left\{  \int dx^{4}\frac{1}{2}m^{2}\left(
\phi +\varphi\right)  ^{2}+\frac{1}{2}\partial_{\mu}\left(
\phi+\varphi\right)
\partial^{\mu}\left(  \phi+\varphi\right)  +\right. \nonumber\\
&  \left.  +\frac{\lambda}{4!}\left(  \phi+\varphi\right)
\star\left( \phi+\varphi\right)  \star\left(  \phi+\varphi\right)
\star\left(
\phi+\varphi\right)  \right\}  \label{traslact}%
\end{align}

The second step is the one of retaining from (\ref{traslact}) only
cubic, and higher, terms in $\phi.$ So doing one obtains that the
interaction vertices
are given by the action%
\[
S_{int}(\phi;\varphi)=\frac{\lambda}{4!}\int
dx^{4}\phi\star\phi\star\phi \star\phi+\frac{\lambda}{6}\int
dx^{4}\phi\star\phi\star\phi\star\varphi
\]
where the cyclitity of the $\star$-product (\ref{mp}) under
integration has been used.

To proceed further we now need to adopt ansatz for the form of
$G(x,y).$ Here we want to consider only transitionally invariant
configurations so that the
more general ansatz we can consider takes the form%
\begin{equation}
G(x,y)=G(x-y)=\int d\alpha^{4}\frac{\exp i\alpha(x-y)}{\alpha^{2}+M^{2}%
(\alpha)} \label{GA}%
\end{equation}
where $M^{2}(\alpha)$ is to be determined. Once the ansatz
(\ref{GA}) has been done we can start calculating the various
terms in the left-hand-side of
(\ref{gammae}). The first term is trivial. The second term is%
\[
Tr\ln(D_{0}^{-1}G)=\int dx^{4}\int dp^{4}\ln\left\{  \frac{p^{2}+m^{2}}%
{p^{2}+M^{2}(p)}\right\}  .
\]
The third term is%
\begin{align*}
Tr\left\{  D^{-1}G-1\right\}   &  =\int dx^{4}\left\{  \int
dk^{4}\frac
{m^{2}-M^{2}(k)}{k^{2}+M^{2}(k)}+\right. \\
&  \left.  +\dfrac{\lambda}{3!}\int dp_{1}^{4}dp_{2}^{4}dp_{3}^{4}%
F(\underline{p})\exp
ix(p_{1}+p_{2})\widetilde{\varphi}(p_{1})\widetilde
{\varphi}(p_{2})\dfrac{1}{p_{3}^{2}+M^{2}(p_{3})}\right\}  ,
\end{align*}
where $F(\underline{p})=\sum_{i=1}^{3}\exp F_{i}(\underline{p}).$

Therefore we have that (\ref{gammae}) with the ansatz (\ref{GA})
takes the
form%
\begin{align}
\Gamma(\varphi,G)  &  =\frac{1}{2}\int dx^{4}\left\{  (\partial_{\mu}%
\varphi)^{2}+m^{2}\varphi^{2}\right\}  +\frac{\lambda}{4!}\int dx^{4}%
\varphi\star\varphi\star\varphi\star\varphi+\label{GammaF}\\
&  +\frac{1}{2}\int dx^{4}\int dp^{4}\ln\left\{  \frac{p^{2}+M^{2}(k)}%
{p^{2}+m^{2}}\right\}  +\nonumber\\
&  +\frac{1}{2}\int dx^{4}\int dk^{4}\frac{m^{2}-M^{2}(k)+\dfrac{\lambda}%
{3!}\int dp_{1}^{4}dp_{2}^{4}F(\underline{p},k)\exp ix(p_{1}+p_{2}%
)\widetilde{\varphi}(p_{1})\widetilde{\varphi}(p_{2})}{k^{2}+M^{2}%
(k)}+\nonumber\\
&  +\Gamma^{2}(\varphi,G)\nonumber
\end{align}

Now we must evaluate $\Gamma^{2}(\varphi,G)$ with vertices given
by $S_{int}(\phi;\varphi)$ and propagator given by $G(x,y).$

As in the commutative case the CJT effective action can be used to
obtain the 1PI effective action. In particular the one-loop 1PI
effective action is obtained setting $\Gamma^{2}=0.$ In this case
the gap equation (\ref{gap2})
reduces to%
\begin{equation}
G^{-1}(x,y)=D^{-1}(x,y).
\end{equation}

Using this expression one can easily compute the one-loop 1PI
effective action in the 2PI CJT formalism. One has only to
calculate
\[
\left[  D^{-1}D_{0}\right]  (x,y)=\delta^{4}(x-y)+K(x,y)
\]

where%
\begin{equation}
K(x,y)=\frac{\lambda}{3!}\int dz\left[
^{4}\delta^{4}(x-z)\star\varphi
\star\varphi+\varphi\star\delta^{4}(x-z)\star\varphi+\varphi\star\varphi
\star\delta^{4}(x-z)\right]  D(z-y). \label{K}%
\end{equation}

Then from the usual expansion%
\begin{align*}
\Gamma(\varphi)  &  =S_{cl}(\varphi)+\frac{1}{2}Tr\ln\left[
\delta
^{4}(x-y)+K(x,y)\right]  +O(\hslash^{2})=\\
&  =S_{cl}(\varphi)+\frac{1}{2}\sum_{i=1}^{\infty}\frac{\left(
-1\right)
^{n-1}}{n}Tr\left\{  \left[  K(x,y)\right]  ^{n}\right\}  +O(\hslash^{2})=\\
&
=S_{cl}(\varphi)+\underset{1-loop,2legs}{\underbrace{\frac{1}{2}\int
dx^{4}K(x,x)}}-\underset{1-loop,4legs\text{ ecc}}{\underbrace{...}}%
+O(\hslash^{2}),
\end{align*}
and from (\ref{K}) follows that%
\begin{equation}
\int dx^{4}K(x,x)=\dfrac{\lambda}{3!}\int
dp_{1}^{4}dp_{3}^{4}\left\{ 2+\exp\left[
-i\theta_{\mu\nu}p_{1}^{\mu}p_{3}^{\nu}\right]  \right\}
\widetilde{\varphi}(p_{1})\widetilde{\varphi}(-p_{1})\dfrac{1}{p_{3}^{2}%
+M^{2}},
\end{equation}
so that%
\[
\Gamma(\varphi)=S_{cl}(\varphi)+\dfrac{\lambda}{3!}\int dp_{1}^{4}dp_{3}%
^{4}\widetilde{\varphi}(p_{1})\widetilde{\varphi}(-p_{1})\left\{
2+\exp\left[  -i\theta_{\mu\nu}p_{1}^{\mu}p_{3}^{\nu}\right]
\right\} \dfrac{1}{p_{3}^{2}+M^{2}}+...
\]

That is the 1PI-effective action up to the one-loop corrected
quadratic terms.

\section{The effective potential}

\qquad While neglecting $\Gamma^{2}$ completely simply gives us
back the one-loop 1PI effective action, a more interesting results
is obtained by approximating $\Gamma^{2}(\varphi,G)$ including
only the ``eight'' diagram (A in Fig.\ref{fig:vvdiagrams}). In
this approximation one has that
\begin{equation}
\Gamma^{2}(\varphi,G)=\frac{1}{4!}\lambda\delta^{4}(0)\int d\alpha^{4}%
d\alpha^{\prime4}\frac{1}{\alpha^{2}+M^{2}(\alpha)}\frac{1}{\alpha^{\prime
2}+M^{2}(\alpha^{\prime})}\left\{  1+2\cos^{2}(\frac{\alpha\theta
\alpha^{\prime}}{2})\right\}  . \label{G2}%
\end{equation}

We observe that differently from the commutative case, where the
momenta circulating in each of the two loops do not mix, in the
noncommutative case this mixing occurs.

From the effective action(\ref{GammaF}) and (\ref{G2}), assuming
that, as consequence of the translational invariance of the
vacuum, $\varphi
(x)=\varphi$ one can extract the potential $V(\varphi,G)=\Gamma^{2PI}%
(\varphi,G)/\int dx^{4}$%
\begin{align}
V(\varphi)  &
=\frac{1}{2}m^{2}\varphi^{2}+\frac{\lambda}{4!}\varphi
^{4}+\label{pp}\\
&  +\frac{1}{2}\int dp^{4}\ln\left\{  \frac{p^{2}+M^{2}(p)}{p^{2}+m^{2}%
}\right\}  +\nonumber\\
&  +\frac{1}{2}\int dk^{4}\frac{m^{2}-M^{2}(k)+\dfrac{\lambda}{2}\varphi^{2}%
}{k^{2}+M^{2}(k)}+\nonumber\\
&  +\frac{1}{4!}\lambda\int
d\alpha^{4}d\alpha^{\prime4}\frac{1}{\alpha
^{2}+M^{2}(\alpha)}\frac{1}{\alpha^{\prime2}+M^{2}(\alpha^{\prime})}\left\{
2+\cos(\alpha\theta\alpha^{\prime})\right\}  .\nonumber
\end{align}

The stationarity conditions (\ref{peq}) and (\ref{gap}) in this case read%

\begin{equation}
0=\frac{\partial V(\varphi,G)}{\partial\varphi}=\varphi\left[  m^{2}%
+\frac{\lambda}{3!}\varphi^{2}+\frac{\lambda}{2}\int dk^{4}\frac{1}%
{k^{2}+M^{2}(k)}\right]  , \label{poteq}%
\end{equation}%

\begin{equation}
0=\frac{\partial V(\varphi,G)}{\partial M^{2}}=M^{2}(\alpha)-m^{2}%
-\dfrac{\lambda}{3!}\varphi^{2}-\frac{\lambda}{6}\int db^{4}\frac{1}%
{b^{2}+M^{2}(b)}\left\{  2+\cos\left(  b\theta\alpha\right)
\right\}  .
\label{gapquation}%
\end{equation}

The first equation has the solutions $\varphi=0$ and
$m^{2}=-\frac{\lambda }{3!}\varphi^{2}-\frac{\lambda}{2}\int
dk^{4}\frac{1}{k^{2}+M^{2}(k)}$ which correspond respectively to
the symmetric phase and to the broken-symmetric phase.
Substituting the gap equation (\ref{gapquation}) in the expression
of
the potential (\ref{pp}) we obtain%

\begin{align}
V(\varphi,G)  &
=\frac{1}{2}m^{2}\varphi^{2}+\frac{\lambda}{4!}\varphi
^{4}+\frac{1}{2}\int dp^{4}\ln\left\{  \frac{p^{2}+M^{2}(p)}{p^{2}+m^{2}%
}\right\}  +\label{potential}\\
&  -\frac{\lambda}{24}\int dk^{4}\frac{1}{k^{2}+M^{2}(k)}\int
db^{4}\frac {1}{b^{2}+M^{2}(b)}\left\{  2+\cos\left(  b\theta
k\right)  \right\} \nonumber
\end{align}

The term in the second row of the above expression is generated by
the ``bubble summation''. The terms appearing in the first row are
already present at the tree level and at one loop but what is
different here is that they now must be evaluated for $M^{2}(p)$
solution of the gap equation (\ref{gapquation}), which under the
hypothesis of translational invariance,
takes the form%
\begin{equation}
M^{2}(\alpha)=m^{2}+\dfrac{\lambda}{2}\varphi^{2}+\frac{\lambda}{6}\int
db^{4}\frac{1}{b^{2}+M^{2}(b)}\left\{  2+\cos\left(
b\theta\alpha\right)
\right\}  . \label{gapeq}%
\end{equation}

Both (\ref{potential}) and (\ref{gapeq}) are ultraviolet divergent
and they both are considered to be regularized with a cutoff
$\Lambda$ on the loop-momenta. In the next section we will deal
with the problem of their renormalization.

\subsection{Commutativity limit}

\qquad In this section we want discuss the commutativity limit
($\theta \rightarrow0$) and the strong noncommutativity limit
($\theta\rightarrow \infty$). In the commutativity limit
eqs.(\ref{gapquation}) and
(\ref{potential}) become respectively%
\begin{equation}
V(\varphi,G)=\frac{1}{2}m^{2}\varphi^{2}+\frac{\lambda}{4!}\varphi^{4}%
+\frac{1}{2}\int dp^{4}\ln\left\{
\frac{p^{2}+M^{2}(p)}{p^{2}+m^{2}}\right\}
-\frac{\lambda}{8}\int dk^{4}\frac{1}{k^{2}+M^{2}(k)}\int db^{4}\frac{1}%
{b^{2}+M^{2}(b)} \label{vcom}%
\end{equation}

and%
\begin{equation}
M^{2}(\alpha)=m^{2}+\dfrac{\lambda}{2}\varphi^{2}+\frac{\lambda}{2}\int
db^{4}\frac{1}{b^{2}+M^{2}(b)}, \label{gcom}%
\end{equation}
which is the well known result of \cite{hep-ph/9211211}. We recall
the procedure that one can use to renormalize (\ref{vcom}) and
(\ref{gcom}) since we will use it widely in the rest of the
chapter. The gap equation (\ref{gcom}) can be renormalized in the
following way \cite{hep-ph/9211211}
\[
\frac{M^{2}}{\lambda}=\frac{m^{2}}{\lambda}+\dfrac{1}{2}\varphi^{2}+\frac
{1}{2}\int db^{4}\frac{1}{b^{2}+M^{2}},
\]%
\[
\int db^{4}\frac{1}{b^{2}+M^{2}}=I_{1}-I_{2}M^{2}+G_{R}(M),
\]%
\[
M^{2}\left(  \frac{1}{\lambda}+\frac{1}{2}\int
db^{4}\frac{1}{b^{4}}\right) =\frac{m^{2}}{\lambda}+\int
db^{4}\frac{1}{b^{2}}+\dfrac{1}{2}\varphi
^{2}+\frac{1}{3}G_{R}(M),
\]
where $G_{R}(M)$ is the finite part of $G(M)$, $I_{1}=\int
db^{4}\frac {1}{b^{2}},$and $I_{2}=\int db^{4}\frac{1}{b^{4}}$
Introducing the
renormalized parameters%
\begin{align}
\dfrac{1}{\lambda_{R}}  &  =\lim_{\Lambda\rightarrow\infty}\left[
\frac
{1}{\lambda}+\frac{1}{2}I_{1}\right]  ,\label{lrc}\\
\dfrac{m_{R}^{2}}{\lambda_{R}}  &
=\lim_{\Lambda\rightarrow\infty}\left[
\frac{m^{2}}{\lambda}+\frac{1}{2}I_{2}\right]  , \label{mrc}%
\end{align}
one gets the renormalized gap equation%
\[
M^{2}=m_{R}^{2}+\dfrac{\lambda_{R}}{2}\varphi^{2}+\frac{\lambda_{R}}{3}%
G_{R}(M).
\]
One would like obtain a renormalized effective potential written
in terms of M$^{2}$. Using (\ref{gcom}) one can write (\ref{vcom})
as the sum of the three
contributions%
\begin{align}
V^{0}  &  =\frac{1}{2}m^{2}\varphi^{2}+\frac{\lambda}{4!}\varphi
^{4},\label{vc1}\\
V^{I}  &  =\frac{1}{2}\int dk^{4}\ln(k^{2})+\frac{1}{2}I_{1}M^{2}-\frac{1}%
{4}I_{2}M^{4}+T\label{vc2}\\
V^{II}  &  =-\frac{1}{2}GM^{2}+\frac{1}{2\lambda}M^{4}-\frac{1}{2\lambda}%
m^{4}-\frac{1}{2}m^{2}\varphi^{2}-\frac{\lambda}{8}\varphi^{4}, \label{vc3}%
\end{align}
and one finds that up to a $\varphi$-independent term which of
course can we
ignored in effective potential analyses%
\begin{equation}
V=V^{II}+V^{I}+V^{0}=-\frac{\lambda}{12}\varphi^{4}+\frac{1}{2}\frac{M^{4}%
}{\lambda_{R}}-\frac{1}{2}M^{2}G_{R}+\frac{1}{2}\int
dk^{4}\ln(k^{2})+T
\label{vcr}%
\end{equation}
where $T$ is the finite part of the expansion of $V^{I}$ (which of
course does not play an important role in renormalization).

We observe that (\ref{vcr}) is finite if written in terms of the
renormalized parameters defined by (\ref{lrc}) and (\ref{mrc}). In
particular it is worth noticing the exact cancellation of the
divergent terms in $m^{2}\varphi^{2}$ which appear with opposite
signs in the tree-level contribution $V^{0}$ and in the loop
correction $V^{II}.$

\subsection{Strong noncommutativity limit}

\qquad Now we analyze the limit of strong noncommutativity
($\theta \rightarrow\infty$). In this limit the strong
oscillations in the phases, which are present in the integrands of
(\ref{potential}) and (\ref{gapeq}), induce the vanishing of the
corresponding integrals. The effective potential
and the gap equation in this case become respectively%
\begin{equation}
V(\varphi,G)=\frac{1}{2}m^{2}\varphi^{2}+\frac{\lambda}{4!}\varphi^{4}%
+\frac{1}{2}\int dp^{4}\ln\left\{
\frac{p^{2}+M^{2}}{p^{2}+m^{2}}\right\}
-\frac{\lambda}{12}\int dk^{4}\frac{1}{k^{2}+M^{2}}\int db^{4}\frac{1}%
{b^{2}+M^{2}} \label{vsnc}%
\end{equation}

and%

\begin{equation}
M^{2}=m^{2}+\dfrac{\lambda}{2}\varphi^{2}+\frac{\lambda}{3}\int
db^{4}\frac
{1}{b^{2}+M^{2}}. \label{gsnc}%
\end{equation}

We observe that these expressions are formally similar to the ones
of the commutative case (\ref{vcom}), (\ref{gcom}). The only
differences are in the terms in front of the eight-diagram
contribution: the nonplanar the eight-diagram contribution in fact
becomes negligible in the $\theta \rightarrow\infty$ limit and
only contributions of the planar diagrams survive. It is also
important to notice that the dressed mass $M^{2}(k)$ become in
this limit momentum independent and one can follow exactly the
same
procedure as in the commutative case. One obtains for the gap equation%
\[
M^{2}=m_{R}^{2}+\dfrac{\lambda_{R}}{2}\varphi^{2}+\frac{\lambda_{R}}{3}%
G_{R}(M).
\]

where now%
\begin{align}
\dfrac{1}{\lambda_{R}}  &  =\lim_{\Lambda\rightarrow\infty}\left[
\frac
{1}{\lambda}+\frac{1}{3}\int db^{4}\frac{1}{b^{4}}\right]  ,\\
\dfrac{m_{R}^{2}}{\lambda_{R}}  &
=\lim_{\Lambda\rightarrow\infty}\left[
\frac{m^{2}}{\lambda}+\frac{1}{3}\int db^{4}\frac{1}{b^{2}}\right]
.
\end{align}

The effective potential is obtained again as the sum of the terms%
\begin{align}
V^{0}  &  =\frac{1}{2}m^{2}\varphi^{2}+\frac{\lambda}{4!}\varphi
^{4}\label{v1snc}\\
V^{I}  &  =\frac{1}{2}\int dk^{4}\ln(k^{2})+\frac{1}{2}I_{1}M^{2}-\frac{1}%
{4}I_{2}M^{4}+T\label{v2snc}\\
V^{II}  &
=-\frac{3}{4\lambda}m^{4}-\dfrac{3\lambda}{16}\varphi^{4}+\frac
{3}{4\lambda}M^{4}-\dfrac{1}{2}M^{2}G-\frac{3}{4}m^{2}\varphi^{2}.
\label{v3snc}%
\end{align}

which gives%

\[
V=V^{II}+V^{I}+V^{0}=-\frac{1}{4}m^{2}\varphi^{2}-\dfrac{7\lambda}{48}%
\varphi^{4}+\frac{3}{4}\dfrac{M^{4}}{\lambda_{R}}+\frac{1}{3}\int dk^{4}%
\frac{1}{k^{4}}+\dfrac{1}{2}\int dk^{4}\ln[k^{2}]-\frac{3}{4\lambda}m^{4}%
\]

We observe that in the case of strong noncommutativity the
cancellation between the $m^{2}\varphi^{2}$ terms in (\ref{v1snc})
and (\ref{v3snc}) does not occur and the resulting potential does
not renormalize.

\subsection{Effective potential in the general case}

\qquad We have so far discussed the limits of commutativity
($\theta \rightarrow0$) and strong noncommutativity
($\theta\rightarrow\infty$). We have seen that in both of these
cases the unknown function $M^{2}(k)$, that appears in the
denominator of $G(M),$ is momentum independent, although it
satisfies different gap equations in the two different cases. Now
we want to address the problem of calculating the effective
potential for general values
of the noncommutativity parameter $\theta.$ We start by defining $M^{2}%
(\alpha)=M^{2}+\Pi(\alpha)$ so that the gap equation (\ref{gapeq})
can be
rewritten as%
\[
\Pi(\alpha)=-M^{2}+m^{2}+\dfrac{\lambda}{2}\varphi^{2}+\frac{\lambda}{3}\int
db^{4}\frac{1}{b^{2}+M^{2}+\Pi(b)}+\frac{\lambda}{6}\int
db^{4}\frac {\cos\left(  b\theta\alpha\right)
}{b^{2}+M^{2}+\Pi(b)}.
\]

The last equation must hold for every value of $\alpha$ and
$\theta.$ Thus we
must have that separately%
\begin{equation}
-M^{2}+m^{2}+\dfrac{\lambda}{2}\varphi^{2}+\frac{\lambda}{3}\int db^{4}%
\frac{1}{b^{2}+M^{2}+\Pi(b)}=\text{C,}%
\end{equation}
and%
\begin{equation}
\Pi(\alpha)-\frac{\lambda}{6}\int db^{4}\frac{\cos\left(  b\theta
\alpha\right)  }{b^{2}+M^{2}+\Pi(b)}=\text{C,}%
\end{equation}
where C is $\alpha$ and $\theta$ independent.

We observe that we can always choose C=0, modulo the redefinitions
$\Pi (\alpha)\rightarrow\Pi(\alpha)+$C, $M^{2}\rightarrow$
$M^{2}-$C. In this way
we obtain%
\begin{equation}
M^{2}=m^{2}+\dfrac{\lambda}{2}\varphi^{2}+\frac{\lambda}{3}\int
db^{4}\frac
{1}{b^{2}+M^{2}+\Pi(b)}, \label{eqM}%
\end{equation}
which determines $M^{2}$, and%
\begin{equation}
\Pi(\alpha)=\frac{\lambda}{6}\int db^{4}\frac{\cos\left(  b\theta
\alpha\right)  }{b^{2}+M^{2}+\Pi(b)} \label{pie}%
\end{equation}
which determines $\Pi(\alpha).$

We see from (\ref{pie}) that $\Pi(\alpha)\rightarrow\infty$ for
$\alpha \rightarrow0$ and for $\theta\rightarrow0,$ and that
$\Pi(\alpha)\rightarrow0$ for $\alpha\rightarrow\infty$ and for
$\theta\rightarrow\infty.$ Equation (\ref{eqM}) can be
renormalized by the following procedure, similar to the one
we have discussed previously%
\[
\frac{M^{2}}{\lambda}=\frac{m^{2}}{\lambda}+\dfrac{1}{2}\varphi^{2}+\frac
{1}{3}\int db^{4}\frac{1}{b^{2}+M^{2}+\Pi(b)},
\]%

\[
\int db^{4}\frac{1}{b^{2}+M^{2}+\Pi(b)}=\int
db^{4}\frac{1}{b^{2}+\Pi (b)}-M^{2}\int
db^{4}\frac{1}{[b^{2}+\Pi(b)]^{2}}+G_{R}(M),
\]%

\[
M^{2}\left(  \frac{1}{\lambda}+\frac{1}{3}\int db^{4}\frac{1}{[b^{2}%
+\Pi(b)]^{2}}\right)  =\frac{m^{2}}{\lambda}+\int
db^{4}\frac{1}{b^{2}+\Pi
(b)}+\dfrac{1}{2}\varphi^{2}+\frac{1}{3}G_{R}(M),
\]
where $G_{R}(M)$ is the finite part of the divergent expression in
(\ref{eqM}). We can now introduce the renormalized parameters in the form%
\begin{align}
\dfrac{1}{\lambda_{R}}  &
=\lim_{\Lambda\rightarrow\infty}\frac{1}{\lambda
}+\frac{1}{3}\int db^{4}\frac{1}{[b^{2}+\Pi(b)]^{2}},\label{rc}\\
\dfrac{m_{R}^{2}}{\lambda_{R}}  &
=\lim_{\Lambda\rightarrow\infty}\frac
{m^{2}}{\lambda}+\int db^{4}\frac{1}{b^{2}+\Pi(b)}, \label{rm}%
\end{align}
and we get the renormalized gap equation%
\[
M^{2}=m_{R}^{2}+\dfrac{\lambda_{R}}{2}\varphi^{2}+\frac{\lambda_{R}}{3}%
G_{R}(M)
\]

We come now to the important issue of the renormalization of the
effective potential. We have seen that the way in which the gap
equation renormalizes fixes uniquely the renormalization of the
bare mass and the renormalization of the coupling. We must check
if the same renormalization conditions provide us with a finite
effective potential. We can use (\ref{eqM}) and (\ref{pie}) in
the expression (\ref{potential}) to obtain for the effective potential%
\begin{align}
V  &  =V^{0}+V^{I}+V^{II}=\nonumber\\
&  =\dfrac{M^{4}}{4\lambda_{R}}+M^{2}\left\{  -\dfrac{1}{2}\int dk^{4}%
R(k)+\dfrac{1}{4}\int dk^{4}\frac{\Pi(k)}{\left[  k^{2}+\Pi(k)\right]  ^{2}%
}\right\}  +\nonumber\\
&  -\dfrac{1}{4}\int
dk^{4}\frac{\Pi(k)}{k^{2}+\Pi(k)}-\dfrac{1}{4}\int
dk^{4}\Pi(k)R(k)+\dfrac{1}{2}\int dk^{4}\ln[1+\frac{\Pi(k)}{k^{2}%
}]+T+\nonumber\\
&  \text{\
}-\dfrac{3m^{4}}{4\lambda}-\dfrac{1}{4}m^{2}\varphi^{2}-\dfrac
{7}{48}\lambda\varphi^{4}+\dfrac{1}{2}\int dk^{4}\ln[k^{2}], \label{vrg}%
\end{align}
where we used%
\begin{align}
V^{0}  &  =\frac{1}{2}m^{2}\varphi^{2}+\frac{\lambda}{4!}\varphi
^{4},\label{v1nc}\\
V^{I}  &  =\frac{1}{2}\int dk^{4}\ln(k^{2})+\frac{1}{2}I_{1}M^{2}-\frac{1}%
{4}I_{2}M^{4}+T,\label{v2nc}\\
V^{II}  &  =M^{4}\left\{  \dfrac{1}{2}\int db^{4}\frac{1}{\left[  k^{2}%
+\Pi(k)\right]  ^{2}}+\dfrac{3}{4}\dfrac{1}{\lambda}\right\}  +\nonumber\\
&  +M^{2}\left\{  -\dfrac{1}{2}\int dk^{4}R(k)-\dfrac{1}{2}\int
dk^{4}\frac
{1}{k^{2}+\Pi(k)}+\dfrac{1}{4}\int dk^{4}\frac{\Pi(k)}{\left[  k^{2}%
+\Pi(k)\right]  ^{2}}\right\}  +\nonumber\\
&  -\dfrac{1}{4}\int
dk^{4}\frac{\Pi(k)}{k^{2}+\Pi(k)}-\dfrac{1}{4}\int
dk^{4}\Pi(k)R(k)-\dfrac{3m^{4}}{4\lambda}-\dfrac{3}{4}m^{2}\varphi^{2}%
-\dfrac{3}{16}\lambda\varphi^{4}, \label{v3nc}%
\end{align}
and we defined%
\begin{align*}
T  &  =\dfrac{1}{2}\int dk^{4}\ln[k^{2}+M^{2}+\Pi(k)]+\\
&  -\left\{  \dfrac{1}{2}\int
dk^{4}\ln[k^{2}+\Pi(k)]+\tfrac{M^{2}}{2}\int
dk^{4}\tfrac{1}{k^{2}+\Pi(k)}-\tfrac{M^{4}}{4}\int
dk^{4}\tfrac{1}{\left[ k^{2}+\Pi(k)\right]  ^{2}}\right\}
\end{align*}
and%
\begin{equation}
R(k)=\tfrac{1}{k^{2}+M^{2}+\Pi(k)}-\left\{  \tfrac{1}{k^{2}+\Pi(k)}%
-\tfrac{M^{2}}{\left[  k^{2}+\Pi(k)\right]  ^{2}}\right\}  .
\end{equation}

We observe that in (\ref{vrg}) all the field-dependent terms, with
the exception of $-\tfrac{1}{4}m^{2}\varphi^{2}$, are finite
thanks to the fact that $\Pi(k)$ vanishes exponentially in the
limit $k\rightarrow\infty$. As in the case of strong
noncommutativity limit, the presence of the divergent term
$-\tfrac{1}{4}m^{2}\varphi^{2}$ is due to the fact that the
corresponding contributions from $V^{0}$ and $V^{II}$ do not
cancel each other. The cancellation occurs only in the commutative
limit.

\section{Remarks on the structure of the CJT effective potential in canonical
noncommutative spacetime}

\qquad Whereas in commutative spacetime the CJT effective
potential can be renormalized and gives a satisfactory description
of the vacua of a given field theory, in our
canonical-noncommutativity analysis the CJT effective potential
(in the bubble-resummation approximation) was found not to be
renormalizable. From a conservative standpoint we should then
assume that in this type of theories the CJT effective potential
cannot provide reliable nonperturbative insight on the phase
structure. This negative conclusion is also supported by the
realization that canonical noncommutativity affects strongly the
structure of the UV divergences of a field theory, and this might
be particularly significant for those techniques that effectively
rely on resummations of contributions from all orders in the
coupling constant. When we establish that a field theory is
renormalizable, we actually verify that it is ``perturbatively
renormalizable'': the divergences at any given order in
coupling-expansion perturbation theory can be reabsorbed in
redefinitions of the parameters of the Lagrangian density. The
fact that the CJT technique gives rise to a renormalizable
effective potential in the commutative-spacetime case is highly
nontrivial, since we are not consistently summing all
contributions up to a given order in the coupling constant (a
calculation which would be ``protected'' by peturbative
renormalizability), we are instead selectively summing a certain
subset of the contributions at each order in the coupling
constant. It is therefore plausible that the fact that our CJT
effective potential cannot be renormalized is simply a sign of an
inadequacy of this technique to the canonical-noncommutativity
context. On the other hand it appears reasonable to explore an
alternative, more optimistic, perspective, which is based on the
observation that the only contribution to the CJT potential that
ends up not being expressed in terms of renormalized quantities
does not is the term $\frac{1}{4}m^{2}\varphi^{2}$. This term
however vanishes in the disordered phase $\varphi=0$. In a certain
sense we have a renormalizable effective potential in the
disordered phase, and our results of nonrenormalizability in the
translationally-invariant ordered phase $\varphi=C$ could be
interpreted as a manifestation of the fact that this phase is not
admissable for these theories in canonical noncommutative
spacetime. This hypothesis finds some support in the arguments
presented in Ref. \cite{hep-th/0006119}, which also concluded that
the only admissable phases for these theories are the disordered
phase and a
(non-translationally-invariant) stripe phase with $\tilde{\varphi}%
(p)=C\delta(p-p_{c})$ (where $\tilde{\varphi}(p)$\ is the Fourier
transform of $\varphi(x)$) and $p_{c}$ is a characteristic
momentum scale of the stipe phase). This argument of
inadmissability of the translationally-invariant ordered phase
might be related with the delicate IR structure of these theories:
$\varphi(x)=C$ means $\tilde{\varphi}(p)=C\delta(p)$, so the
concept of a translationally-invariant ordered phase is closely
connected with the zero-momentum structure of the theory of
interest. To explore these issues it would be necessary to
consider the CJT effective action, which explores the more general
class of candidate vacua $\varphi(x)$, rather than stopping, as we
did here, at the level of the CJT effective potential (which
assumes from the beginning a translationally-invariant vacuum).
With the CJT effective action one could investigate the
renormalizability of the stripe phase (which is not
translationally invariant, and therefore cannot be studied with
the effective potential). Moreover, while the effective potential
is the generating functional of Green functions at zero external
momentum, could be particularly sensitive the effective action is
the generating functional of generic Green functions and might be
less sensitive to the troublesome IR sector of these theories. The
analysis of the CJT effective action is postponed to a future
study. Even in commutative-spacetime theories the evaluation of
the CJT effective action turns out to be very complex, basically
intractable analytically, and a troublesome calculation even
numerically. It is likely that in the canonical-noncommutativity
context the evaluation of the CJT effective action may prove even
more troublesome, but from the indications that emerged from our
analysis of the CJT effective potential it appears that such an
analysis is well motivated, as it could provide insight for the
understanding of some key physical predictions of these theories.



\chapter{Conclusions}

\qquad It this thesis, we have explored the hypothesis that
nonclassical effects of spacetime may manifest through the
noncommutativity of spacetime at short distances. We focused on
the two most popular examples of noncommutative spacetimes:
canonical spacetimes, which have been at the center of an intense
scientific debate over the last few years (mostly because of their
relevance for the description of string theory in certain
backgrounds) and $\kappa $-Minkowski spacetime, which, being the
only fully-worked-out example of spacetime requiring a
Planck-scale ``deformation'' of Poincar\'{e} symmetries, is also
being investigated by a large number of research groups.

We focused on some issues that provide key physical
characterizations of these spacetimes. In the light of the fact
that plans for experimental searches of a possible dependence of
the group velocity on the Planck scale are already at an advanced
stage~\cite{astro-ph/0009271,GLAST,AMS}, we analyzed wave
propagation both in canonical and in $\kappa$-Minkowski spacetime.
The idea that the Planck-scale (quantum) structure of spacetime
might affect the group-velocity/wavelength relation is plausible
(and in some cases inevitable) in most quantum-gravity approaches,
including phenomenological models of spacetime
foam~\cite{astro-ph/9712103}, loop quantum gravity (see,
\textit{e.g.}, Ref.~\cite{gr-qc/9809038}), superstring theory
(see, \textit{e.g.}, Ref.~\cite{hep-th/0002075}), and
noncommutative geometry. While a detailed careful description of
wave propagation is beyond the reach of the present technical
understanding of most quantum-gravity scenarios, we showed here
that wave propagation in certain noncommutative spacetimes can be
rigorously analyzed. We have shown that the features of the
propagating waves strongly depend on the type of noncommutative
spacetime one is considering. In the case of waves in canonical
spacetime we found no observable departure from the classical
picture of propagating wave. Instead, in the case of waves in
$\kappa$-Minkowski spacetime, our analysis showed that the group
velocity is affected by noncommutativity. We found that the
formula $v=dE(p)/dp$, where $E(p)$ is fixed by the
$\kappa$-Poincar\'{e} dispersion relation, still holds in
$\kappa$-Minkowski spacetime (just like $v=dE(p)/dp$ holds in the
Galilei/Minkowski classical spacetimes and in the canonical
noncommutative spacetime) but it actually sets a new type of
relation between group velocity and momentum as a result of the
fact that in $\kappa$-Minkowski the dispersion relation $E(p)$ is
fixed by the $\kappa$-Poincar\'{e} mass Casimir, which differs
from the familiar Poincar\'{e} mass Casimir at the level of
Planck-scale-suppressed effects.

The validity of $v=dE(p)/dp$ in $\kappa$-Minkowski had been
largely expected in the literature, even before our direct
analysis, but such a direct analysis had become more urgent after
the appearance of some recent
articles~\cite{hep-th/0107054,gr-qc/0111056} which had argued in
favor of alternatives to $v=dE(p)/dp$ for $\kappa$-Minkowski. We
have shown that these recent claims were incorrect: the analysis
reported in Ref.~\cite{gr-qc/0111056} was based on erroneous
implementation of the $\kappa$-Minkowski differential calculus,
while the analysis in Ref.~\cite{hep-th/0107054} interpreted as
momenta some quantities which cannot be properly described in
terms of translation generators.

Of course, one is interested in going much beyond the description
of wave propagation: a key objective for this research field is
the construction and analysis of quantum field theories in these
noncommutative spacetimes. As discussed in Chapter 3, the path
toward the construction of a sensible quantum field theory in
$\kappa$-Minkowski spacetime appears to be still confronted with a
large number of delicate obstacles. Our analysis of wave
propagation in $\kappa$-Minkowski is therefore the (very limited)
``state of the art'' in the analysis of the physical predictions
of this noncommutative spacetime. Since a key ingredient of a
quantum field theory is essentially an expansion of fields in
plane waves, perhaps the results that emerged from our rigorous
analysis of wave propagation in $\kappa$-Minkowski spacetime could
prove useful for an improved formulation of QFT in this spacetime
but at present we do not see an obvious way to approach this
project.

For canonical noncommutative spacetimes there is instead a much
studied approach to the construction of quantum field theories,
based on the $\star $-product technique. In this framework it
appears that several important issues are entangled with the
peculiar failure of Wilson decoupling between infrared and
ultraviolet degrees of freedom. In particular, as we showed in
Chapter 4, the IR/UV mixing has wide implications for the
strategies that should be adopted in order to falsify/verify these
theories. Theories that (according to our conventional
perspective) differ only in an experimentally unaccessible range
of momenta may give rise to different predictions in the
low-energy regime. In fact we found that predictions for the
low-energy (i.e. experimentally accessible) physics depend
strongly not only on the low-energy structure of the theory but
also on its high-energy structure. Therefore the bounds on
parameter space that one usually is able to set using low-energy
data are here only ``conditional''. A comparison between
low-energy data and the low-energy sector of theory can be
reliably done only after having fully specified the high-energy
sector.

While in Chapter 4 we explored the consequences of the IR/UV
mixing using a standard perturbative Feynman-diagrammatic
approach, in Chapter 5 we used the Cornwall-Jackiw-Tomboulis
formalism, a nonperturbative technique (effectively resumming
infinite series of 1PI Feynman diagrams) which is usually very
fruitful in contexts in which the infrared sector is problematic
(as in Thermal Quantum Field Theories
\cite{CJT,hep-ph/9311324,hep-ph/9211211}). We studied the CJT
effective potential in the canonical-noncommutative
``$\lambda\varphi^{4}$'' theory, adopting the so-called
``bubble-resummation approximation'', which includes all
contributions from daisy and super-daisy diagrams. We found that
the effective potential is in general nonrenormalizable, but the
left-over UV divergences disappear in the disordered phase
$\varphi=0$. We argued that the nonrenormalizability of the
effective potential might be another manifestation of the IR/UV
mixing, and that the problems with the translationally-invariant
ordered phase $\varphi=$C might also be due to the delicate IR
structure. Probably the infrared structure of these theories
prohibits the condensation of the zero-momentum modes which is
necessary requirement for a transition to a
transitionally-invariant vacuum. A characteristic prediction of
these theories might therefore be the presence of
energetically-favored transitionally-invariant phases, such as the
stripe phases mentioned in Sec.~5.4.

Among the investigations that could find motivation in the results
here reported we emphasize the study of the CJT effective action
in theories in canonical noncommutative spacetime, which could
bring key insight on the phase structure of these theories.

It would also be interesting to explore new strategies for setting
limits on the canonical-noncommutativity parameters, since, as
shown in Chapter 4, the standard strategy is unreliable. Perhaps
the only possibility is the one of ``building a case'' in favor or
against consistency with observations (whereas in commutative
spacetime a single experiment can give conclusive unconditional
indications). The case would be built by considering a variety of
data, and observing that they are all consistent with the
characteristic structure of theories in canonical noncommutative
spacetime. Because of the nature of these characteristic features
of canonical noncommutativity, it might be useful to rely on data
that concern a wide range of energy scales. The astrophysical
studies analyzed, for what concerns canonical noncommutativity, in
Ref.~\cite{hep-th/0109191} could play an important role in this
programme.






\appendix

\chapter{Hopf algebras}

\label{ap:ap1}

\qquad In this appendix we report the basic definitions and the
basic properties of the Hopf algebras that have been used in this
thesis.

\begin{definition}
An algebra $\mathcal{A}$ is a vector space (over the field
$\mathcal{C}$) in which two linear applications m:\ $\mathcal{A}$
$\otimes\mathcal{A}$ $\rightarrow\mathcal{A}$ \ and
$\eta:\mathcal{C\rightarrow A}$ are defined
such that%
\begin{align}
m\circ(id\otimes m) &  =m\circ(m\otimes id)\label{al1}\\
m\circ(\eta\otimes id) &  =m\circ(id\otimes\eta)=id\label{al2}%
\end{align}
where $\circ$ indicates the composition of applications and $id$
is the identity application.
\end{definition}

The property (\ref{al1}) is the associativity of the product $m$.
The property (\ref{al2}) is the neutrality of the identity with
respect to the product $m$.

\begin{definition}
A coalgebra $\mathcal{B}$ is a vector space (over the field
$\mathcal{C}$) in which two linear maps
$\Delta:\mathcal{B\rightarrow B\otimes B}$ (the `coproduct') and
$\varepsilon:\mathcal{B\rightarrow C}$ (the `counit') are defined
such that
\begin{align}
(\Delta\otimes id)\circ\Delta &  =(id\otimes\Delta)\circ\Delta,\label{c1}\\
(\varepsilon\otimes id)\circ\Delta &  =(id\otimes\varepsilon)\circ
\Delta=id.\label{c2}%
\end{align}
\end{definition}

The property (\ref{c1}) is the (co-)associativity of the coproduct
$\Delta$. The property (\ref{c2}) is the (co-)neutrality of the
identity with respect to the coproduct $\Delta$.

It is worth noticing that a coalgebra ($\mathcal{B}$) can always
be constructed from an algebra ($\mathcal{A}$) by duality (i.e.
using a nondegenerate map $\left\langle ,\right\rangle
:\mathcal{A}$ $\otimes \mathcal{B}\rightarrow\mathcal{C}$) in the
following way.

\begin{enumerate}
\item  First, one renders $\mathcal{B}$ a vector space by the definitions%
\begin{align}
\left\langle a,b+c\right\rangle  &  :=\left\langle
a,b\right\rangle
+\left\langle a,c\right\rangle \text{ \ \ \ \ \ }\forall a\in\mathcal{A}%
,\forall b,c\in\mathcal{B}\text{,}\\
\left\langle a,\lambda b\right\rangle  &  :=\lambda\left\langle
a,b\right\rangle \ \ \ \ \ \ \ \forall a\in\mathcal{A},\forall
b\in
\mathcal{B},\ \forall\lambda\in\mathcal{C}\text{.}%
\end{align}

\item  Secondly, one defines the coproduct
$\Delta:\mathcal{B\rightarrow B\otimes B}$ and the counit
$\varepsilon:\mathcal{B\rightarrow C}$ by the
relations%
\begin{align}
\left\langle a\otimes b,\Delta c\right\rangle  &  :=\left\langle
ab,c\right\rangle ,\label{d1}\\
\varepsilon(c) &  :=\left\langle
\mathbb{I}_{\mathcal{A}},c\right\rangle
,\label{d2}%
\end{align}
where $\left\langle ,\right\rangle $ extends to tensor products
pairwise (i.e.
$\left\langle a\otimes b,\Delta c\right\rangle =\left\langle a,c_{(1)}%
\right\rangle \left\langle b,c_{(2)}\right\rangle $ ).
\end{enumerate}

It is easy to verify that $\Delta$ and $\varepsilon$ defined by (\ref{d1}%
,\ref{d2}) satisfy (\ref{c1},\ref{c2}), so that $\mathcal{B}$ is a
coalgebra.

\begin{definition}
($\mathcal{H},m,\eta,\Delta,\varepsilon)$ is a bialgebra if

\begin{enumerate}
\item $\mathcal{H}$ is an algebra with respect to $m,\eta.$

\item $\mathcal{H}$ is a coalgebra with respect to
$\Delta,\varepsilon.$

\item $\Delta$ and $\varepsilon$ obey the following relations:%
\begin{align}
\Delta(ab) &  =\Delta(a)\Delta(b)\text{ \ \ \ \ \ \ \ \ }\forall
a,b\in\mathcal{A}\text{,}\label{b1}\\
\varepsilon(ab) &  =\varepsilon(a)\varepsilon(b)\text{ \ \ \ \ \ \ \ \ \ \ }%
\forall a,b\in\mathcal{A}\text{,}\label{b2}%
\end{align}

where the tensor product $(h\otimes g)(h^{\prime}\otimes g^{\prime
}):=m(h,h^{\prime})\otimes m(g,g^{\prime})$ is defined $\forall
h,h^{\prime },g,g^{\prime}\in\mathcal{H}$.
\end{enumerate}
\end{definition}

The property (\ref{b1}) means that the coproduct ($\Delta$)
furnishes a representation of the product ($m$) over
$\mathcal{H}\otimes\mathcal{H}$. The property (\ref{b2}) means
that the counit ($\varepsilon$) furnishes a representation of the
product ($m$) over $\mathcal{C}$.

\begin{definition}
An Hopf algebra is a bialgebra
$(\mathcal{H},m,\eta,\Delta,\varepsilon)$ with a map
$S:\mathcal{H\rightarrow H}$ (the `antipode') such that
\begin{equation}
m\circ(S\otimes
id)\circ\Delta=\eta\circ\varepsilon=m\circ(id\otimes
S)\circ\Delta.\label{h1}%
\end{equation}
\end{definition}

From (\ref{h1}) it follows that
\begin{equation}
S(ab)=S(a)S(b)\text{ \ \ \ \ \ \ \ \ }\forall a,b\in\mathcal{H}, \label{s1}%
\end{equation}
and that%
\begin{equation}
\Delta(S(a))=S(a_{(2)})\otimes S(a_{(1)})\text{ \ \ \ \ \ \ \ \
}\forall
a\in\mathcal{H}. \label{s2}%
\end{equation}

The property (\ref{s1}) says that the antipode ($S$) furnishes a
representation of the product ($m$) over $\mathcal{H}$. The
property (\ref{s2}) says that the antipode ($S$) furnishes an
(anti-)representation of the coproduct ($\Delta$) over
$\mathcal{H}\otimes\mathcal{H}$.

\begin{definition}
An Hopf algebra is commutative if it is commutative as an algebra.
It is `cocommutative' if it is cocommutative as a coalgebra (i.e.
if $\tau \circ\Delta=\Delta,$ where $\tau$ is the `flip operator'
defined by $\tau(a_{(1)}\otimes a_{(2)}):=a_{(2)}\otimes
a_{(1)}).$
\end{definition}

It is also easy to verify that

\begin{corollary}
If $\mathcal{H}$ is a commutative or cocommutative Hopf algebra,
then $S^{2}=id.$
\end{corollary}

We notice that the key difference between a ``classical'' group
and a truly ``quantum'' group regards the coalgebra sector: while
the former is cocommutative, the latter is noncocommutative.

\begin{definition}
Two Hopf algebras $\mathcal{H}$ and $\mathcal{H}$' are `dually
paired' by a
map $\left\langle ,\right\rangle :\mathcal{H}$ $\otimes\mathcal{H}%
$'$\rightarrow\mathcal{C}$ if%
\begin{align}
\left\langle \phi\psi,h\right\rangle  &  =\left\langle
\phi\otimes\psi,\Delta h\right\rangle ,\text{ \ \ \ \ \ \ \
}\left\langle 1,h\right\rangle
=\epsilon(h),\\
\left\langle \Delta\phi,h\otimes g\right\rangle  &  =\left\langle
\phi,hg\right\rangle ,\text{ \ \ \ \ \ \ \ }\left\langle
\phi,1\right\rangle
=\epsilon(\phi),\\
\left\langle S\phi,h\right\rangle  &  =\left\langle
\phi,Sh\right\rangle ,
\end{align}

for all $\phi,\psi\in\mathcal{H}$' and $h,g\in\mathcal{H}$. Here
$\left\langle ,\right\rangle $ extends pairwise to tensor
products.
\end{definition}

This definition implies that the product of $\mathcal{H}$ and the
coproduct of $\mathcal{H}^{\prime}$ are adjoint to each other
under $\left\langle ,\right\rangle $, and vice-versa. Likewise,
the units and counits are mutually adjoint, and the antipodes are
adjoint.

\begin{definition}
A bialgebra or Hopf algebra $\mathcal{H}$ acts\footnote{Here we
only consider left actions.} on an algebra $\mathcal{A}$ if a
linear map (the action ) is
defined $\triangleright:\mathcal{H\otimes A\rightarrow A}$ such that%
\begin{equation}
(ab)\triangleright c=a\triangleright(b\triangleright c),\text{ \ \ \ \ \ }%
\forall a,b\in\mathcal{H},\text{ \ \ }\forall c\in\mathcal{A}\text{.}%
\end{equation}
\end{definition}

The notion of action is the generalization of the notion of linear
transformation to the Hopf algebras.

\begin{definition}
Given two Hopf algebras $\mathcal{H}$ and $\mathcal{A},$ an action
$\triangleright$ of $\mathcal{H}$ on $\mathcal{A}$ is called
covariant if the conditions hold

\begin{enumerate}
\item $\triangleright$ commutes with the product map $m$
(i.e.$h\triangleright (ab)=(h_{(1)}\triangleright
a)(h_{(2)}\triangleright b),$ \ \ \ $\forall h\in\mathcal{H},$ \ \
\ $\forall a,b\in\mathcal{A}$),

\item $\triangleright$ commutes with the unit map $\eta$ (i.e.
$h\triangleright\mathbb{I}_{\mathcal{A}}=\epsilon(h)\mathbb{I}_{\mathcal{A}},$
$\forall h\in\mathcal{H}$).
\end{enumerate}
\end{definition}

We observe that a covariant action involves the coalgebric sector
of the Hopf algebra$.$ Moreover a covariant action preserves both
the algebraic and the coalgebraic structure of the Hopf algebra.
Two examples of covariant actions are:

the canonical action%
\begin{equation}
a\overset{can}{\triangleright}b:=b_{(1)}\left\langle
a,b_{(2)}\right\rangle ,
\end{equation}

the adjoint action\footnote{The adjoint action is defined in the
case
$\mathcal{H}$=$\mathcal{A}$.}%
\begin{equation}
a\overset{ad}{\triangleright}b:=a_{(1)}bS(a_{(2)}).
\end{equation}

One example of a noncovariant action is%
\begin{equation}
a\triangleright b:=ab,
\end{equation}
that does not preserve the coalgebraic structure of the Hopf
algebra.

This completes the review of the definitions and of the properties
regarding Hopf algebras which have been used in this thesis.

%
%
%
%
\addcontentsline{toc}{chapter}{Bibliography}


\begin{thebibliography}{100}
\expandafter\ifx\csname url\endcsname\relax
  \def\url#1{\texttt{#1}}\fi
\expandafter\ifx\csname
urlprefix\endcsname\relax\def\urlprefix{URL }\fi
\providecommand{\selectlanguage}[1]{\relax}
\providecommand{\eprint}[2][]{\url{#2}}

\bibitem{hep-th/9711162}
\textsc{A.~Connes}, \textsc{M.~R. Douglas}, \textsc{A.~Schwarz},
  \emph{Noncommutative geometry and matrix theory: Compactification on tori},
  JHEP, \textbf{02}: (1998), 003,
  \eprint[http://arXiv.org/abs]{hep-th/9711162.}

\bibitem{hep-th/9908142}
\textsc{N.~Seiberg}, \textsc{E.~Witten}, \emph{String theory and
noncommutative
  geometry}, JHEP, \textbf{09}: (1999), 032,
  \eprint[http://arXiv.org/abs]{hep-th/9908142.}

\bibitem{peierls}
\textsc{R.~Peierls}, Z. Phys., \textbf{80}: (1933), 763.

\bibitem{physics/0209108}
\textsc{R.~Jackiw}, \emph{Lochlainn O'Raifeartaigh, fluids, and
noncommuting fields},
\eprint[http://arXiv.org/abs]{physics/0209108.}

\bibitem{dunnejack}
\textsc{R.~J. G.~Dunne}, \emph{'Peierls Substitution' and
Chern-Simons Quantum
  Mechanics}, Nucl. Phys. B. (Proc. Suppl.), \textbf{33C}: (1933), 114.

\bibitem{lnrt}
\textsc{J.~Lukierski}, \textsc{A.~Novicki}, \textsc{H.~Ruegg},
  \textsc{V.~Tolstoy}, \emph{q-Deformation of Poincare algebra}, Phys. Lett.,
  \textbf{B268}: (1991), 331.

\bibitem{hep-th/9312153}
\textsc{J.~Lukierski}, \textsc{H.~Ruegg}, \textsc{W.~J.
Zakrzewski},
  \emph{Classical quantum mechanics of free kappa relativistic systems}, Ann.
  Phys., \textbf{243}: (1995), 90--116,
  \eprint[http://arXiv.org/abs]{hep-th/9312153.}

\bibitem{hep-th/9405107}
\textsc{S.~Majid}, \textsc{H.~Ruegg}, \emph{Bicrossproduct
structure of kappa
  Poincare group and noncommutative geometry}, Phys. Lett., \textbf{B334}:
  (1994), 348--354, \eprint[http://arXiv.org/abs]{hep-th/9405107.}

\bibitem{lukie}
\textsc{J.~Lukierski}, \textsc{A.Nowicki}, \emph{Heisenberg double
description
  of kappa Poincare algebra and kappa deformed phase space},
  \eprint{q-alg/9702003.}

\bibitem{hep-th/9907110}
\textsc{G.~Amelino-Camelia}, \textsc{S.~Majid}, \emph{Waves on
noncommutative
  spacetime and gamma-ray bursts}, Int. J. Mod. Phys., \textbf{A15}: (2000),
  4301--4324, \eprint[http://arXiv.org/abs]{hep-th/9907110.}

\bibitem{hep-th/0207022}
\textsc{J.~Lukierski}, \textsc{A.~Nowicki}, \emph{kappa-deformed
kinematics and
  addition law for deformed velocities},
  \eprint[http://arXiv.org/abs]{hep-th/0207022.}

\bibitem{gr-qc/9910089}
\textsc{G.~Amelino-Camelia}, \emph{Are we at the dawn of
quantum-gravity
  phenomenology?}, Lect. Notes Phys., \textbf{541}: (2000), 1--49,
  \eprint[http://arXiv.org/abs]{gr-qc/9910089.}

\bibitem{astro-ph/0004225}
\textsc{N.~E. Mavromatos}, \emph{The quest for quantum gravity:
Testing times
  for theories?}, \eprint[http://arXiv.org/abs]{astro-ph/0004225.}

\bibitem{gr-qc/0204051}
\textsc{G.~Amelino-Camelia}, \emph{Quantum-gravity phenomenology:
Status and
  prospects}, Mod. Phys. Lett., \textbf{A17}: (2002), 899--922,
  \eprint[http://arXiv.org/abs]{gr-qc/0204051.}

\bibitem{gr-qc/0204092}
\textsc{S.~Sarkar}, \emph{Possible astrophysical probes of quantum
gravity},
  Mod. Phys. Lett., \textbf{A17}: (2002), 1025--1036,
  \eprint[http://arXiv.org/abs]{gr-qc/0204092.}

\bibitem{gr-qc/0205121}
\textsc{D.~V. Ahluwalia}, \emph{Interface of gravitational and
quantum realms},
  Mod. Phys. Lett., \textbf{A17}: (2002), 1135,
  \eprint[http://arXiv.org/abs]{gr-qc/0205121.}

\bibitem{astro-ph/9712103}
\textsc{G.~Amelino-Camelia}, \textsc{J.~R. Ellis}, \textsc{N.~E.
Mavromatos},
  \textsc{D.~V. Nanopoulos}, \textsc{S.~Sarkar}, \emph{Potential Sensitivity of
  Gamma-Ray Burster Observations to Wave Dispersion in Vacuo}, Nature,
  \textbf{393}: (1998), 763--765,
  \eprint[http://arXiv.org/abs]{astro-ph/9712103.}

\bibitem{gr-qc/9810044}
\textsc{S.~D. Biller}, \textsc{et~al.}, \emph{Limits to quantum
gravity effects
  from observations of TeV flares in active galaxies}, Phys. Rev. Lett.,
  \textbf{83}: (1999), 2108--2111,
  \eprint[http://arXiv.org/abs]{gr-qc/9810044.}

\bibitem{gr-qc/9808029}
\textsc{G.~Amelino-Camelia}, \emph{Gravity-wave interferometers as
  quantum-gravity detectors}, Nature, \textbf{398}: (1999), 216--218,
  \eprint[http://arXiv.org/abs]{gr-qc/9808029.}

\bibitem{astro-ph/9904164}
\textsc{T.~Kifune}, \emph{Invariance violation extends the cosmic
ray
  horizon?}, Astrophys. J., \textbf{518}: (1999), L21--L24,
  \eprint[http://arXiv.org/abs]{astro-ph/9904164.}

\bibitem{astro-ph/0008107}
\textsc{G.~Amelino-Camelia}, \textsc{T.~Piran}, \emph{Planck-scale
deformation
  of Lorentz symmetry as a solution to the UHECR and the TeV-gamma paradoxes},
  Phys. Rev., \textbf{D64}: (2001), 036005,
  \eprint[http://arXiv.org/abs]{astro-ph/0008107.}

\bibitem{gr-qc/0107086}
\textsc{G.~Amelino-Camelia}, \emph{Space-time quantum solves three
experimental
  paradoxes}, Phys. Lett., \textbf{B528}: (2002), 181--187,
  \eprint[http://arXiv.org/abs]{gr-qc/0107086.}

\bibitem{gr-qc/0006061}
\textsc{C.~Rovelli}, \emph{Notes for a brief history of quantum
gravity},
  \eprint[http://arXiv.org/abs]{gr-qc/0006061.}

\bibitem{gr-qc/0110034}
\textsc{T.~Thiemann}, \emph{Introduction to modern canonical
quantum general
  relativity}, \eprint[http://arXiv.org/abs]{gr-qc/0110034.}

\bibitem{astro-ph/0009271}
\textsc{A.~de~Angelis}, \emph{GLAST, the gamma-ray large area
space telescope},
  \eprint[http://arXiv.org/abs]{astro-ph/0009271.}

\bibitem{hep-th/0109191}
\textsc{G.~Amelino-Camelia}, \textsc{L.~Doplicher}, \textsc{S.-k.
Nam},
  \textsc{Y.-S. Seo}, \emph{Phenomenology of particle production and
  propagation in string-motivated canonical noncommutative spacetime},
  \eprint[http://arXiv.org/abs]{hep-th/0109191.}

\bibitem{hep-th/0107054}
\textsc{J.~Kowalski-Glikman}, \emph{Planck-scale relativity from
quantum
  kappa-Poincare algebra}, Mod. Phys. Lett., \textbf{A17}: (2002), 1--12,
  \eprint[http://arXiv.org/abs]{hep-th/0107054.}

\bibitem{gr-qc/0111056}
\textsc{T.~Tamaki}, \textsc{T.~Harada}, \textsc{U.~Miyamoto},
  \textsc{T.~Torii}, \emph{Have we already detected astrophysical symptoms of
  space- time noncommutativity ?}, Phys. Rev., \textbf{D65}: (2002), 083003,
  \eprint[http://arXiv.org/abs]{gr-qc/0111056.}

\bibitem{hep-th/0005129}
\textsc{J.~Gomis}, \textsc{T.~Mehen}, \emph{Space-time
noncommutative field
  theories and unitarity}, Nucl. Phys., \textbf{B591}: (2000), 265--276,
  \eprint[http://arXiv.org/abs]{hep-th/0005129.}

\bibitem{hep-th/0005040}
\textsc{N.~Seiberg}, \textsc{L.~Susskind}, \textsc{N.~Toumbas},
\emph{Strings
  in background electric field, space/time noncommutativity and a new
  noncritical string theory}, JHEP, \textbf{06}: (2000), 021,
  \eprint[http://arXiv.org/abs]{hep-th/0005040.}

\bibitem{hep-th/0201222}
\textsc{D.~Bahns}, \textsc{S.~Doplicher}, \textsc{K.~Fredenhagen},
  \textsc{G.~Piacitelli}, \emph{On the unitarity problem in space/time
  noncommutative theories}, Phys. Lett., \textbf{B533}: (2002), 178--181,
  \eprint[http://arXiv.org/abs]{hep-th/0201222.}

\bibitem{hep-th/0206011}
\textsc{Y.~Liao}, \textsc{K.~Sibold}, \emph{Time-ordered
perturbation theory on
  noncommutative spacetime. II. Unitarity}, Eur. Phys. J., \textbf{C25}:
  (2002), 479--486, \eprint[http://arXiv.org/abs]{hep-th/0206011.}

\bibitem{hep-th/0201144}
\textsc{C.-S. Chu}, \textsc{J.~Lukierski}, \textsc{W.~J.
Zakrzewski},
  \emph{Hermitian analyticity, IR/UV mixing and unitarity of noncommutative
  field theories}, Nucl. Phys., \textbf{B632}: (2002), 219--239,
  \eprint[http://arXiv.org/abs]{hep-th/0201144.}

\bibitem{hep-th/9912072}
\textsc{S.~Minwalla}, \textsc{M.~Van~Raamsdonk},
\textsc{N.~Seiberg},
  \emph{Noncommutative perturbative dynamics}, JHEP, \textbf{02}: (2000), 020,
  \eprint[http://arXiv.org/abs]{hep-th/9912072.}

\bibitem{hep-th/0002075}
\textsc{A.~Matusis}, \textsc{L.~Susskind}, \textsc{N.~Toumbas},
\emph{The IR/UV
  connection in the non-commutative gauge theories}, JHEP, \textbf{12}: (2000),
  002, \eprint[http://arXiv.org/abs]{hep-th/0002075.}

\bibitem{hep-th/0005272}
\textsc{H.~O. Girotti}, \textsc{M.~Gomes}, \textsc{V.~O.
Rivelles},
  \textsc{A.~J. da~Silva}, \emph{A consistent noncommutative field theory: The
  Wess-Zumino model}, Nucl. Phys., \textbf{B587}: (2000), 299--310,
  \eprint[http://arXiv.org/abs]{hep-th/0005272.}

\bibitem{hep-th/0011218}
\textsc{V.~V. Khoze}, \textsc{G.~Travaglini}, \emph{Wilsonian
effective actions
  and the IR/UV mixing in noncommutative gauge theories}, JHEP, \textbf{01}:
  (2001), 026, \eprint[http://arXiv.org/abs]{hep-th/0011218.}

\bibitem{hep-th/0105082}
\textsc{S.~M. Carroll}, \textsc{J.~A. Harvey}, \textsc{V.~A.
Kostelecky},
  \textsc{C.~D. Lane}, \textsc{T.~Okamoto}, \emph{Noncommutative field theory
  and Lorentz violation}, Phys. Rev. Lett., \textbf{87}: (2001), 141601,
  \eprint[http://arXiv.org/abs]{hep-th/0105082.}

\bibitem{hep-ph/0112003}
\textsc{J.~L. Hewett}, \textsc{F.~J. Petriello}, \textsc{T.~G.
Rizzo},
  \emph{Non-commutativity and unitarity violation in gauge boson scattering},
  Phys. Rev., \textbf{D66}: (2002), 036001,
  \eprint[http://arXiv.org/abs]{hep-ph/0112003.}

\bibitem{hep-ph/0106356}
\textsc{A.~Anisimov}, \textsc{T.~Banks}, \textsc{M.~Dine},
  \textsc{M.~Graesser}, \emph{Comments on non-commutative phenomenology}, Phys.
  Rev., \textbf{D65}: (2002), 085032,
  \eprint[http://arXiv.org/abs]{hep-ph/0106356.}

\bibitem{hep-ph/0205040}
\textsc{I.~Hinchliffe}, \textsc{N.~Kersting}, \emph{Review of the
phenomenology
  of noncommutative geometry}, \eprint[http://arXiv.org/abs]{hep-ph/0205040.}

\bibitem{CJT}
\textsc{J.~Cornwall}, \textsc{R.~Jackiw}, \textsc{E.~Tomboulis},
  \emph{Effective action for composite operators}, Phys. Rev., \textbf{D10}:
  (1974), 2428--2445.

\bibitem{hep-ph/9311324}
\textsc{R.~Jackiw}, \textsc{G.~Amelino-Camelia}, \emph{Field
theoretical
  background for thermal physics},
  \eprint[http://arXiv.org/abs]{hep-ph/9311324.}

\bibitem{hep-ph/9211211}
\textsc{G.~Amelino-Camelia}, \textsc{S.-Y. Pi},
\emph{Selfconsistent
  improvement of the finite temperature effective potential}, Phys. Rev.,
  \textbf{D47}: (1993), 2356--2362,
  \eprint[http://arXiv.org/abs]{hep-ph/9211211.}

\bibitem{hep-ph/9305222}
\textsc{G.~Amelino-Camelia}, \emph{Selfconsistently improved
finite temperature
  effective potential for gauge theories}, Phys. Rev., \textbf{D49}: (1994),
  2740--2751, \eprint[http://arXiv.org/abs]{hep-ph/9305222.}

\bibitem{hep-th/0211022}
\textsc{G.~Amelino-Camelia}, \textsc{F.~D'Andrea},
\textsc{G.~Mandanici},
  \emph{Group velocity in noncommutative spacetime},
  \eprint[http://arXiv.org/abs]{hep-th/0211022.}

\bibitem{hep-th/0209254}
\textsc{G.~Amelino-Camelia}, \textsc{G.~Mandanici},
\textsc{K.~Yoshida},
  \emph{On the IR / UV mixing and experimental limits on the parameters of
  canonical noncommutative spacetimes},
  \eprint[http://arXiv.org/abs]{hep-th/0209254.}

\bibitem{gmgacld}
\textsc{G.~Mandanici}, \emph{Cornwall-Jackiw-Tomboulis effective
potential for canonical noncommutative field theories},
  \eprint[http://arXiv.org/abs]{hep-th/0304090.}

\bibitem{sny}
\textsc{H.~Snyder}, \emph{Quantized space-time}, Phys. Rev.,
\textbf{71}:
  (1947), 38.

\bibitem{ven}
\textsc{G.~Veneziano}, \emph{A stringy nature needs just two
constants},
  Europhys. Lett., \textbf{2}: (1986), 199.

\bibitem{grmen}
\textsc{D.~J. Gross}, \textsc{P.~F. Mende}, \emph{String theory
beyond the Planck scale}, Nucl. Phys., \textbf{B303}: (1988), 407.

\bibitem{amcive}
\textsc{D.~Amati}, \textsc{M.~Ciafaloni}, \textsc{G.~Veneziano},
\emph{Can space-time be probed below the string size?}, Phys.
Lett., \textbf{B216}:
  (1989), 41.

\bibitem{gr-qc/9403008}
\textsc{L.~J. Garay}, \emph{Quantum gravity and minimum length},
Int. J. Mod.
  Phys., \textbf{A10}: (1995), 145--166,
  \eprint[http://arXiv.org/abs]{gr-qc/9403008.}

\bibitem{hep-th/9301067}
\textsc{M.~Maggiore}, \emph{A Generalized uncertainty principle in
quantum
  gravity}, Phys. Lett., \textbf{B304}: (1993), 65--69,
  \eprint[http://arXiv.org/abs]{hep-th/9301067.}

\bibitem{dofrero}
\textsc{S.~Doplicher}, \textsc{K.~Fredenhagen}, \textsc{J.~E.
Roberts},
  \emph{The Quantum structure of space-time at the Planck scale and quantum
  fields}, Commun. Math. Phys., \textbf{172}: (1995), 187--220.

\bibitem{hep-th/0004074}
\textsc{T.~Yoneya}, \emph{String theory and space-time uncertainty
principle},
  Prog. Theor. Phys., \textbf{103}: (2000), 1081--1125,
  \eprint[http://arXiv.org/abs]{hep-th/0004074.}

\bibitem{hep-th/0110057}
\textsc{R.~Jackiw}, \emph{Physical instances of noncommuting
coordinates},
  Nucl. Phys. Proc. Suppl., \textbf{108}: (2002), 30--36,
  \eprint[http://arXiv.org/abs]{hep-th/0110057.}

\bibitem{hep-th/0012238}
\textsc{G.~Amelino-Camelia}, \emph{Testable scenario for
relativity with
  minimum-length}, Phys. Lett., \textbf{B510}: (2001), 255--263,
  \eprint[http://arXiv.org/abs]{hep-th/0012238.}

\bibitem{gr-qc/0106004}
\textsc{G.~Amelino-Camelia}, \emph{Status of Relativity with
  observer-independent length and velocity scales},
  \eprint[http://arXiv.org/abs]{gr-qc/0106004.}

\bibitem{hep-th/0203040}
\textsc{J.~Kowalski-Glikman}, \textsc{S.~Nowak}, \emph{Doubly
special
  relativity theories as different bases of kappa-Poincare algebra}, Phys.
  Lett., \textbf{B539}: (2002), 126--132,
  \eprint[http://arXiv.org/abs]{hep-th/0203040.}

\bibitem{gr-qc/0012051}
\textsc{G.~Amelino-Camelia}, \emph{Relativity in space-times with
  short-distance structure governed by an observer-independent (Planckian)
  length scale}, Int. J. Mod. Phys., \textbf{D11}: (2002), 35--60,
  \eprint[http://arXiv.org/abs]{gr-qc/0012051.}

\bibitem{hep-th/0102098}
\textsc{J.~Kowalski-Glikman}, \emph{Observer independent quantum
of mass},
  Phys. Lett., \textbf{A286}: (2001), 391--394,
  \eprint[http://arXiv.org/abs]{hep-th/0102098.}

\bibitem{hep-th/0107039}
\textsc{N.~R. Bruno}, \textsc{G.~Amelino-Camelia},
  \textsc{J.~Kowalski-Glikman}, \emph{Deformed boost transformations that
  saturate at the Planck scale}, Phys. Lett., \textbf{B522}: (2001), 133--138,
  \eprint[http://arXiv.org/abs]{hep-th/0107039.}

\bibitem{B293}
\textsc{J.~Lukierski}, \textsc{A.~Nowicki}, \textsc{H.~Ruegg},
\emph{New
  quantum Poincare algebra and k deformed field theory}, Phys. Lett.,
  \textbf{B293}: (1992), 344--352.

\bibitem{majidBOOK}
\textsc{S.~Majid}, \emph{Foundations of quantum group theory},
C.U.P.,1995.

\bibitem{hep-th/0105120}
\textsc{G.~Amelino-Camelia}, \textsc{M.~Arzano}, \emph{Coproduct
and star
  product in field theories on Lie-algebra non-commutative space-times}, Phys.
  Rev., \textbf{D65}: (2002), 084044,
  \eprint[http://arXiv.org/abs]{hep-th/0105120.}

\bibitem{astro-ph/0006250}
\textsc{J.~Kowalski-Glikman}, \emph{Testing dispersion relations
of quantum
  kappa-Poincare algebra on cosmological ground}, Phys. Lett., \textbf{B499}:
  (2001), 1--8, \eprint[http://arXiv.org/abs]{astro-ph/0006250.}

\bibitem{gzk1}
\textsc{K.~Greisen}, Phys. Rev. Lett., \textbf{16}: (1966), 748.

\bibitem{gzk2}
\textsc{G.~Zatsepin}, \textsc{V.~Kuzmin}, Sov. Phys. JEPT Lett.,
\textbf{4}:
  (1966), 78.

\bibitem{astro-ph/9807193}
\textsc{M.~Takeda}, \textsc{et~al.}, \emph{Extension of the
cosmic-ray energy
  spectrum beyond the predicted Greisen-Zatsepin-Kuzmin cutoff}, Phys. Rev.
  Lett., \textbf{81}: (1998), 1163--1166,
  \eprint[http://arXiv.org/abs]{astro-ph/9807193.}

\bibitem{astro-ph/9903159}
\textsc{F.~A. Aharonian}, \textsc{P.~S. Coppi}, \emph{Simultaneous
X-Ray and
  Gamma-Ray Observations of TeV Blazars: Testing Synchro-Compton Emission
  Models and Probing the Infrared Extragalactic Background}, Astrophys. J.,
  \textbf{521}: (1999), L33, \eprint[http://arXiv.org/abs]{astro-ph/9903159.}

\bibitem{AMS}
\textsc{S.~P. Ahlen}, \textsc{et~al.}, \emph{An Antimatter
spectrometer in
  space}, Nucl. Instrum. Meth., \textbf{A350}: (1994), 351--367.

\bibitem{GLAST}
\textsc{E.~D. Bloom} (GLAST team), \emph{GLAST}, Space Sci. Rev.,
\textbf{75}:
  (1996), 109--125.

\bibitem{astro-ph/9912136}
\textsc{J.~P. Norris}, \textsc{J.~T. Bonnell}, \textsc{G.~F.
Marani},
  \textsc{J.~D. Scargle}, \emph{GLAST, GRBs, and quantum gravity},
  \eprint[http://arXiv.org/abs]{astro-ph/9912136.}

\bibitem{hep-th/9402037}
\textsc{A.~Kempf}, \textsc{S.~Majid}, \emph{Algebraic q
integration and Fourier
  theory on quantum and braided spaces}, J. Math. Phys., \textbf{35}: (1994),
  6802--6837, \eprint[http://arXiv.org/abs]{hep-th/9402037.}

\bibitem{gr-qc/0208002}
\textsc{T.~Tamaki}, \textsc{T.~Harada}, \textsc{U.~Miyamoto},
  \textsc{T.~Torii}, \emph{Particle velocity in noncommutative space-time},
  \eprint[http://arXiv.org/abs]{gr-qc/0208002.}

\bibitem{WeylMoyal2}
\textsc{H.~Weyl}, \emph{The Theory of Groups and Quantum
Mechanics}, Dover,
  1931.

\bibitem{WeylMoyal}
\textsc{J.~Moyal}, \emph{Quantum Mechanics as a Statistical
Theory}, Proc.
  Cambr. Phil. Soc., \textbf{45}: (1949), 99.

\bibitem{hep-th/0003160}
\textsc{R.~Gopakumar}, \textsc{S.~Minwalla},
\textsc{A.~Strominger},
  \emph{Noncommutative solitons}, JHEP, \textbf{05}: (2000), 020,
  \eprint[http://arXiv.org/abs]{hep-th/0003160.}

\bibitem{hep-th/0008057}
\textsc{A.~Micu}, \textsc{M.~M. Sheikh~Jabbari},
\emph{Noncommutative phi**4
  theory at two loops}, JHEP, \textbf{01}: (2001), 025,
  \eprint[http://arXiv.org/abs]{hep-th/0008057.}

\bibitem{hep-th/9902037}
\textsc{P.~Kosinski}, \textsc{J.~Lukierski}, \textsc{P.~Maslanka},
\emph{Local
  D = 4 field theory on kappa-deformed Minkowski space}, Phys. Rev.,
  \textbf{D62}: (2000), 025004, \eprint[http://arXiv.org/abs]{hep-th/9902037.}

\bibitem{hep-th/0205047}
\textsc{G.~Amelino-Camelia}, \textsc{M.~Arzano},
\textsc{L.~Doplicher},
  \emph{Field theories on canonical and Lie-algebra noncommutative spacetimes},
  \eprint[http://arXiv.org/abs]{hep-th/0205047.}

\bibitem{hep-th/0112252}
\textsc{J.~Lukierski}, \emph{From noncommutative space-time to
quantum
  relativistic symmetries with fundamental mass parameter},
  \eprint[http://arXiv.org/abs]{hep-th/0112252.}

\bibitem{hep-th/0103127}
\textsc{P.~Kosinski}, \textsc{J.~Lukierski}, \textsc{P.~Maslanka},
  \emph{kappa-deformed Wigner construction of relativistic wave functions and
  free fields on kappa-Minkowski space}, Nucl. Phys. Proc. Suppl.,
  \textbf{102}: (2001), 161--168,
  \eprint[http://arXiv.org/abs]{hep-th/0103127.}

\bibitem{hep-th/0104153}
\textsc{B.~Jurco}, \textsc{L.~Moller}, \textsc{S.~Schraml},
\textsc{P.~Schupp},
  \textsc{J.~Wess}, \emph{Construction of non-Abelian gauge theories on
  noncommutative spaces}, Eur. Phys. J., \textbf{C21}: (2001), 383--388,
  \eprint[http://arXiv.org/abs]{hep-th/0104153.}

\bibitem{hep-th/0005208}
\textsc{A.~Armoni}, \emph{Comments on perturbative dynamics of
non-commutative
  Yang- Mills theory}, Nucl. Phys., \textbf{B593}: (2001), 229--242,
  \eprint[http://arXiv.org/abs]{hep-th/0005208.}

\bibitem{hep-th/9903187}
\textsc{T.~Krajewski}, \textsc{R.~Wulkenhaar}, \emph{Perturbative
quantum gauge
  fields on the noncommutative torus}, Int. J. Mod. Phys., \textbf{A15}:
  (2000), 1011--1030, \eprint[http://arXiv.org/abs]{hep-th/9903187.}

\bibitem{hep-th/9903077}
\textsc{C.~P. Martin}, \textsc{D.~Sanchez-Ruiz}, \emph{The
one-loop UV
  divergent structure of U(1) Yang-Mills theory on noncommutative R**4}, Phys.
  Rev. Lett., \textbf{83}: (1999), 476--479,
  \eprint[http://arXiv.org/abs]{hep-th/9903077.}

\bibitem{hep-th/0001182}
\textsc{H.~Grosse}, \textsc{T.~Krajewski}, \textsc{R.~Wulkenhaar},
  \emph{Renormalization of noncommutative Yang-Mills theories: A simple
  example}, \eprint[http://arXiv.org/abs]{hep-th/0001182.}

\bibitem{hep-th/0002084}
\textsc{S.~Ferrara}, \textsc{M.~A. Lledo}, \emph{Some aspects of
deformations
  of supersymmetric field theories}, JHEP, \textbf{05}: (2000), 008,
  \eprint[http://arXiv.org/abs]{hep-th/0002084.}

\bibitem{hep-th/0002119}
\textsc{S.~Terashima}, \emph{A note on superfields and
noncommutative
  geometry}, Phys. Lett., \textbf{B482}: (2000), 276--282,
  \eprint[http://arXiv.org/abs]{hep-th/0002119.}

\bibitem{hep-th/0007050}
\textsc{A.~A. Bichl}, \textsc{et~al.}, \emph{The superfield
formalism applied
  to the noncommutative Wess-Zumino model}, JHEP, \textbf{10}: (2000), 046,
  \eprint[http://arXiv.org/abs]{hep-th/0007050.}

\bibitem{hep-th/0012009}
\textsc{D.~Zanon}, \emph{Noncommutative N = 1,2 super U(N)
Yang-Mills: UV/IR
  mixing and effective action results at one loop}, Phys. Lett., \textbf{B502}:
  (2001), 265--273, \eprint[http://arXiv.org/abs]{hep-th/0012009.}

\bibitem{WeBa}
\textsc{J.~Wess}, \textsc{J.~Bagger}, \emph{Supersymmetry and
supergravity},
  Princeton University Press, 1983.

\bibitem{hep-th/0009043}
\textsc{Y.~Yoshida}, \emph{Nonperturbative aspect in N = 2
supersymmetric
  noncommutative Yang-Mills theory},
  \eprint[http://arXiv.org/abs]{hep-th/0009043.}

\bibitem{hep-th/0009174}
\textsc{D.~Bellisai}, \textsc{J.~M. Isidro}, \textsc{M.~Matone},
\emph{On the
  structure of noncommutative N = 2 super Yang-Mills theory}, JHEP,
  \textbf{10}: (2000), 026, \eprint[http://arXiv.org/abs]{hep-th/0009174.}

\bibitem{hep-th/0203141}
\textsc{A.~A. Bichl}, \textsc{et~al.}, \emph{Non-commutative U(1)
  super-Yang-Mills theory: Perturbative self-energy corrections},
  \eprint[http://arXiv.org/abs]{hep-th/0203141.}

\bibitem{hep-th/0102007}
\textsc{A.~Armoni}, \textsc{R.~Minasian}, \textsc{S.~Theisen},
\emph{On
  non-commutative N = 2 super Yang-Mills}, Phys. Lett., \textbf{B513}: (2001),
  406--412, \eprint[http://arXiv.org/abs]{hep-th/0102007.}

\bibitem{hep-th/0009196}
\textsc{D.~Zanon}, \emph{Noncommutative perturbation in
superspace}, Phys.
  Lett., \textbf{B504}: (2001), 101--108,
  \eprint[http://arXiv.org/abs]{hep-th/0009196.}

\bibitem{hep-th/0010275}
\textsc{A.~Santambrogio}, \textsc{D.~Zanon}, \emph{One-loop
four-point function
  in noncommutative N = 4 Yang- Mills theory}, JHEP, \textbf{01}: (2001), 024,
  \eprint[http://arXiv.org/abs]{hep-th/0010275.}

\bibitem{hep-th/9907166}
\textsc{A.~Hashimoto}, \textsc{N.~Itzhaki}, \emph{Non-commutative
Yang-Mills
  and the AdS/CFT correspondence}, Phys. Lett., \textbf{B465}: (1999),
  142--147, \eprint[http://arXiv.org/abs]{hep-th/9907166.}

\bibitem{hep-th/0005015}
\textsc{N.~Seiberg}, \textsc{L.~Susskind}, \textsc{N.~Toumbas},
  \emph{Space/time non-commutativity and causality}, JHEP, \textbf{06}: (2000),
  044, \eprint[http://arXiv.org/abs]{hep-th/0005015.}

\bibitem{hep-th/0209253}
\textsc{H.~Bozkaya}, \textsc{et~al.}, \emph{Space/time
noncommutative field
  theories and causality}, \eprint[http://arXiv.org/abs]{hep-th/0209253.}

\bibitem{wilson}
\textsc{K.~G. Wilson}, \textsc{J.~B. Kogut}, \emph{The
Renormalization group
  and the epsilon expansion}, Phys. Rept., \textbf{12}: (1974), 75--200.

\bibitem{polchi}
\textsc{J.~Polchinski}, \emph{Renormalization and effective
lagrangians}, Nucl.
  Phys., \textbf{B231}: (1984), 269--295.

\bibitem{hep-th/9607188}
\textsc{C.~Becchi}, \emph{On the construction of renormalized
gauge theories
  using renormalization group techniques},
  \eprint[http://arXiv.org/abs]{hep-th/9607188.}

\bibitem{hep-th/0012111}
\textsc{S.~Arnone}, \textsc{S.~Chiantese}, \textsc{K.~Yoshida},
  \emph{Applications of exact renormalization group techniques to the
  non-perturbative study of supersymmetric gauge field theory}, Int. J. Mod.
  Phys., \textbf{A16}: (2001), 1811,
  \eprint[http://arXiv.org/abs]{hep-th/0012111.}

\bibitem{hep-th/0104217}
\textsc{L.~Griguolo}, \textsc{M.~Pietroni}, \emph{Wilsonian
renormalization
  group and the non-commutative IR/UV connection}, JHEP, \textbf{05}: (2001),
  032, \eprint[http://arXiv.org/abs]{hep-th/0104217.}

\bibitem{hep-th/0103199}
\textsc{E.~T. Akhmedov}, \textsc{P.~DeBoer}, \textsc{G.~W.
Semenoff},
  \emph{Non-commutative Gross-Neveu model at large N}, JHEP, \textbf{06}:
  (2001), 009, \eprint[http://arXiv.org/abs]{hep-th/0103199.}

\bibitem{hep-th/0003137}
\textsc{B.~A. Campbell}, \textsc{K.~Kaminsky},
\emph{Noncommutative field
  theory and spontaneous symmetry breaking}, Nucl. Phys., \textbf{B581}:
  (2000), 240--256, \eprint[http://arXiv.org/abs]{hep-th/0003137.}

\bibitem{hep-th/0102022}
\textsc{B.~A. Campbell}, \textsc{K.~Kaminsky},
\emph{Noncommutative linear
  sigma models}, Nucl. Phys., \textbf{B606}: (2001), 613--635,
  \eprint[http://arXiv.org/abs]{hep-th/0102022.}

\bibitem{hep-th/0202011}
\textsc{F.~Ruiz~Ruiz}, \emph{UV/IR mixing and the Goldstone
theorem in
  noncommutative field theory}, Nucl. Phys., \textbf{B637}: (2002), 143--167,
  \eprint[http://arXiv.org/abs]{hep-th/0202011.}

\bibitem{hep-th/0104106}
\textsc{S.~Sarkar}, \textsc{B.~Sathiapalan}, \emph{Comments on the
  renormalizability of the broken symmetry phase in noncommutative scalar field
  theory}, JHEP, \textbf{05}: (2001), 049,
  \eprint[http://arXiv.org/abs]{hep-th/0104106.}

\bibitem{hep-th/0202171}
\textsc{S.~Sarkar}, \emph{On the UV renormalizability of
noncommutative field
  theories}, JHEP, \textbf{06}: (2002), 003,
  \eprint[http://arXiv.org/abs]{hep-th/0202171.}

\bibitem{hep-th/0006119}
\textsc{S.~S. Gubser}, \textsc{S.~L. Sondhi}, \emph{Phase
structure of
  non-commutative scalar field theories}, Nucl. Phys., \textbf{B605}: (2001),
  395--424, \eprint[http://arXiv.org/abs]{hep-th/0006119.}

\bibitem{hep-ph/0010354}
\textsc{J.~L. Hewett}, \textsc{F.~J. Petriello}, \textsc{T.~G.
Rizzo},
  \emph{Signals for non-commutative interactions at linear colliders}, Phys.
  Rev., \textbf{D64}: (2001), 075012,
  \eprint[http://arXiv.org/abs]{hep-ph/0010354.}

\bibitem{gr-qc/0205125}
\textsc{G.~Amelino-Camelia}, \emph{On the fate of Lorentz symmetry
in loop
  quantum gravity and noncommutative spacetimes},
  \eprint[http://arXiv.org/abs]{gr-qc/0205125.}

\bibitem{hep-th/0205153}
\textsc{J.~M. Grimstrup}, \textsc{R.~Wulkenhaar},
\emph{Quantisation of
  theta-expanded non-commutative QED},
  \eprint[http://arXiv.org/abs]{hep-th/0205153.}

\bibitem{hep-ph/0112320}
\textsc{N.~G. Deshpande}, \textsc{X.-G. He}, \emph{Triple neutral
gauge boson
  couplings in noncommutative standard model}, Phys. Lett., \textbf{B533}:
  (2002), 116--120, \eprint[http://arXiv.org/abs]{hep-ph/0112320.}

\bibitem{hep-th/0205214}
\textsc{P.~Aschieri}, \textsc{B.~Jurco}, \textsc{P.~Schupp},
\textsc{J.~Wess},
  \emph{Non-commutative GUTs, standard model and C, P, T},
  \eprint[http://arXiv.org/abs]{hep-th/0205214.}

\bibitem{hep-ph/0202223}
\textsc{X.-G. He}, \emph{Strong, electroweak interactions and
their unification
  with noncommutative space-time},
  \eprint[http://arXiv.org/abs]{hep-ph/0202223.}

\bibitem{hep-ph/0107291}
\textsc{C.~E. Carlson}, \textsc{C.~D. Carone}, \textsc{R.~F.
Lebed},
  \emph{Bounding noncommutative QCD}, Phys. Lett., \textbf{B518}: (2001),
  201--206, \eprint[http://arXiv.org/abs]{hep-ph/0107291.}

\bibitem{hep-ph/0202121}
\textsc{W.~Behr}, \textsc{et~al.}, \emph{The Z --> gamma gamma, g
g decays in
  the noncommutative standard model},
  \eprint[http://arXiv.org/abs]{hep-ph/0202121.}

\bibitem{hep-th/9912075}
\textsc{I.~Y. Aref'eva}, \textsc{D.~M. Belov}, \textsc{A.~S.
Koshelev},
  \emph{Two-loop diagrams in noncommutative phi**4(4) theory}, Phys. Lett.,
  \textbf{B476}: (2000), 431--436,
  \eprint[http://arXiv.org/abs]{hep-th/9912075.}

\bibitem{gr-qc/9809038}
\textsc{R.~Gambini}, \textsc{J.~Pullin}, \emph{Nonstandard optics
from quantum
  spacetime}, Phys. Rev., \textbf{D59}: (1999), 124021,
  \eprint[http://arXiv.org/abs]{gr-qc/9809038.}

\end{thebibliography}

\end{document}